\documentclass[a4paper,fleqn,usenatbib,,threeparttable]{mnras}

\usepackage[T1]{fontenc}
\usepackage{ae,aecompl}

\usepackage{graphicx}	
\usepackage{amsmath}	
\usepackage{amssymb}	
\usepackage{xspace}

\usepackage{natbib}
\usepackage[usenames]{color}
\usepackage{subfig} 
\usepackage{multirow}
\usepackage{array}
\usepackage{verbatim}
\usepackage{booktabs,caption,fixltx2e} 
\usepackage[flushleft]{threeparttable}
\usepackage{color}

\title[Fidelity of Reconstructed Lensed AGN Hosts]{H0LiCOW VI. Testing the fidelity of lensed quasar host galaxy reconstruction}

\author[Xuheng Ding et al.]{\parbox{\textwidth}{
Xuheng Ding,$^{1,2}$\thanks{E-mail: dxh@astro.ucla.edu}
Kai Liao,$^{2}$
Tommaso Treu,$^{2}$
Sherry H. Suyu,$^{3,4,5}$
Geoff C.-F. Chen, $^{6,4}$
Matthew W. Auger, $^{7}$
Philip J. Marshall,$^{8}$
Adriano Agnello,$^{2,9}$
Frederic Courbin, $^{10}$
Anna M. Nierenberg,$^{11}$
Cristian E. Rusu,$^{6}$
Dominique Sluse,$^{12}$
Alessandro Sonnenfeld,$^{13}$
Kenneth C. Wong$^{14,4}$
}
\\
\\
\parbox{\textwidth}{
$^{1}$Department of Astronomy, Beijing Normal University, Beijing
100875, China\\
$^{2}$Department of Physics and Astronomy, University of California,
Los Angeles, CA, 90095-1547, USA\\
$^{3}$Max-Planck-Institut f{\"u}r Astrophysik, Karl-Schwarzschild-Str.~1,
85748 Garching, Germany\\
$^{4}$Institute of Astronomy and Astrophysics, Academia Sinica,
P.O.~Box 23-141, Taipei 10617, Taiwan\\
$^{5}$Physik-Department, Technische Universit\"at M\"unchen, James-Franck-Stra\ss{}e~1, 85748 Garching, Germany\\
$^{6}$Department of Physics, University of California, Davis, CA 95616, USA\\
$^{7}$Institute of Astronomy, University of Cambridge, Madingley Road,
Cambridge CB3 0HA, UK\\
$^{8}$Kavli Institute for Particle Astrophysics and Cosmology,
Stanford University, 452 Lomita Mall, Stanford, CA 94035, USA\\
$^{9}$European Southern Observatories, Karl-Schwarzschild-Str.~2,
85748 Garching, Germany\\
$^{10}$Laboratoire d'Astrophysique, Ecole Polytechnique
F{\'e}d{\'e}rale de Lausanne (EPFL), Observatoire de Sauverny, CH-1290
Versoix, Switzerland\\
$^{11}$Center for Cosmology and Astro-Particle Physics, The Ohio
State University, Columbus OH 43210, USA \\
$^{12}$STAR Institute, Quartier Agora - All\'ee du six Ao\^ut, 19c
B-4000 Li\`ege, Belgium\\
$^{13}$Kavli IPMU (WPI), UTIAS, The University of Tokyo, Kashiwa,
Chiba 277-8583, Japan\\
$^{14}$National Astronomical Observatory of Japan, 2-21-1 Osawa,
Mitaka, Tokyo 181-8588, Japan\\
}}

\date{Accepted XXX. Received YYY; in original form ZZZ}

\pubyear{2016}

\begin{document}

\def\mbh{{$\mathcal M_{\rm BH}$}}
\def\efr{{$R_{\mathrm{eff}}$}}
\def\mstar{{$\mathcal M_*$}}
\def\lbulge{{$L_{\rm bulge}$}}
\def\lhost{{$L_{\rm host}$}}
\def\glee{{\sc Glee}}
\def\hst{\textit{HST}}

\label{firstpage}
\pagerange{\pageref{firstpage}--\pageref{lastpage}}
\maketitle

\begin{abstract}
The empirical correlation between the mass of a super-massive black
hole (\mbh) and its host galaxy properties is widely considered to be
evidence of their co-evolution. A powerful way to test the
co-evolution scenario and learn about the feedback processes linking
galaxies and nuclear activity is to measure
these correlations as a function of redshift. Unfortunately, currently \mbh\ can only be estimated in active galaxies at cosmological distances. At these
distances, bright active galactic nuclei (AGN) can outshine the host
galaxy, making it extremely difficult to measure the host's
luminosity. Strongly lensed AGNs provide in principle a great
opportunity to improve the sensitivity and accuracy of the host galaxy
luminosity measurements as the host galaxy is magnified and more
easily separated from the point source, provided the lens model is
sufficiently accurate. In order to measure
the \mbh $-L$ correlation
with strong lensing, it is necessary to ensure that the lens modelling
is accurate, and that the host galaxy luminosity can be recovered to
at least a precision and accuracy better than that of the typical
\mbh\ measurement. We carry out extensive and realistic simulations
of deep \textit{Hubble Space Telescope} observations of lensed AGNs
obtained by our collaboration. We show that the host galaxy luminosity
can be recovered with better accuracy and precision than the typical
uncertainty on \mbh ($\sim0.5$ dex) for hosts as faint as 2$-$4
magnitudes dimmer than the AGN itself. Our simulations will be used to
estimate bias and uncertainties on the actual measurements to be
presented in a future paper.
\end{abstract}




\section{Introduction}

In the past two decades, tight correlations have been discovered
between supermassive black hole mass (\mbh) and the properties of the
host galaxy, including luminosity (\lbulge), stellar mass (\mstar) and
stellar velocity dispersion ($\sigma_*$)
\citep[e.g.,][]{Mag++98,F+M00,Geb++01b,M+H03,H+R04,Gul++09,Gra++2011,Beifi2012,Park15,Kormendy:2013p30943}.

These tight correlations are usually considered to be the result of
physical coupling between supermassive BH growth and galaxy evolution
\cite[see, however,][for a different view]{Peng:2007p22771}. 
The origin of this co-evolution is still an open question, central for
our understanding of the role of active galactic nuclei (AGN) feedback
in galaxy formation and evolution. One key to study this coupling is
to trace the correlations to higher redshift, determining how and when
they emerged and evolved
\citep[e.g.,][]{TMB04,Sal++06,Woo++06,Jah++09,DeG++15}.

Unfortunately, there are challenges in mapping these correlations
beyond the local universe (i.e. redshift $z\sim0.1$). At cosmologically
significant look-back times, the sphere of influence of the black hole
cannot be directly resolved, forcing\footnote{In very rare cases
gravitational lensing can also be used to estimate the black hole mass
using the properties of the central image \citep{Qui++16,Ken++15,WRK04,Tamura2015}.} one to
study active black holes, where \mbh\ can be estimated indirectly
using the properties of the broad emission lines
\citep[see recent reviews
by][]{Shen:2013p29308,Peterson2014SSRv}. However, the presence of a
bright point source makes it very hard to study the properties of the
host galaxy, which is typically barely resolved and faint owing to
cosmological surface brightness dimming. For these reasons, much
attention so far has been devoted to relatively faint AGN (Seyfert 1s)
where the brightness of the point source is comparable to that of the
host \citep{Woo++06,Tre++07,Ben++11,Park15}, and very little is known
about the hosts of the most luminous AGNs, hosting the most massive
black holes \cite[see, e.g.,][for a few exceptions]{Walter:2004p35}.

Strong gravitational lensing \citep[see, e.g.,][for
reviews]{CSS02,SKW06,Tre10} provides a unique opportunity to make
progress in the study of the luminosity of the host galaxies of the
most luminous AGN in the distant universe. Even with lensing
magnification, the AGN is unresolved at \textit{Hubble Space
Telescope} (\hst) resolution, while the host galaxy is typically well 
resolved. \citet{Pen++06qsob} and
\citet{Pen++06qsoa} first studied the \mbh -\lhost\ relation using
gravitational lensed AGN image out to redshift $2-4$, finding that
host galaxies were typically underluminous compared to the local
universe, when taking the evolution of stellar populations into
account, provided that systematic uncertainties can be controlled.

Given the advances in both the quality of the data
and lensing modelling techniques in the past decade, it is time to
revisit this measurement. In this paper we present extensive and
realistic simulations of ultradeep \hst\ images of lensed AGNs
obtained with the goal of measuring cosmological parameters from
gravitational time delays. We then analyze the simulated images with
the state of the art code \glee\footnote{Developed by \citet{S+H10}
based on \citet{Suy++06} and \citet{Halk2008}} as if they were real
images, in order to quantify realistic uncertainties and uncover
biases and systematic errors associated with the measurement of the
lensed host galaxy luminosity. The data analyzed here represent a
dramatic improvement over the NICMOS data used a decade ago: not only
is WFC3-IR at least a factor of 10 faster (in terms of time required
to reach a certain depth), but our exposure times are typically 5$-$10
longer for each object.

Likewise, \glee\ has several advantages with respect to the lensing
code used by \citet{Pen++06qsob}. First, the host galaxy is not
assumed to have a regular surface brightness profile, but is modelled
as a grid of surface brightness pixels obtained through regularized
linear inversion. Thus, it has the flexibility to describe the complex
morphologies observed in some cases \citep{Suy++13,Suy++14}. This
allows the observer to choose the most appropriate parametrized form to
describe the surface brightness distribution of each specific
system. Second, the deflector mass distribution can be described with
models that are more flexible than singular isothermal ellipsoid thus
eliminating a potential source of bias in the host galaxy luminosity
measurement
\citep[e.g.][]{Mar++07}.

In addition to quantifying bias and uncertainties for each system in
our sample, we study trends with lensing configuration, host galaxy
brightness and host galaxy to AGN contrast, which should be general
enough to provide guidance for future studies. We note that the \mbh\
estimates are considered to be uncertain at the $\sim0.5$ dex level
even in the local universe \citep{V+P06}. Thus, the natural goal of
this simulation exercise is to find out whether the random error and
bias on \lhost\ can be constrained or corrected to a level that is
smaller than 1.25 magnitudes. In a future paper we will apply the
results of our simulations to the actual data and combine them with
spectroscopy to construct the
\mbh-\lhost\ relation.

The paper is organized as follows. We briefly summarize the properties
of the data set in Section~\ref{sec:data}. We describe our simulation
pipeline in Section~\ref{sec:sim}. In section~\ref{sec:rest} we
present our main results. Discussion and summary are presented in
Section~\ref{sec:disc} and~\ref{sec:sum}, respectively.
Magnitudes are given in the AB System. 

\section{Data} 
\label{sec:data}

The simulations are intended to reproduce as closely as possible the
deep \hst\ observations available for a sample of eight lenses
currently being analyzed by our team. More information and references
for each system are given in Table~\ref{data_set}. The real data are
taken from three \hst\ programs and are described in detail
elsewhere. Briefly, the data for RXJ1131 were obtained as part of \hst\
program (GO-9744; PI: Kochanek) and, together with data for three
lenses from program \hst-GO-12889 (PI: Suyu), they form the bulk of the
$H_0$ Lenses in COSMOGRAIL's Wellspring (H0LiCOW) sample. We refer to
\citet{Suy++13,H0licow1} for a description of these data. Four systems
have been imaged in cycle 23 as part of Program \hst-GO-14254 (PI:
Treu). The cycle 23 sample was selected to have measured time delays
from the COSMOGRAIL collaboration \citep{Eig++05}
\footnote{\url{http://cosmograil.epfl.ch/}}, confirmed to have
extended host galaxy features in near infrared adaptive optics images
taken at the Keck Telescope (PI: Treu) \citep{Agn++16}, before being
imaged with $HST$. More details on these observations will be given
by Agnello et al. (2017, in preparation). We note that we do not
consider the system B1608+656 in this work even though it is an
integral part of the H0LiCOW sample. That system is different from the
other eight considered here in that the lensed source does not host a
bright AGN, but it is rather an active nucleus without broad emission
lines \citep{Fas++96}. Therefore broad line based \mbh\ estimates
cannot be obtained for this system.


\begin{table*}
\centering
\caption{Summary of lensed AGN properties and observational setup.}\label{data_set}
\resizebox{18cm}{!}{
     \begin{tabular}{crrrrrrrrccr}
        \toprule
        Name & $z_s$ & camera & filter & exposure & RA & DEC &\hst\        & PI & AGN & Approximate & References \\ 
                  &            &              &        &time(s)       &      &          & program && image    &Magnification &   \\ 
        \hline\hline\\     
        HE0435$-$1223  & 1.69 & WFC3-IR & F160W & 9940 & 04:38:14.90 &	-12:17:14.4 & 12889 &  S.~H.~Suyu & 4 & 4.0 & (1), (2)\\
        HE1104$-$1805  & 2.32 & WFC3-IR & F160W & 14700& 11:06:33.45 &	-18:21:24.2 & 12889 &  S.~H.~Suyu & 2 & 2.2& (3), (4)\\
        SDSS1206$+$4332& 1.79 & WFC3-IR & F160W & 8460 & 12:06:29.65 &    +43:32:17.6 & 14254 & T.~Treu & 2/4 & 3.0& (5), (6)\\
        WFI2033$-$4723 & 1.66 & WFC3-IR & F160W & 14470& 20:33:42.08 &	-47:23:43.0 & 12889 &  S.~H.~Suyu & 4 & 3.2 & (7), (2)\\
        HE0047$-$1756  & 1.66 & WFC3-UVIS& F814W& 9710 & 00:50:27.83 &    -17:40:08.8 & 14254 & T.~Treu & 2 & 2.8 & (8), (9)\\
        SDSS0246$-$0825& 1.68 & WFC3-UVIS& F814W& 9680 & 02:46:34.11 &	-08:25:36.2 & 14254 & T.~Treu & 2 & 2.8& (10), (11)\\
        HS 2209$+$1914 & 1.07 & WFC3-UVIS& F814W& 14240& 22:11:30.30 &	+19:29:12.0 & 14254 & T.~Treu & 2 & 2.6& (12), (13)\\
        RX J1131$-$1231& 0.65 & ACS      & F814W& 2085 & 11:31:51.6  &    -12:31:57   &  9744 & C.S.~Kochanek & 4 & 4.3 &(14), (15)\\
        \bottomrule
     \end{tabular}}
     \begin{tablenotes}
      \small 
       \item {Note:$-$ For conciseness, we abbreviate each lens name to four digits (e.g HE0435$-$1223 to HE0435). Approximate magnification is given in magnitudes.\\
       References:$-$ (1) \cite{HE0435_discover}; (2) \cite{Sluse++2012}; (3) \cite{HE1104_discover}; (4) \cite{1104_d_redshift}; (5) \cite{1206_discover}
       (6) \cite{Eul++13}; (7) \cite{2033_discover}; (8) \cite{0047_discover}; (9) \cite{Ofe++06}; (10) \cite{0246_discover}; (11) \cite{0246_redshift};
       (12) \cite{2209_discover};  (13) \cite{2209_more}; (14) \cite{1131_discover}; (15) \cite{1131_redshift}}.
     \end{tablenotes}
\end{table*}

\section{Simulations} 
\label{sec:sim}
The purpose of this section is to describe simulations of lens systems
that have similar host luminosity and lensing parameters to the real
data. Conceptually, the following steps need to be taken in order to
carry out a realistic simulation of strongly lensed AGN.

\begin{enumerate}
\item Compute high resolution simulated images of the lensed AGN, host, and deflector light.
\item Convolve with the telescope point spread function (PSF) and sample according to the pixel size of the camera.
\item Rebin the high resolution image to the actual data resolution.
\item Add noise.
\item Repeat steps (iii) and (iv) at different positions to simulate dithering and drizzle the individual images to recover
some of the resolution lost to undersampling as is done with the real observations. This step introduces correlated noise.
\end{enumerate}

In the remainder of this section we describe these steps in detail,
referring to the specific case of HE0435$-$1223 as an example, when
necessary. More details about this system are available in five papers
detailing the cosmographic analysis
\citep{H0licow1,H0licow2,H0licow3,H0licow4,H0licow5}.

\subsection{Simulating high resolution noiseless images}
\label{ssec:highres}

We start by generating a high resolution PSF
using {\sc Tiny Tim}\footnote{\url{http://www.stsci.edu/hst/observatory/focus/TinyTim}}.
In practice, {\sc Tiny Tim} will not yield a sufficiently good model
of the PSF for our application. However, for the purpose of this
simulation exercise we assume that we know the PSF, and that the
details of its shape are irrelevant as long as they possess a
sufficiently realistic description of \hst-quality PSF profiles. We
have developed an iterative PSF modelling process which can accurately
recover the PSF for real \hst\ observations \citep[][Suyu et al.~in
preparation]{Chen2016,H0licow4}. This iterative process will be used
when analyzing the real lens systems, and the residual uncertainties
taken into account when measuring the host galaxy luminosity. An
alternative scheme for modelling the PSF in ground based data is
described by \citet{Rusu++2016} and applied to adaptive optics
observations of gravitationally lensed AGNs from SDSS Quasar Lens
Search (SQLS).

In the case of HE0435, the high resolution PSF was oversampled by a
linear factor of 4, i.e. using a pixel size of
$0\farcs13/4=0\farcs0325$ in steps (i) and (ii) listed above. The same
factor for pixel size is used for the other lenses imaged with WFC3-IR
except for HE1104, for which we used $0\farcs13/6=0\farcs0217$, in
order to simulate the actual dither pattern used during the
observations (see Section~\ref{ssec:dither} for details).

Once the PSF is generated, we simulate the lensing system itself. The
surface brightness is described as a hybrid model \citep{Suyu12}
comprised of $(1)$ point sources for the lensed AGN in the image
plane, $(2)$ a simply-parametrized model for the deflector surface
brightness in the image plane, and $(3)$ a simply-parametrized model
for the host galaxy in the source plane. 

For simplicity, in this study we adopt a single \citet{Ser68} profile
to describe the surface brightness profile of the host (step (i)) 
and of the deflector. This S\'ersic profile is commonly used in the
literature and provides a reasonably accurate description of the
surface brightness of galaxies, ranging from exponential disks to
\citet{deV48} profiles. In future papers, when analyzing real systems, 
we will examine the pixellated reconstructed surface brightness
profile of each host galaxy and identify the most appropriate
parametrization. Likewise we will adopt different (or multiple)
surface brightness profiles to describe the deflector light, if
warranted by the data as in the case of RXJ1131
\citep{Suy++13,Suy++14}. We note that choosing a simply-parametrized
description of the surface brightness pixels is necessary in order to
obtain results directly comparable with the rest of the literature
and also to extrapolate the observed flux to an infinite aperture, as
it is common practice (the extrapolation is typically 
several tens of percent in total luminosity,
as the reconstructed source image does not extend much further
than the effective radius).

The \citet{Ser68} profile is parametrized as follows:
\begin{eqnarray}
   \label{eq:sersic}
   &I(R) = I_{\mathrm{e}} \exp\left[-k\left(\left(\frac{R}{R_{\mathrm{eff}}}\right)^{1/\mathrm{n}}-1\right)\right] ,\\
   &R(x,y,q) = \sqrt{qx^2+y^2/q}.
\end{eqnarray}
$I_{\mathrm{e}}$ is the pixel surface brightness at the effective
radius \efr. The S\'ersic index $n$ controls the shape of the radial
surface brightness profile; a larger $n$ corresponds to a steeper
inner profile and a highly extended outer wing. The constant $k$
depends on $n$ so as to ensure that the isophote at $R=R_{\rm eff}$
encloses half of the total light, i.e. that $R_{\rm eff}$ is the
half-light radius \citep{C+B99}. The symbol $q$ denotes the axis ratio.

The deflector mass density profile is described by a class of
elliptical power-law models. The surface mass density is parametrized
as
\begin{equation}
 \label{massmodel}
 \Sigma(x,y)=\Sigma_{cr}\frac{3-\gamma^{\prime}}{1+q_m}\left(\frac{\sqrt{ X^{2}+q_m^{-2}Y^{2}}}{R_{\rm E}}\right)^{1-\gamma^{\prime}},
\end{equation}
where $\Sigma_{cr}$ is the \textit{critical density}, depending on the
angular diameter distances to the deflector ($D_{\mathrm{d}}$), to the source
($D_{\mathrm{s}}$) and between the deflector and the source
($D_{\mathrm{ds}}$). The Einstein radius $R_{\rm{E}}$ is chosen such
that, in the spherical limit ($q_m=1$), it encloses a mean surface
density equal to $\Sigma_{cr}.$ This is also the radius of a ring
traced by the host of the AGN when this is exactly aligned with the
lens galaxy. The exponent $\gamma'$ is the so-called slope of the mass
density profile, $\approx 2$ for massive elliptical galaxies \citep{T+K02a,T+K04}. The
projected axis ratio of the mass density profile is described by
$q_m$. We refer the reader to the reviews by \citet{Sch06, Bar10,
Tre10} for more details.

Having specified the intrinsic surface brightness of the host galaxy
and the mass model of the deflector, the lensed images can be computed
by solving the lens equation. In practice, this is done using the code
{\sc
Pylens}\footnote{\url{https://github.com/tommasotreu/HIGHRESOLUTIONIMAGING/tree/master/code/pylens}}
written by one of us (M.W.A.) and functionally tested extensively by
\citet{Men++15}.

In order to create simulations that are as realistic as possible, we
first obtained rough models of the real systems. This set of rough
models enables us to set realistic values and ranges (see
Table~\ref{para_config} and Section \ref{sec:rest} as for details) to
the input parameters needed for {\sc Pylens} simulations of the host
galaxy and deflector surface brightness. The AGN images are added
directly in the image plane as PSFs scaled to match the observed
flux. The procedure is illustrated in Fig.~\ref{fig:sim_pipe}, panels
$a$ and $b$.

\subsection{Simulating dithering}
\label{ssec:dither}

The \hst\ PSF is undersampled by the camera setups that we simulate in
this study. Therefore, a common strategy consists of dithering the
observations by non-integer number of pixels in order to recover some
of the information lost via a process known as
drizzling\footnote{\url{http://www.stsci.edu/~fruchter/dither/\#basics}}.

\begin{figure*}
\centering
\begin{tabular}{cc}
\subfloat[Host arc and lens image, no AGN yet.]{\includegraphics[trim = 0mm 20mm 0mm 0mm, clip,width=0.4\textwidth]{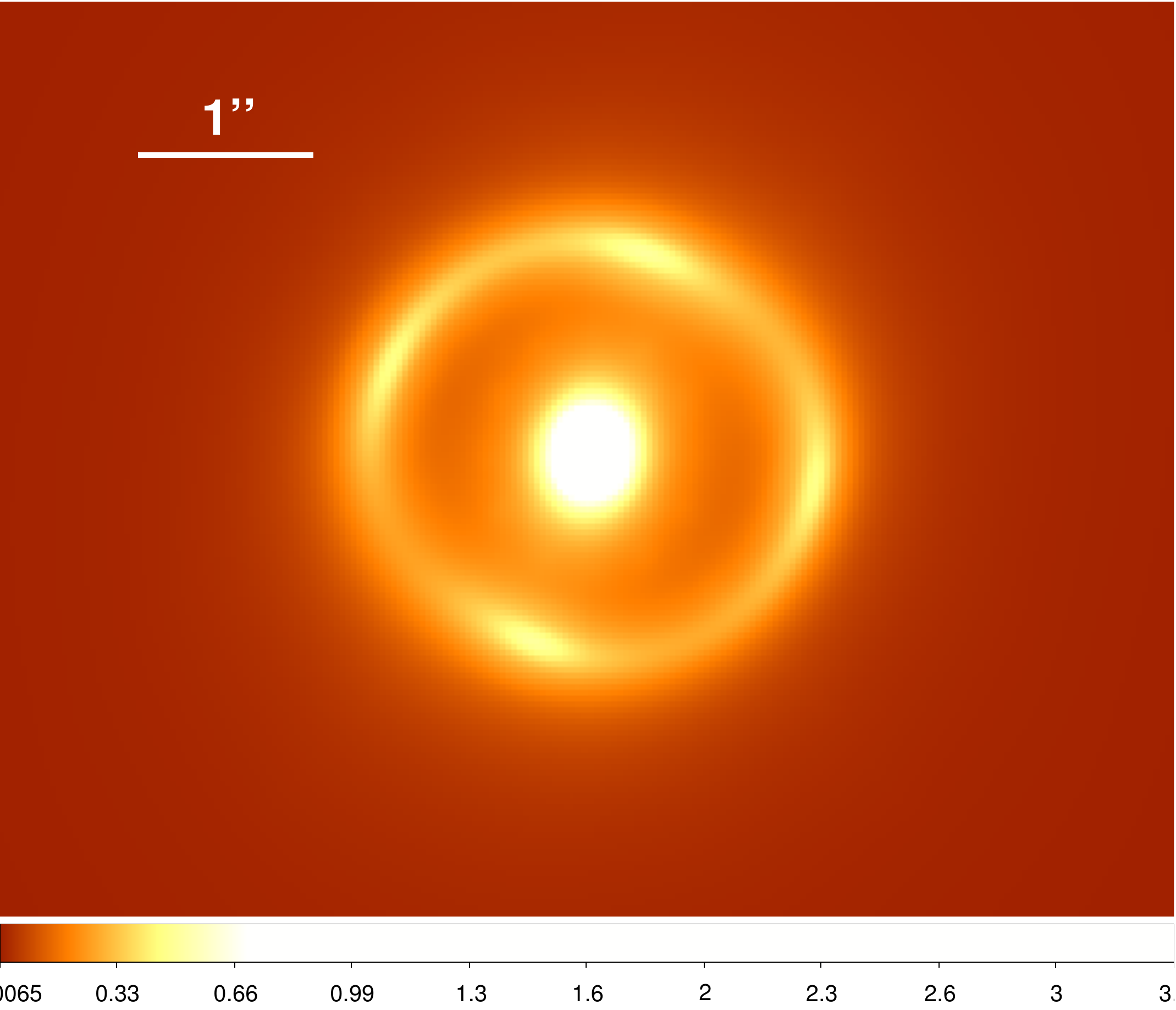}}&
\subfloat[Adding AGN images.]{\includegraphics[trim = 0mm 20mm 0mm 0mm, clip,width=0.4\textwidth]{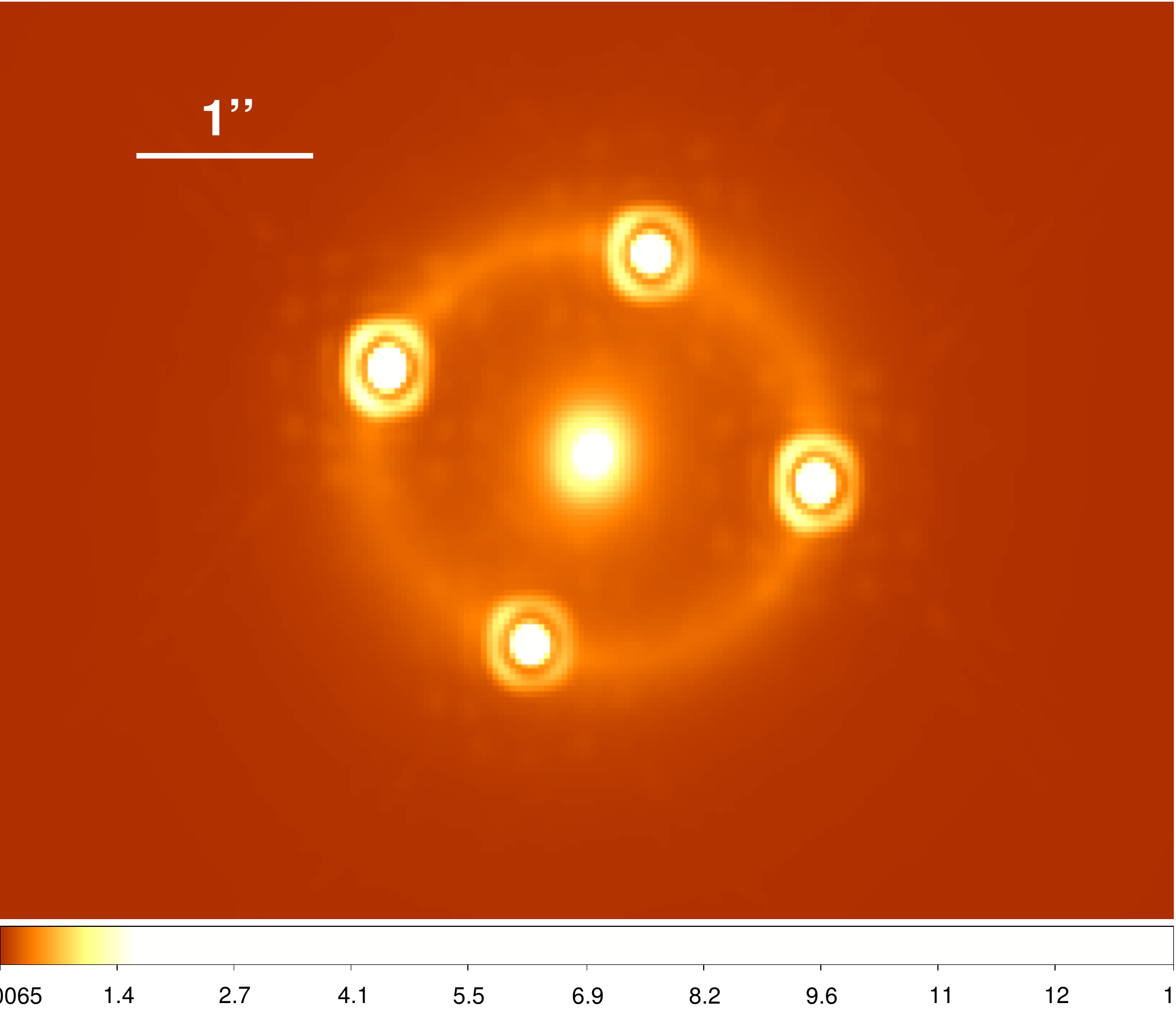}}\\
\subfloat[Binning into 8 images.]{\includegraphics[trim = 0mm 20mm 0mm 0mm, clip,width=0.4\textwidth]{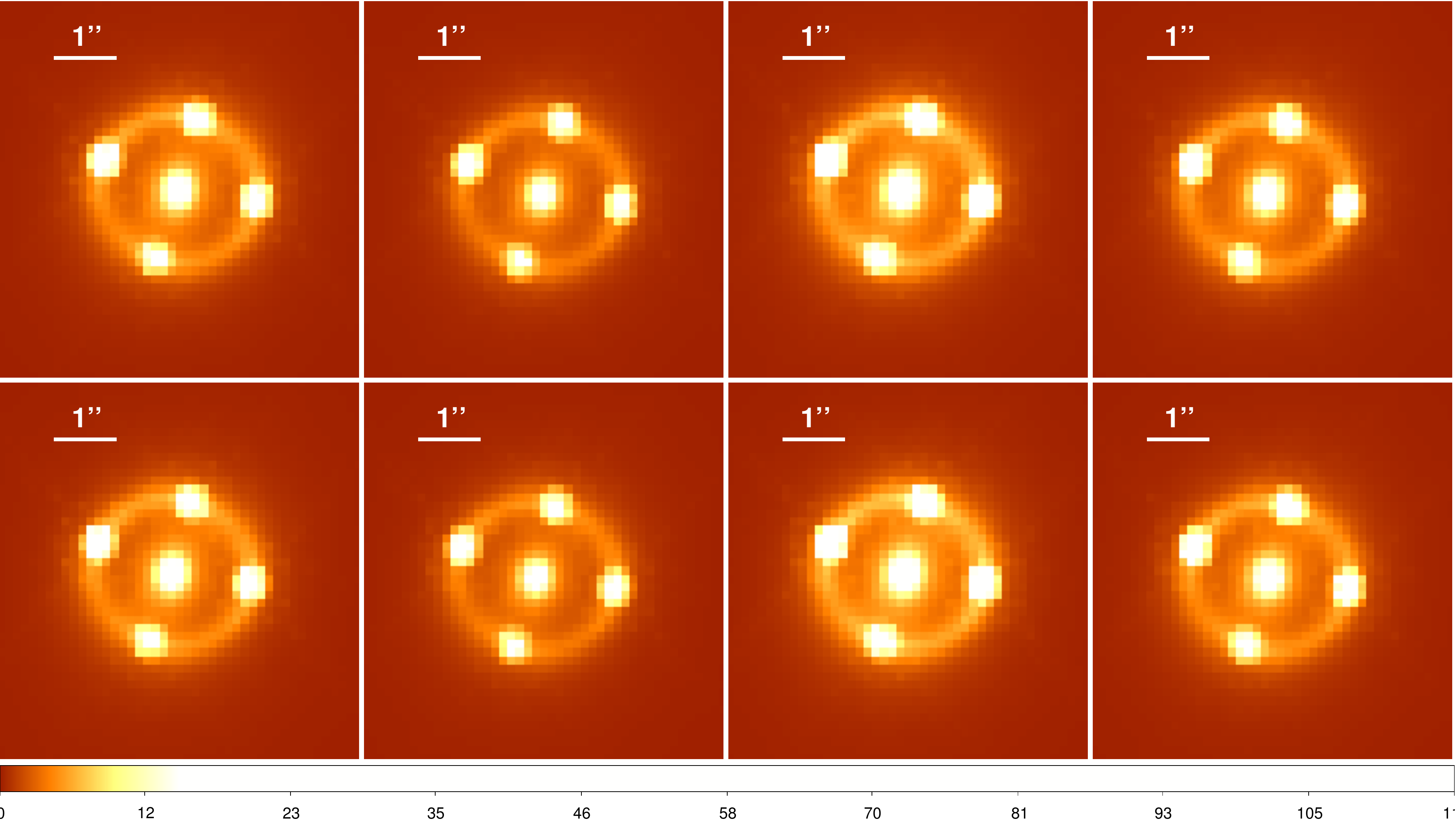}} &
\subfloat[Adding noise.]{\includegraphics[trim = 0mm 20mm 0mm 0mm, clip,width=0.4\textwidth]{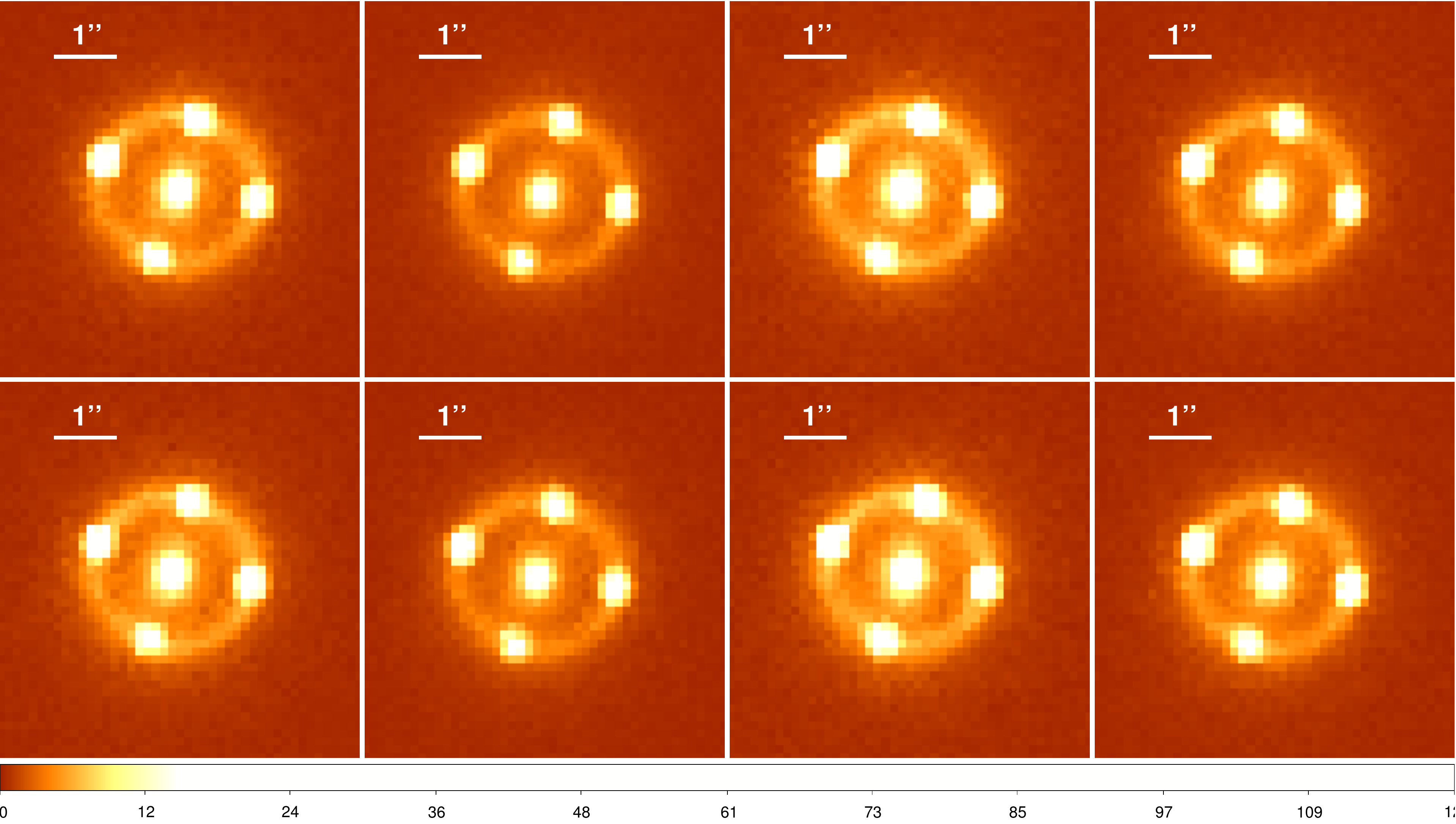}}\\
\subfloat[Final image, after drizzling.]{\includegraphics[trim = 0mm 20mm 0mm 0mm, clip,width=0.4\textwidth]{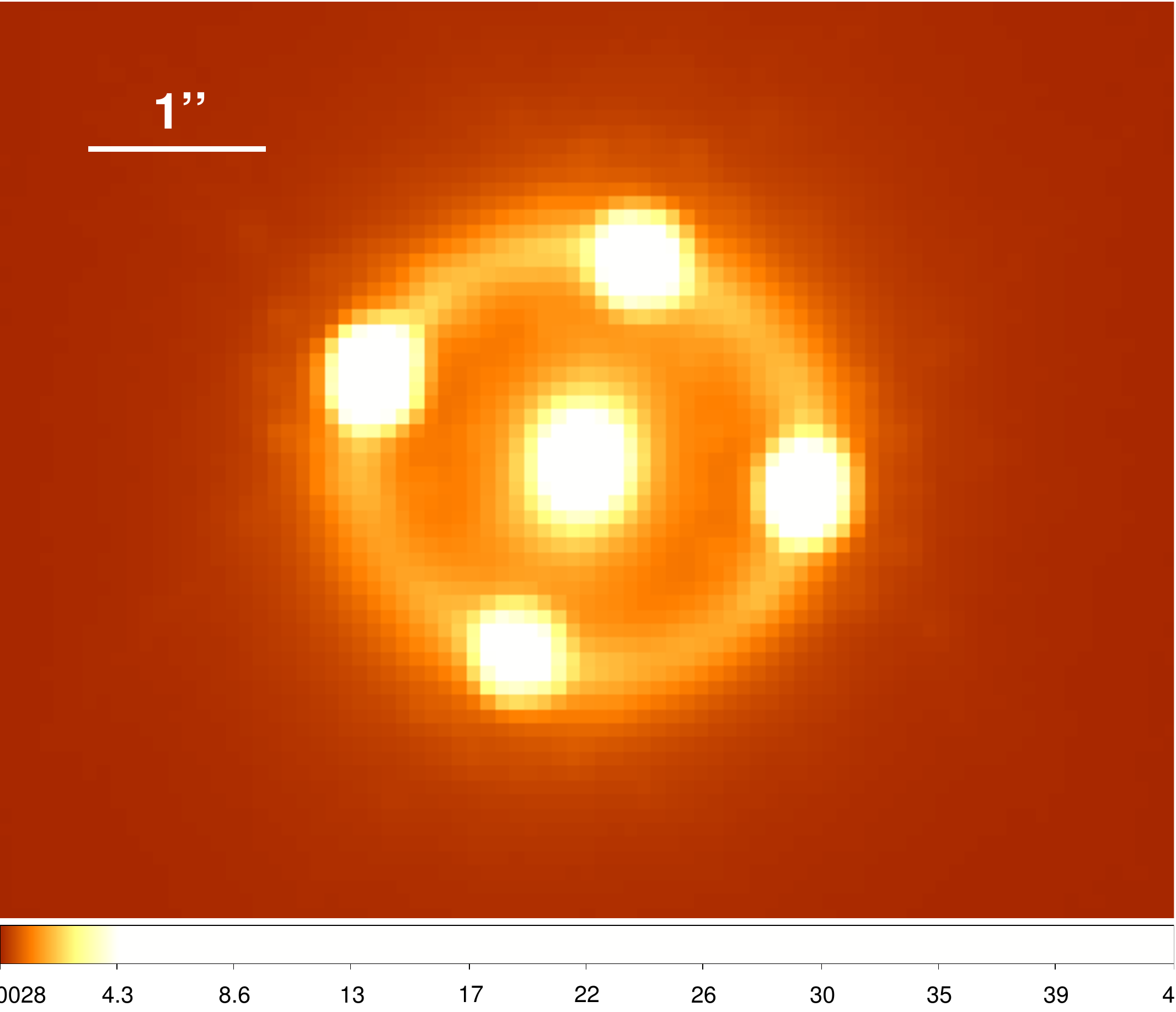}}&
\subfloat[Simulated 48 images VS the real on the bottom right.]{\includegraphics[trim = 0mm 20mm 0mm 0mm, clip,width=0.4\textwidth]{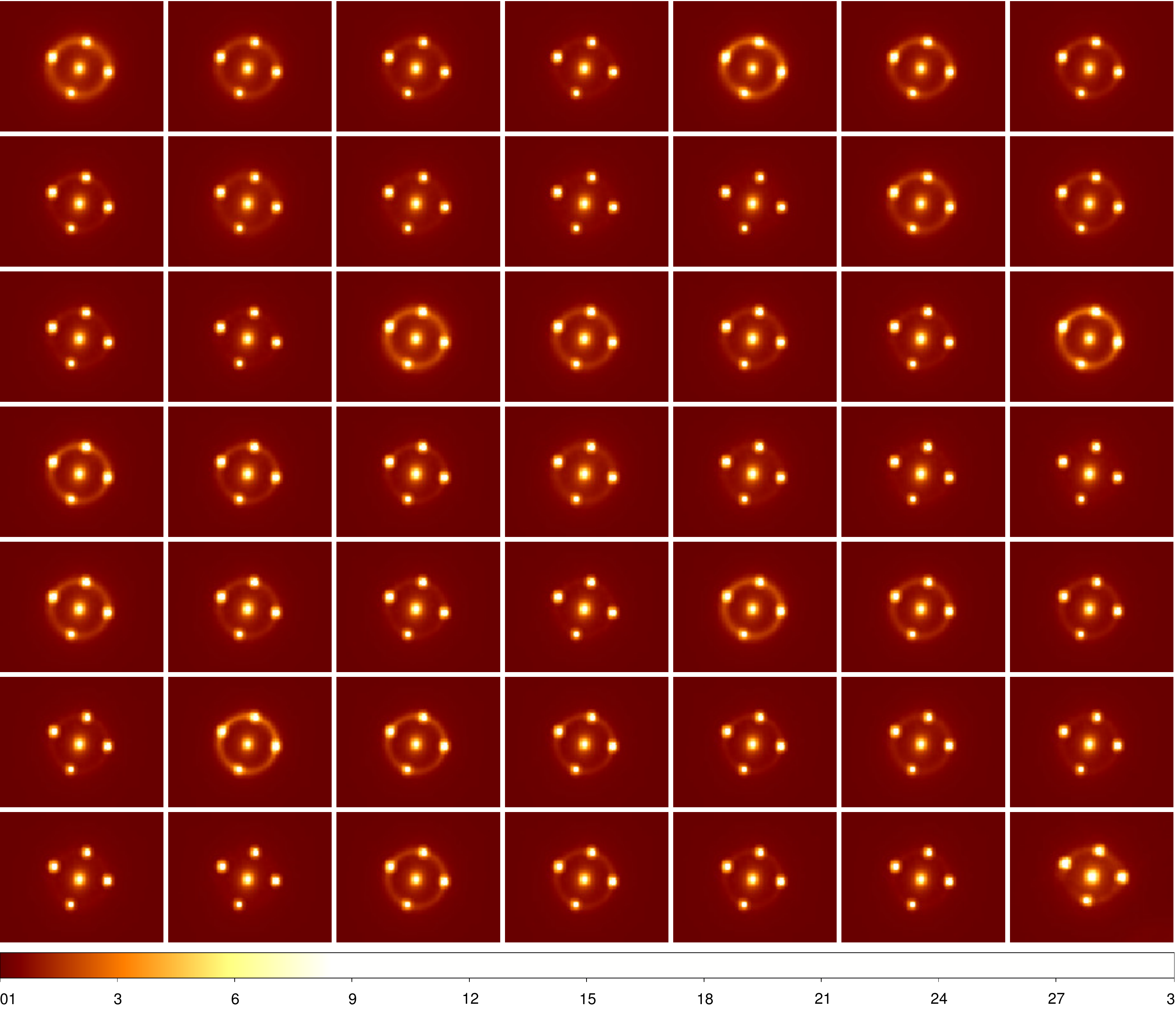}}\\
\end{tabular}
\caption{\label{fig:sim_pipe} Illustration of the simulation process using a linear color scale.
(a) Simulating the host arc and lens image with {\sc Pylens},
(b) Adding AGN image as appropriately normalized PSFs,
(c) Binning into a lower resolution images in 8 different ways following the actual dithering pattern,
(d) Including the noise information,
(e) Drizzling 8 images together to obtain the final image,
(f) Montage of 48 simulated images plus the real one in the bottom right panel, presented on an oversampled grid. It is clearly seen that the arc brightness is varying from case to case and brackets the real data.}
\end{figure*}

\begin{figure}
\centering
\includegraphics[trim = 0mm 20mm 0mm 0mm, clip,width=7cm]{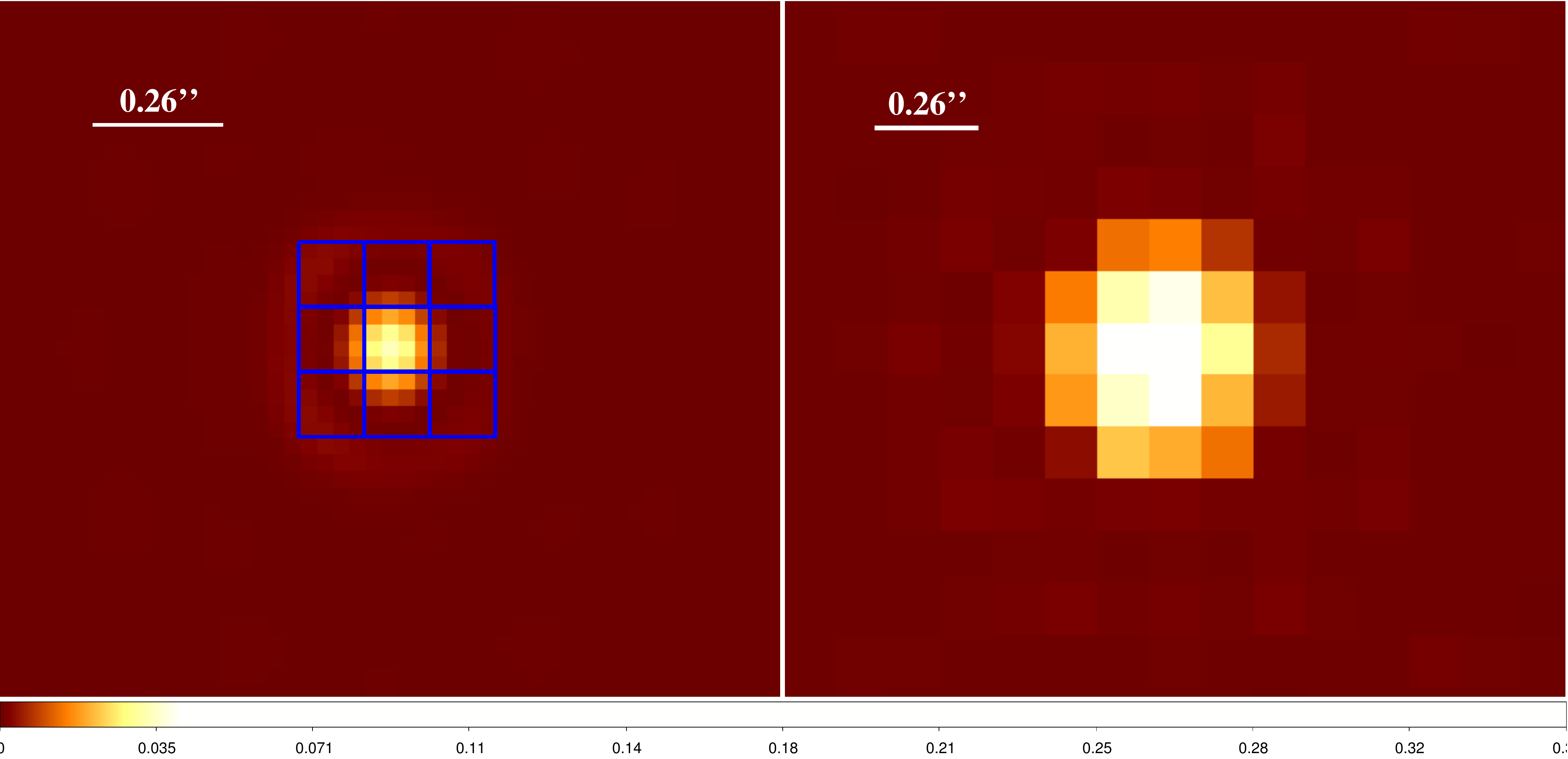}\\
\includegraphics[trim = 0mm 20mm 0mm 0mm, clip,width=7cm]{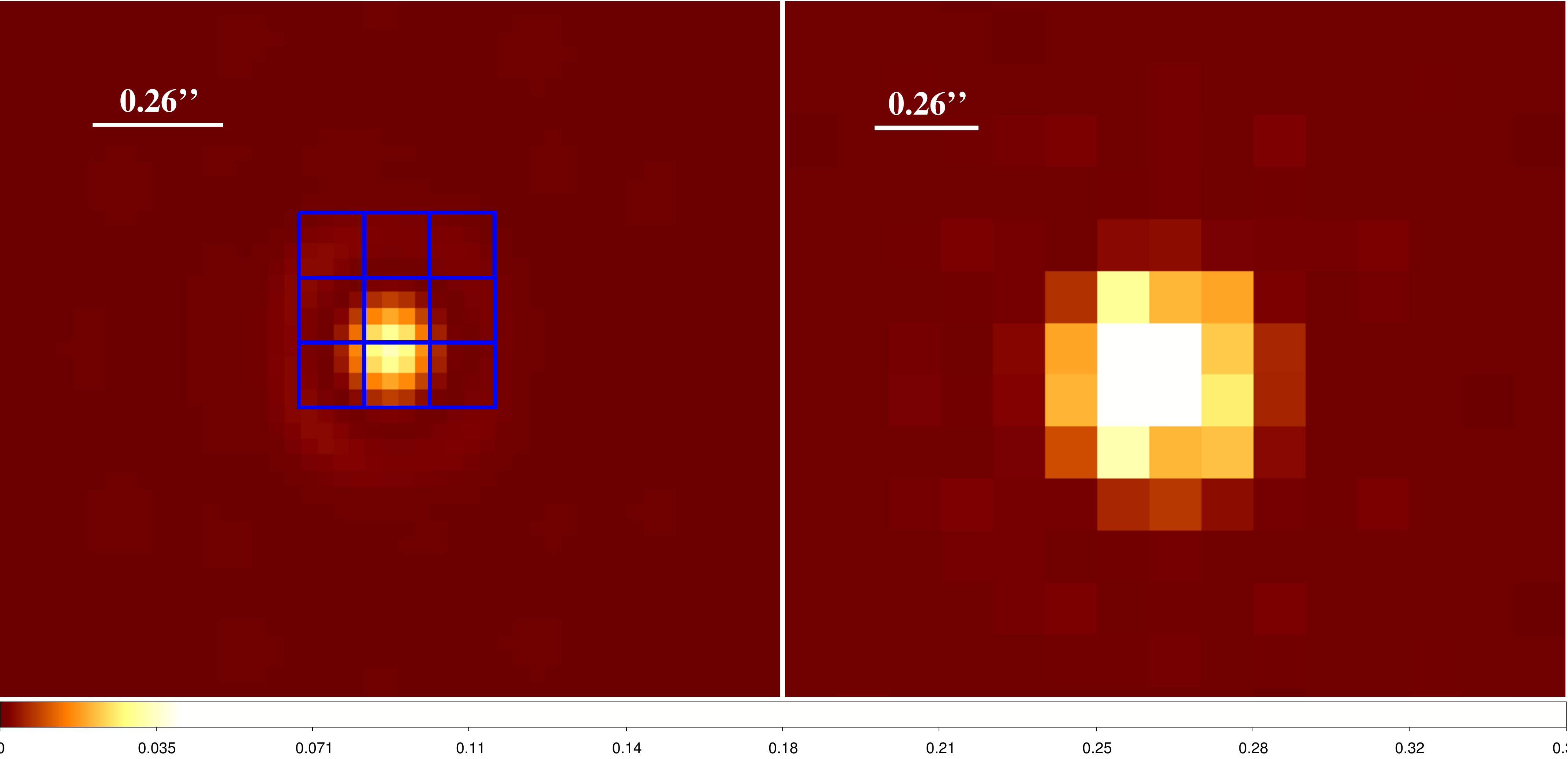}\\
\includegraphics[trim = 0mm 20mm 0mm 0mm, clip,width=7cm]{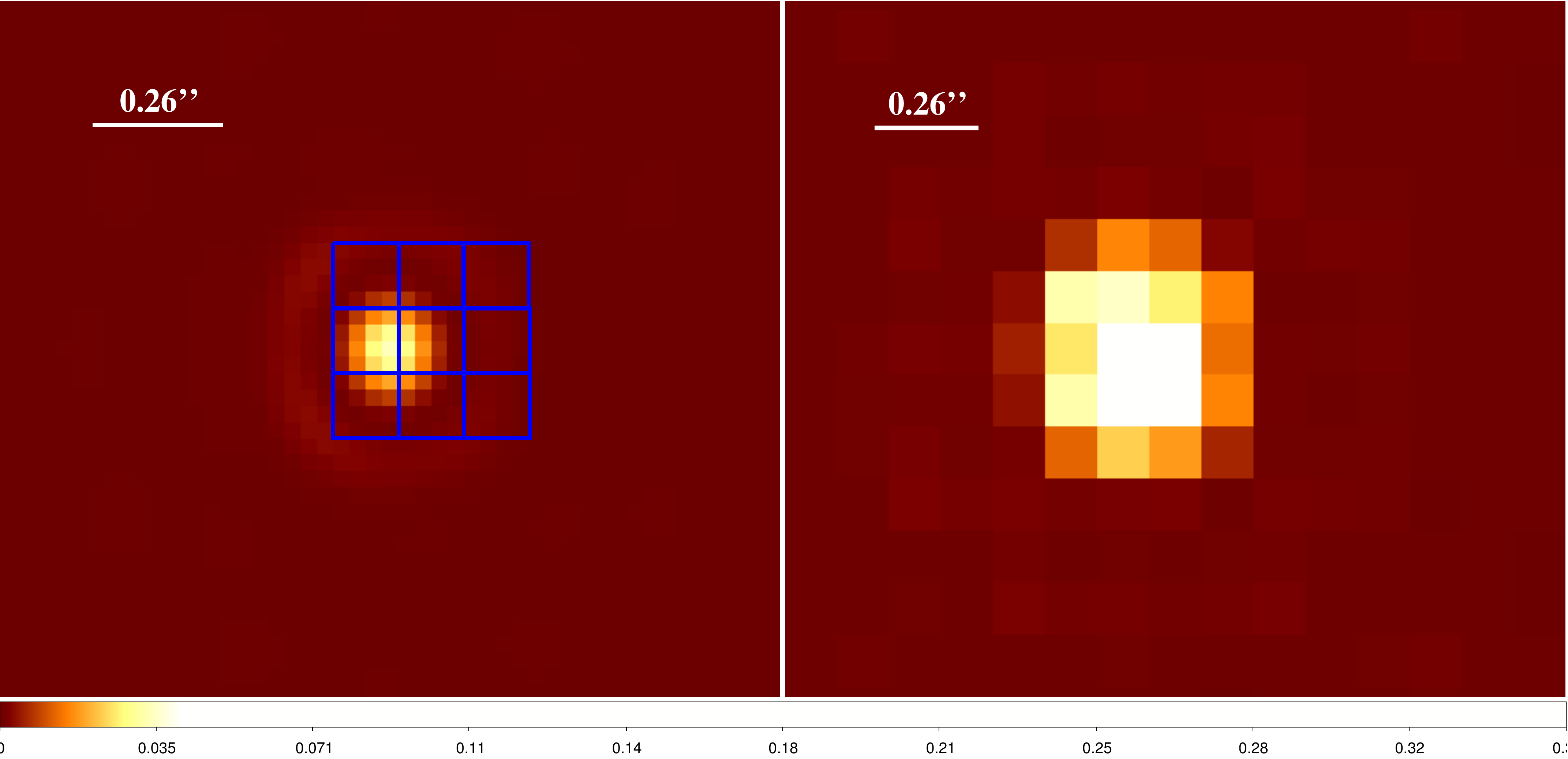}\\
\includegraphics[trim = 0mm 20mm 0mm 0mm, clip,width=7cm]{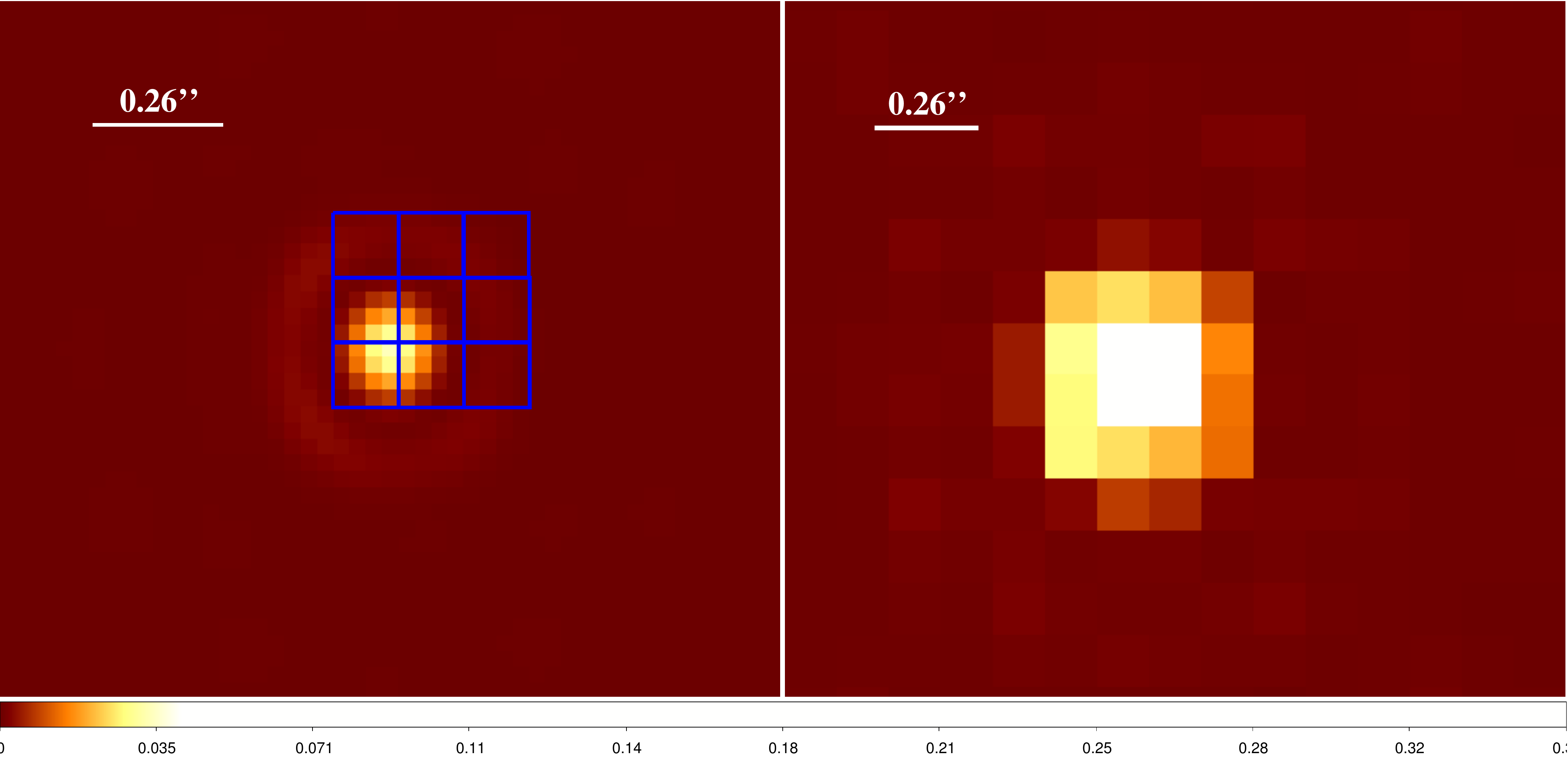}\\
\caption{\label{fig:binpix}
We illustrate 4 different ways to bin high resolution image pixels
into lower resolution, using the same linear color stretch.
Left panels show the high resolution
images before binning. By combining all the pixel inside the blue box, we
get the low resolution images shown in the right panels. To make the
dither pattern more evident, we only show 9 blue boxes,
rather than cover the entire plane. $0\farcs26$ corresponds to 8 and 2 pixels in left and right panels, respectively.}
\end{figure}

In order to simulate the undersampling and dithering, we bin the high
resolution image ($4\times4$ for HE0345 as an example), varying the
location of the binning grid so as to mimic the actual dithering
pattern adopted in the \hst\ observations (see
Fig.~\ref{fig:sim_pipe}-(c), \ref{fig:binpix} for an
illustration).

\subsection{Adding noise}
\label{ssec:noise}

The noise in each pixel is due as usual to read noise, background noise, and
Poisson noise from the astronomical sources themselves. The read
noise and background noise level are measured directly from empty
regions of the real data while the Poisson noise level is related to
the effective exposure time in each pixel. For the systems observed
with WFC3-IR, three exposures were obtained at each dither position:
one short exposure ($\sim43s$) followed by two long exposures
($\sim599s$). This strategy allowed us to obtain unsaturated images of
the brightest pixels in the center of the AGN, while minimizing
overheads and maximizing the signal to noise ratio on the low surface
brightness parts of the images. In order to reproduce this strategy,
for the central AGN area ($3-5$ pixels in the AGN center), we consider
the noise corresponding to the short exposures. For the rest of the
image we consider the full exposure time. By comparing the final simulation
to the real sample, we find that the simulated noise
properties match the noise properties of the observed data very well.
(Fig.~\ref{fig:sim_pipe}-(d) shows the image after adding noise).

\subsection{Drizzling}
\label{ssec:drizzle}
The final step is to combine the 8 images by using {\sc
MultiDrizzle}\footnote{{\sc MultiDrizzle} is a product of the Space
Telescope Science Institute, which is operated by AURA for NASA, see
\url{http://www.stsci.edu/hst/wfpc2/analysis/drizzle.html} for more
information.}. The images are drizzled to a final pixel scale
$0\farcs08$ pixel$^{-1}$ as in real data. The final simulated image is
shown on Fig.~\ref{fig:sim_pipe}-(e). Of course, dithered images of
the PSF are also drizzled following the same exact procedure in order
to produce a realistic PSF for the final combined image.

\subsection{Modifications for WFC3-UVIS and ACS }
Our sample includes systems observed with WFC3-UVIS and the Advanced
Camera for Surveys (ACS). The pipeline used to simulate the lensed AGN
in these bands is similar to that used for WFC3-IR and described
above. The main difference is that the pixel scale is $0\farcs03$ for
WFC-UVIS and $0\farcs05$ for ACS after drizzling. Also, the noise
levels and dither patterns are modified to match the actual
observations.

\section{Results}
\label{sec:rest}

\begin{figure*}
\centering
\begin{tabular}{cc}
\subfloat[HE1104.]{\includegraphics[width=8cm]{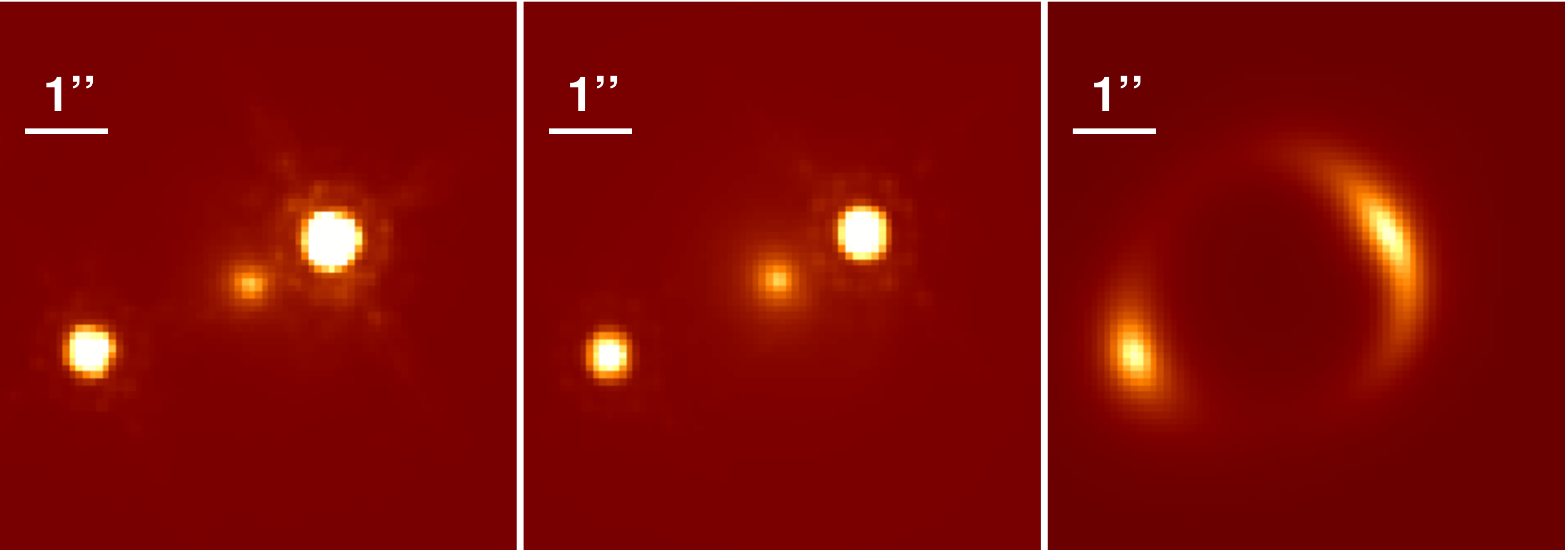}}&
\subfloat[HE0435.]{\includegraphics[width=8cm]{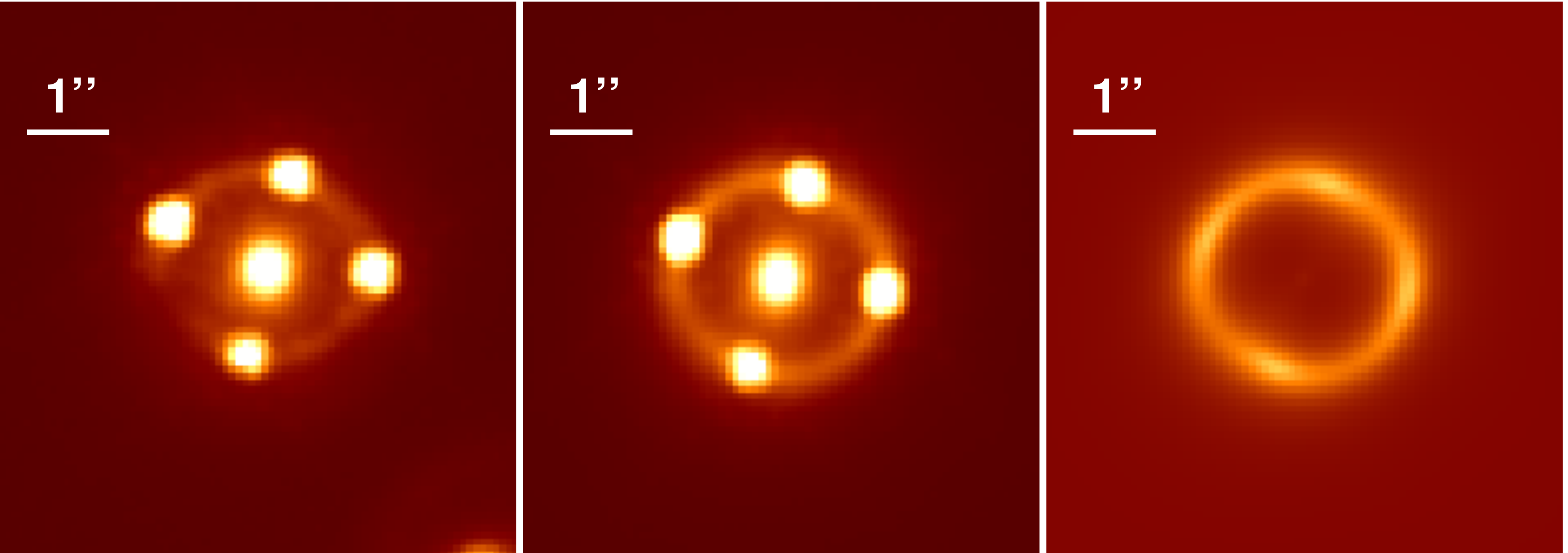}}\\
\subfloat[WFI2033.]{\includegraphics[width=8cm]{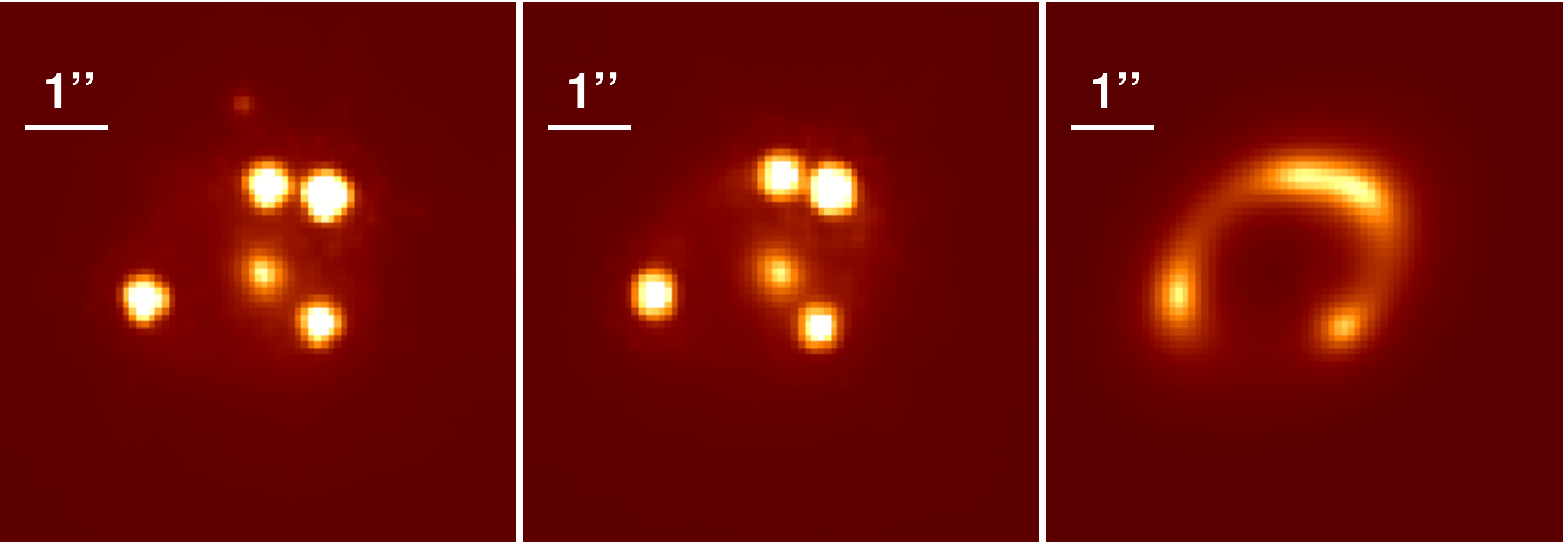}}&
\subfloat[SDSS1206.]{\includegraphics[width=8cm]{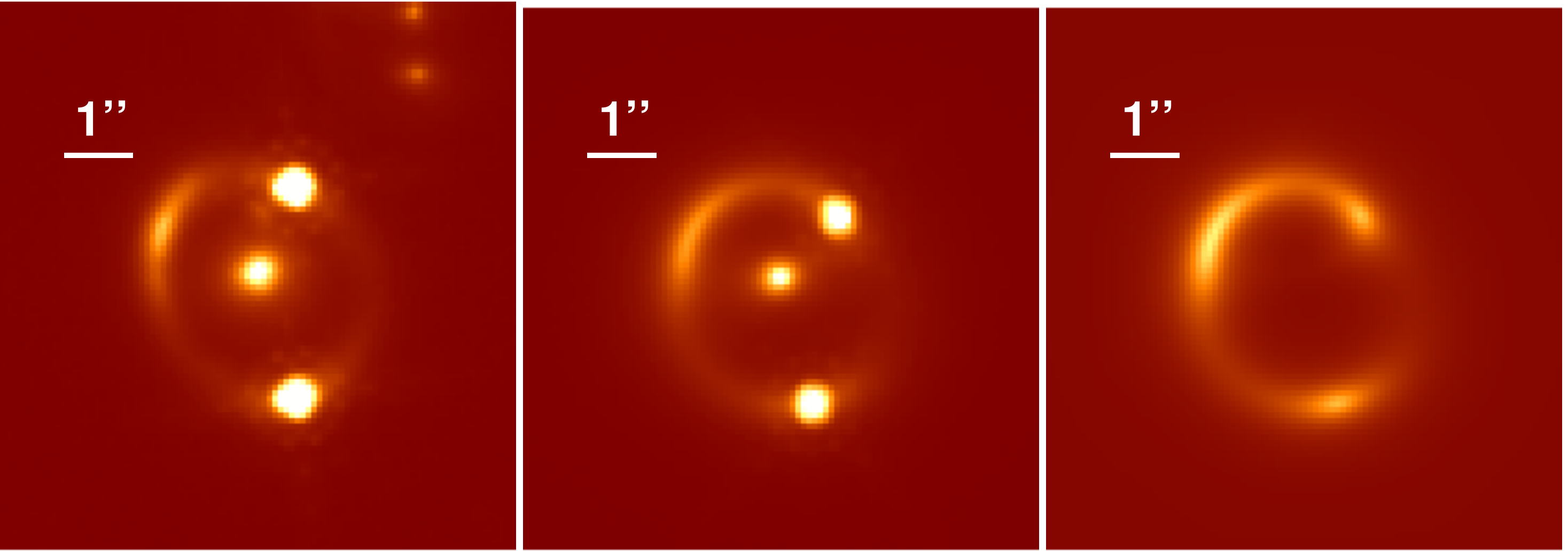}}\\
\subfloat[SDSS0246.]{\includegraphics[width=8cm]{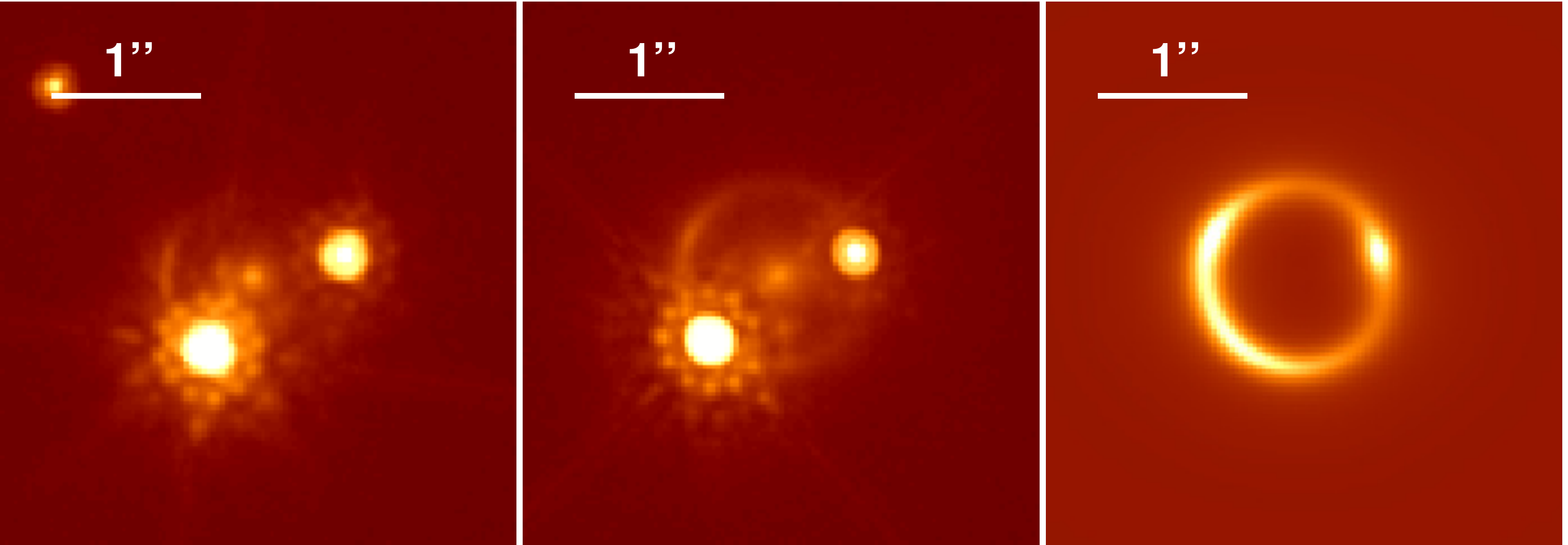}}&
\subfloat[HS2209.]{\includegraphics[width=8cm]{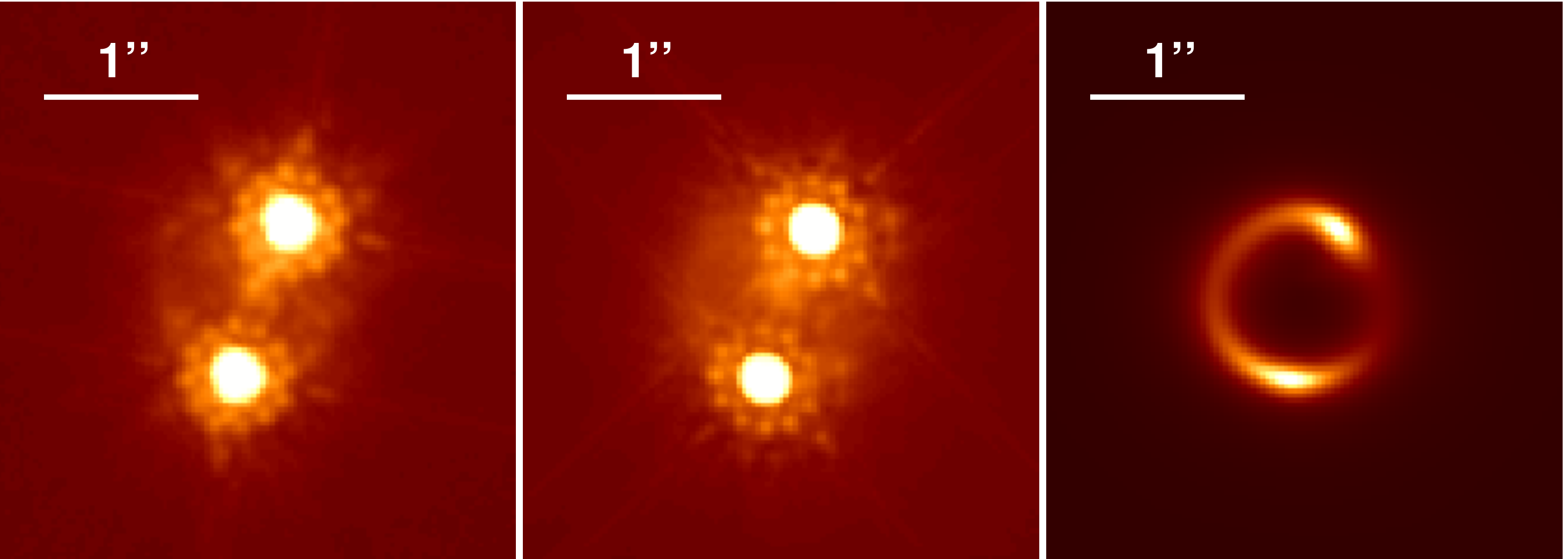}}\\
\subfloat[HE0047.]{\includegraphics[width=8cm]{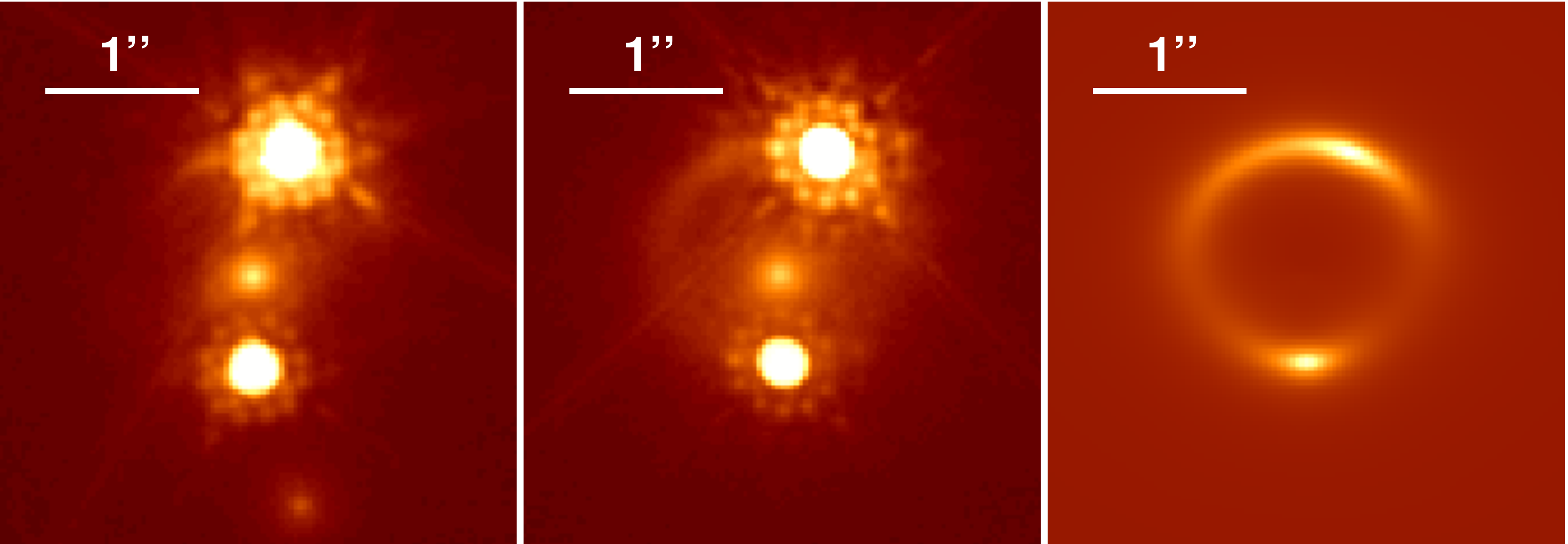}}&
\subfloat[RXJ1131.]{\includegraphics[width=8cm]{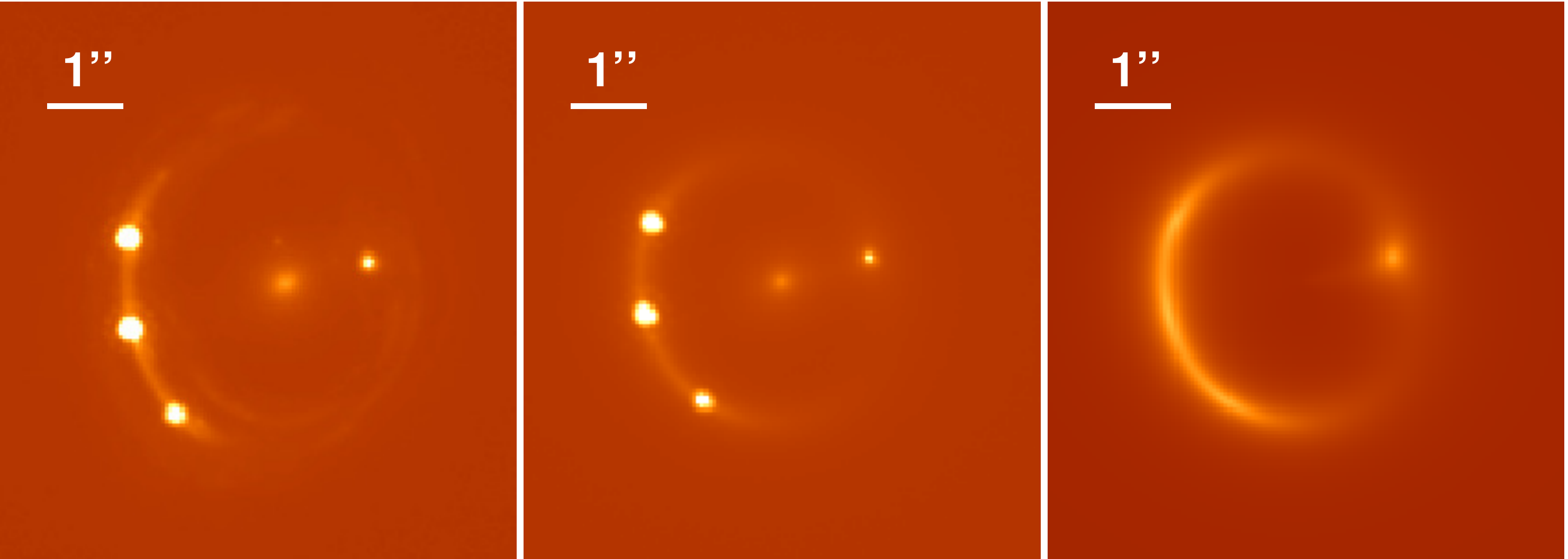}
}

\end{tabular}
\caption{\label{fig:simufig}
For each lens the left panel shows the observed image and the middle
panel shows the simulation with the same stretch. (a)-(d) and (h) are
in linear color scale; (e)-(g) are in log color scale. For clarity,
we also show in the right panels the simulated arc image without the
bright AGN and deflector, using a different stretch. (a)-(d) are from
WFC3-IR observation, (e)-(g) are from WFC-UVIS and (h) is from ACS.}
\end{figure*}

In this section we present the results of our simulations of eight
systems. Four of the systems are imaged with WFC3-IR through filter
F160W, three with WFC3-UVIS through filter F814W, and the remaining
one with ACS through filter F814W. For each system we construct 48
realizations by changing the parameters describing the unknown
properties of the host galaxy and mass distribution of the deflector,
as summarized in Table~\ref{para_config}. As mentioned above, the set
of unknown parameters is chosen as to represent the range of expected
true values in this particular data set. The parameters describing the
surface brightness of the deflector and the brightness of the point
sources are measured directly from the data, with negligible
uncertainty for our purposes and are therefore kept fixed. Of course,
the brightness of the AGNs are the key parameters for this
study. However, the AGN flux can be measured with very high precision
and residual uncertainties have negligible effect on the host galaxy
reconstruction, provided a good PSF model is available. We will return
to this issue and account for for residual uncertainties when we
analyze the actual data in a future paper. Examples of simulated
images are shown in Fig.~\ref{fig:simufig}.

\begin{table}
\centering
  \begin{threeparttable}
    \caption{Parameter grid}\label{para_config}
     \begin{tabular}{ c c}
     \hline
     parameter & simulated values \\
     \hline\hline
     $n$ & 2, 4 \\
     $l_{eta}$ & 0.8, 1.0, 1.2\\ \hline\hline
     WFC-IR images\\ \hline
     $S_{re}$ (arc second) & 0.39, 0.78 \\
     HE1104 $S_{Mag}$ & 21.0, 21.5, 22.0, 22.5\\
     HE0435 $S_{Mag}$ & 20.5, 21.0, 21.5, 22.0\\
     WFI2033 $S_{Mag}$ & 20.5, 21.0, 21.5, 22.0\\
     SDSS1206 $S_{Mag}$ & 20.0, 20.5, 21.0, 21.5\\\hline
     WFC-UVIS images\\ \hline
     $S_{re}$ (arc second) & 0.3, 0.6 \\
     SDSS0246 $S_{Mag}$ & 21.5, 22.0, 22.5, 23.0\\
     HS2209 $S_{Mag}$ & 21.0, 21.5, 22.0, 22.5\\
     HE0047 $S_{Mag}$ & 21.5, 22.0, 22.5, 23.0\\\hline
     ACS images\\ \hline
     $S_{re}$ (arc second) & 0.5, 1.0 \\
     RXJ1131 $S_{Mag}$ & 19.0,19.5, 20.0, 20.5\\
     \hline
     \end{tabular}
     \begin{tablenotes}
      \small
      \item Note:$-$ Grid of simulated parameters. $n$, $S_{re}$ and $S_{Mag}$ are the S\'ersic index, \efr\ and the magnitude of the source brightness respectively. After preliminary tests we found that the fidelity of the reconstruction does not depend significantly on source ellipticity. So, we decided to set all ellipticities to 0.7 to save computation time. The magnitude ranges are selected to bracket the range of observed brightness. $l_{eta}$ describes the slope of the mass density profile of the deflector ($l_{eta}= \gamma' -1$, as defined by Eq.~\ref{massmodel}).
\end{tablenotes}
\end{threeparttable}
\end{table}

\subsection{\glee\ fitting}

We use \glee\ to reconstruct the host galaxy properties from the
simulated lensed AGN and analyze the resulting uncertainty and bias on
the main parameters including magnitude, effective radius and sersic
index of the host galaxy. We are mostly interested in magnitude, which
is the key quantity entering the correlation with
\mbh.

The inputs to \glee\ are the simulated image, a noise map, a mask
identifying the lensed image of the arc, and a PSF. The crucial point
is that \glee\ describes the host galaxy as a set of surface
brightness pixels, not as a S\'ersic profile. This choice enables the
extraction of more information, down to the noise level, even though
one still needs to fit the reconstructed source plane surface
brightness map with a S\'ersic profile in order to obtain the desired
parameters.

\begin{figure}
\centering
\begin{tabular}{cc}
\subfloat[Example of a simulated image of HE0435.]{\includegraphics[trim = 0mm 20mm 0mm 0mm, clip,width=4cm]{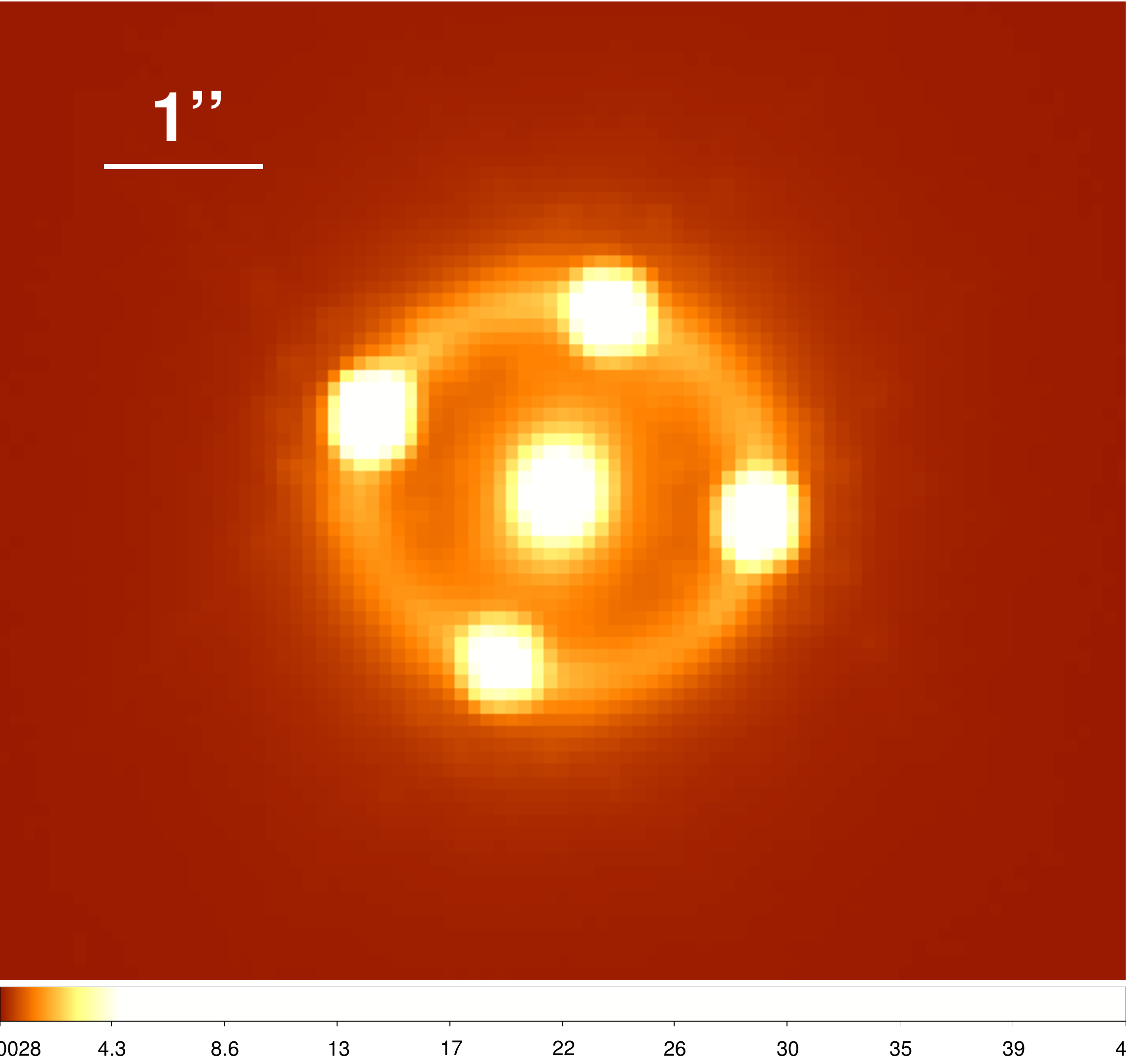}}&
\subfloat[Residual from best fit AGN and deflector brightness, showing the lensed host galaxy.]{\includegraphics[trim = 0mm 20mm 0mm 0mm, clip,width=4cm]{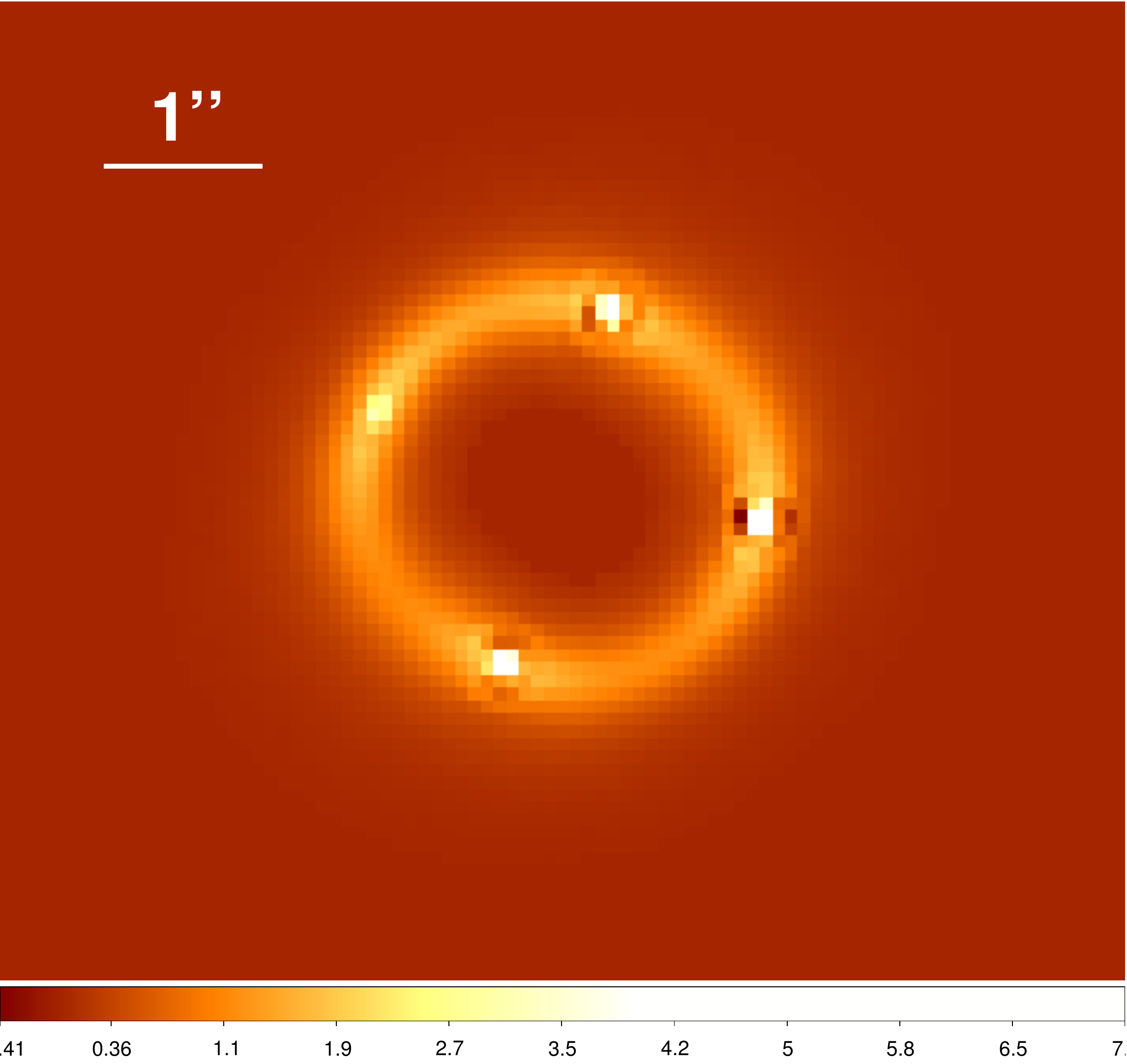}}\\
\subfloat[Best fit model of the host galaxy in the image plane.]{\includegraphics[trim = 0mm 20mm 0mm 0mm, clip,width=4cm]{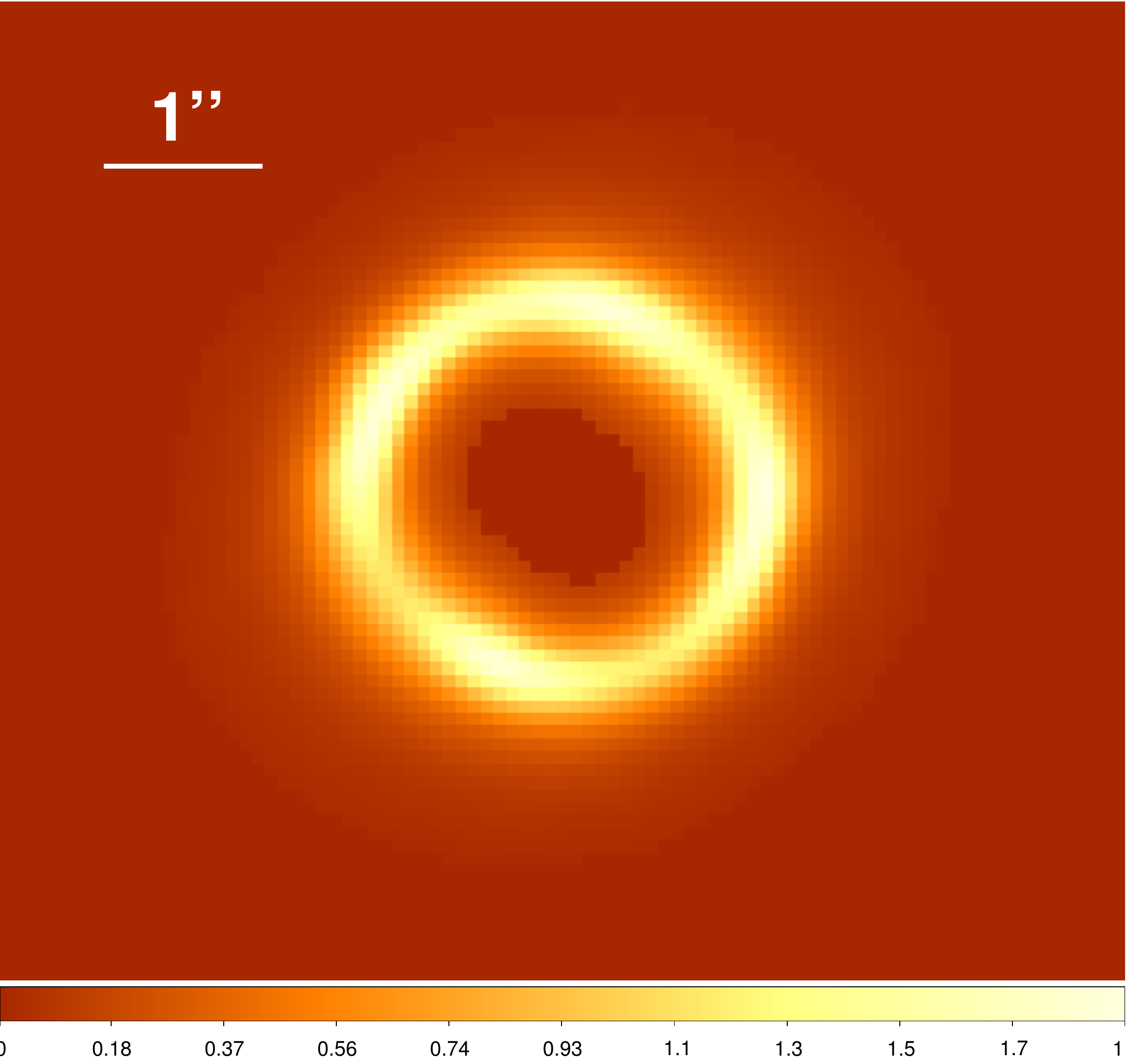}} &
\subfloat[Normalized residuals.]{\includegraphics[trim = 0mm 20mm 0mm 0mm, clip,width=4cm]{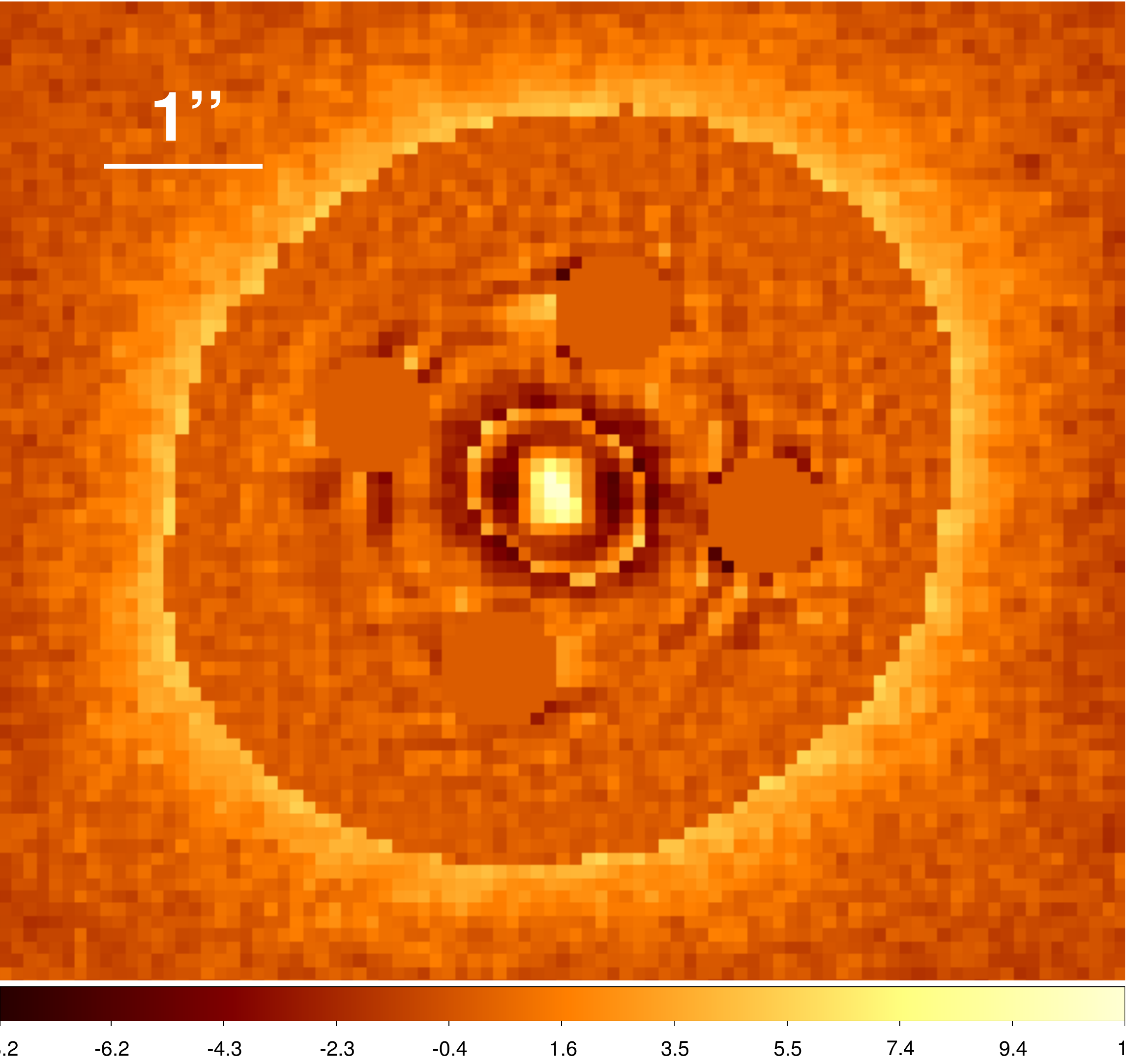}}\\
\subfloat[Reconstructed host in the source plane.]{\includegraphics[trim = 0mm 20mm 0mm 0mm, clip,width=4cm]{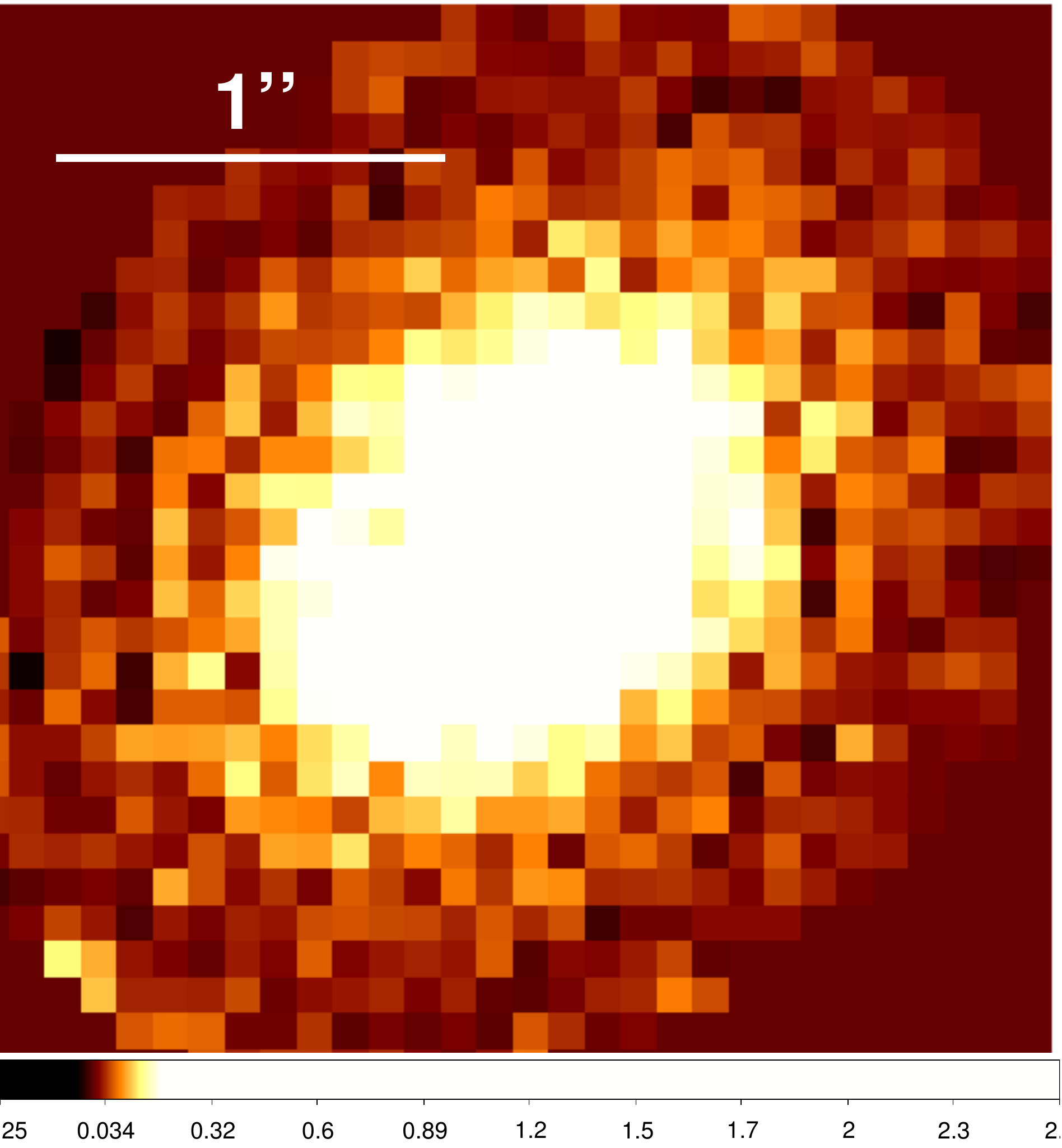}}&
\subfloat[True host image in the source plane.]{\includegraphics[trim = 0mm 20mm 0mm 0mm, clip,width=4cm]{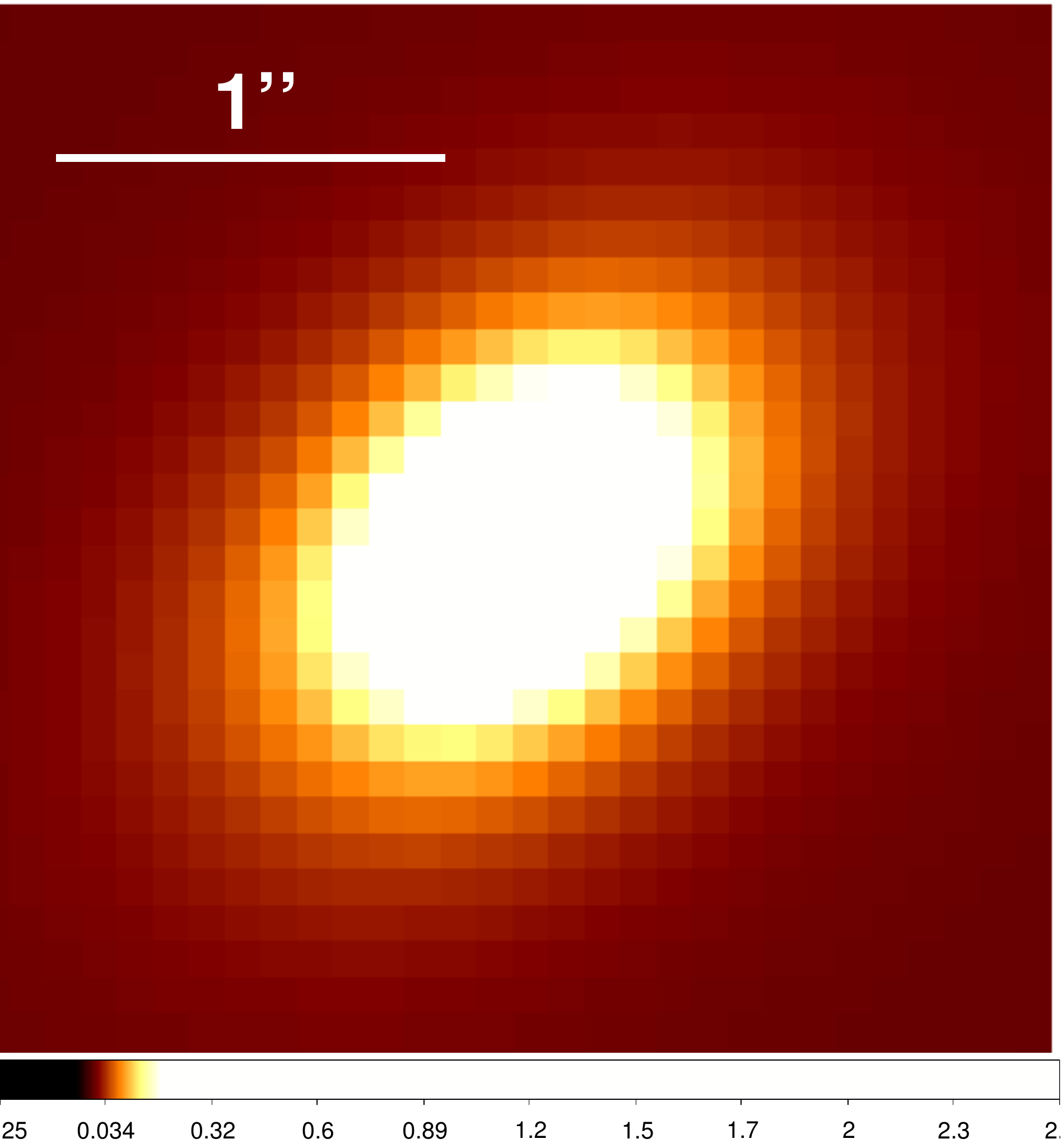}}\\
\end{tabular}
\caption{\label{fig:glee_pip}
Illustration of the key ingredients in \glee\ reconstruction of the host galaxy image in the source plane.
The mask region can be clearly seen in (c) and (d).}
\end{figure}

In principle, the mask containing the lensed host galaxy light should
be selected to be as large as possible, including pixels with surface
brightness close to the noise level. However, these masks would be
extremely large, and thus too computationally expensive.
Thus, in order to keep the run time manageable (of order a week per
system on a multicore linux box), and in order to closely mimic what
is done in the H0LiCOW analysis, we select a mask corresponding to a
thick ring around the Einstein radius (see examples in
Fig.~\ref{fig:glee_pip}).

Importantly, the image pixels used in modelling the AGN host are kept
fixed to those in the selected arc mask.  For a given set of lens mass
parameters, these image pixels in the arc mask are casted back to
the source plane, forming a region on the source plane where we
would have information to reconstruct the AGN host surface brightness
distribution. We then use a minimal rectangular grid with fixed dimensions
that encapsulates this source reconstruction region for our AGN host
surface brightness modelling.  In this way, the data (image pixels) are
kept fixed throughout the modelling as the lens mass parameters vary,
which is important for sampling the posterior of the data and for deriving
lens parameter constraints.  Even though the source grid changes as the
modelling proceeds, the source pixel parameters are marginalized
\citep[see, e.g.,][]{S+H10} when computing the posterior of the data,
so the constraints on the lens parameters account for the variable source
region. To study the AGN host, we then consider the best-fitting source
reconstructions from a set of lens mass models whose parameters are
obtained from the modelling.


In order to avoid overfitting
the noise, regularization is introduced in the framework of Bayesian
statistics \citep{Suyu12,Suy++06}. Following common practice
\citep[e.g.][]{Suy++13,Suy++14,Agn++16}, in order to account 
for systematic errors in the interpolation of the subsampled PSF,
we boost the noise estimate in the central pixels of each AGN image to
be effectively infinite. 
In practice this means that the reconstruction of the host galaxy brightness
in the central pixels is dominated by the regularization.
This procedure is designed to account for PSF modelling
systematic errors and helps avoid underestimating the uncertainties on
the final parameter estimates.

Briefly, \glee\ models each system by simultaneously fitting the
following ingredients through a Markov Chain Monte Carlo process:

\begin{enumerate}
\item A set of scaled PSFs to represent the AGN images;
\item A S\'ersic profile to represent the surface brightness of the deflector;
\item A grid of pixels in the source plane representing the surface brightness of the host galaxy;
\item A mass model for the deflector.
\end{enumerate}

The modelled host image is generated onto a $30\times 30$ pixel grid.
The pixel resolution for each reconstruction is determined by the arc
mask region and light profile of reconstruction.  We have tested
that the results of our analysis are the same if a 40x40 source grid
is used instead of 30x30, using a subset of models. Similar tests on
real data performed by our group show that the effect of
discretization bias \citep{V+K09a,Nig++15} is small provided that
the mask is large enough and the source resolution is high enough
\citep{Suy++13,H0licow4}.  Moreover, we have tested reconstructing
the host galaxy by using the true slope of the mass density profile,
and 
the results on the AGN host flux are unchanged.


We illustrate the steps of \glee\ modelling in
Fig.~\ref{fig:glee_pip}, taking HE0435 as an example. The normalized
residual map (i.e. Fig.~\ref{fig:glee_pip}-(d)) shows that the light
model for the deflector is imperfect near the center. This could be
fixed by adopting a more extended mask, covering the central region of
the image, at a cost of more computing time. In order to test
  whether this residual could cause systematic uncertainties we
  repeated the fit for a subset of objects zeroing the flux associated
  with the deflector and the results on the lensed arc are
  unchanged. Thus, we conclude that this
residual mismatch is not important for the host galaxy property
studies and we do not try to reduce it further in this work.

\subsection{Analysis}

In order to obtain a description of the total flux of the host galaxy,
accounting for the flux lost by masking, we cannot rely on direct
integration of the surface brightness map in the source plane.
Instead we rely on the software {\sc Galfit} \citep{Pen++02} to fit a
S\'ersic profile to the data, thus reproducing the procedure used for
isolated galaxies. Of course, this approach will provide not only the
total magnitude but also the effective radius $S_{re}$ and S\'ersic
index $n$.

In some cases, the reconstructed images are very faint and cover only
a small radial range across an arcsecond. This can lead to unphysical
results from {\sc Galfit} fitting. Thus, we constrain $n$ to be in the
range 1$-$4, typical of normal galaxies. We also test the effects of
fixing $n$, as would be the case if we knew that the host galaxy of
the bright AGN was a pure elliptical or disk galaxy. Errors on the
{\sc Galfit} fit for each simulated image are obtained by repeating
the fit on 30 sets of source reconstructions taken from the MCMC chain
output by \glee, and computing the mean and standard deviation of the
results. We present four sets of comparisons between input and
recovered values including total magnitude ($n$ free and $n$ fixed),
effective radius $S_{re}$ and S\'ersic index $n$.

In the following subsections we describe the results for each system.
The lens names are abbreviated in the rest of the paper. See
Table~\ref{data_set} for the full name.

\subsubsection{HE1104}
\label{sssec:he1104}
\begin{figure*}
\centering
\begin{tabular}{c c}
{\includegraphics[width=9cm]{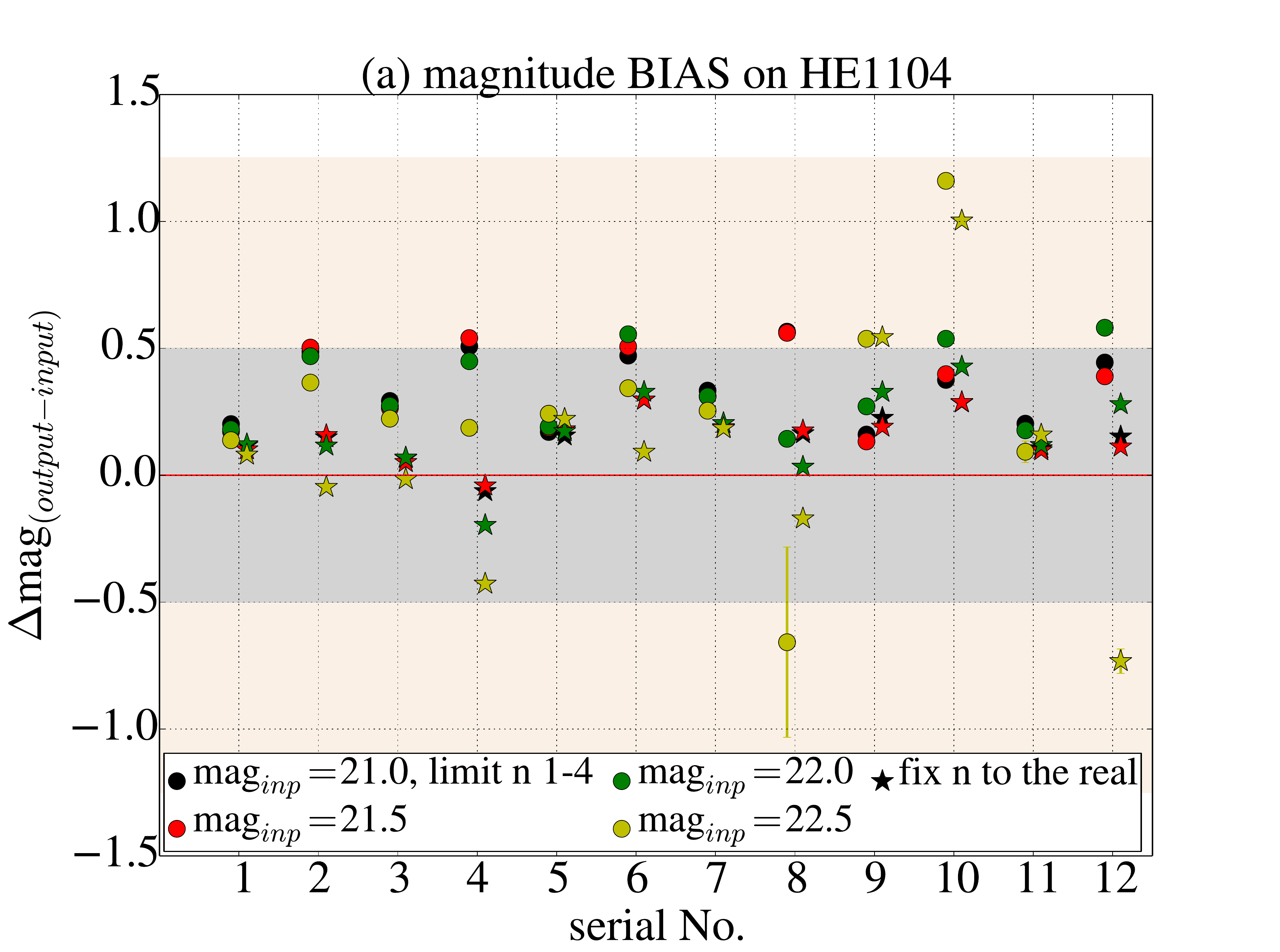}}&
{\includegraphics[width=9cm]{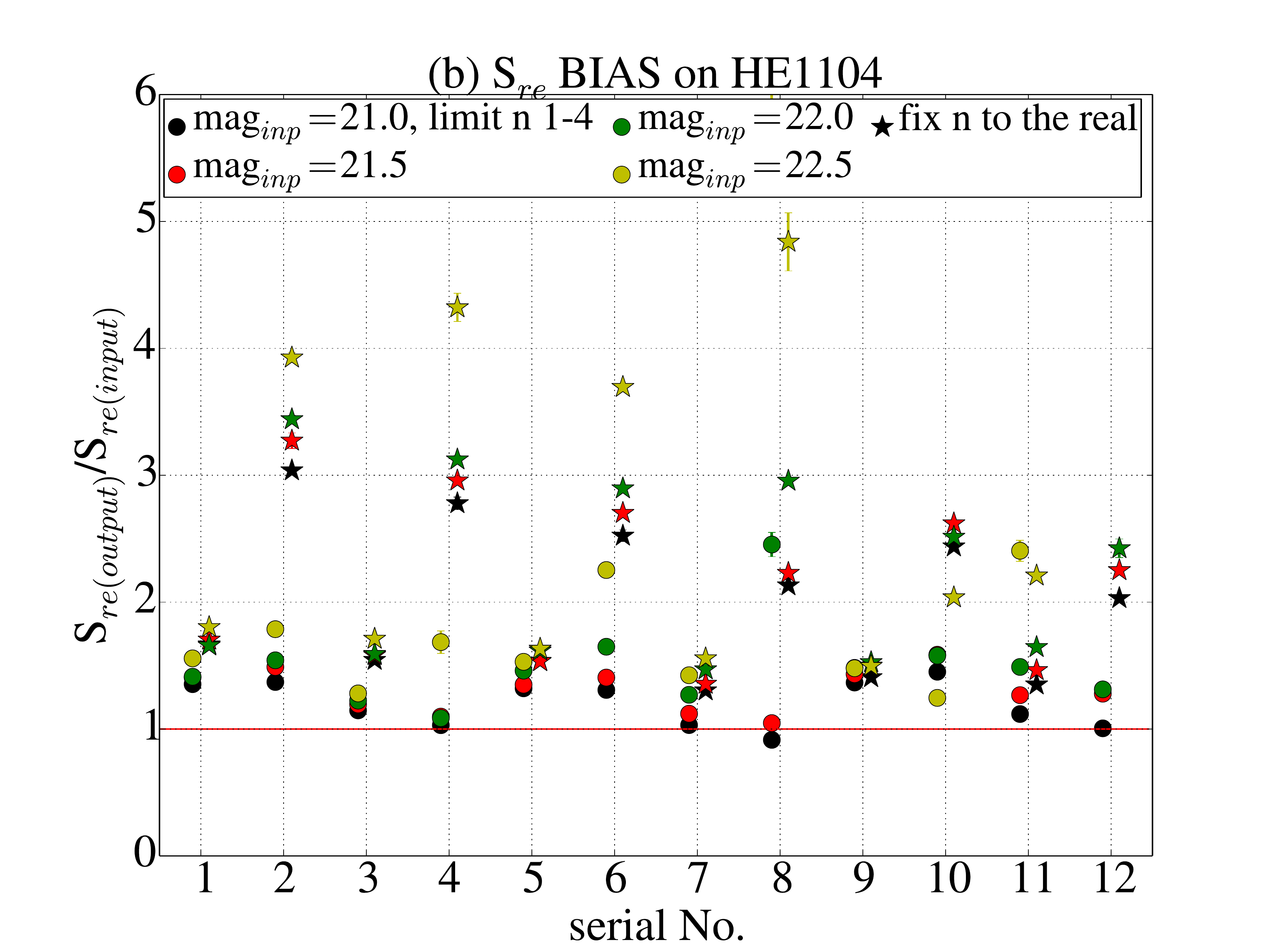}}\\
{\includegraphics[width=9cm]{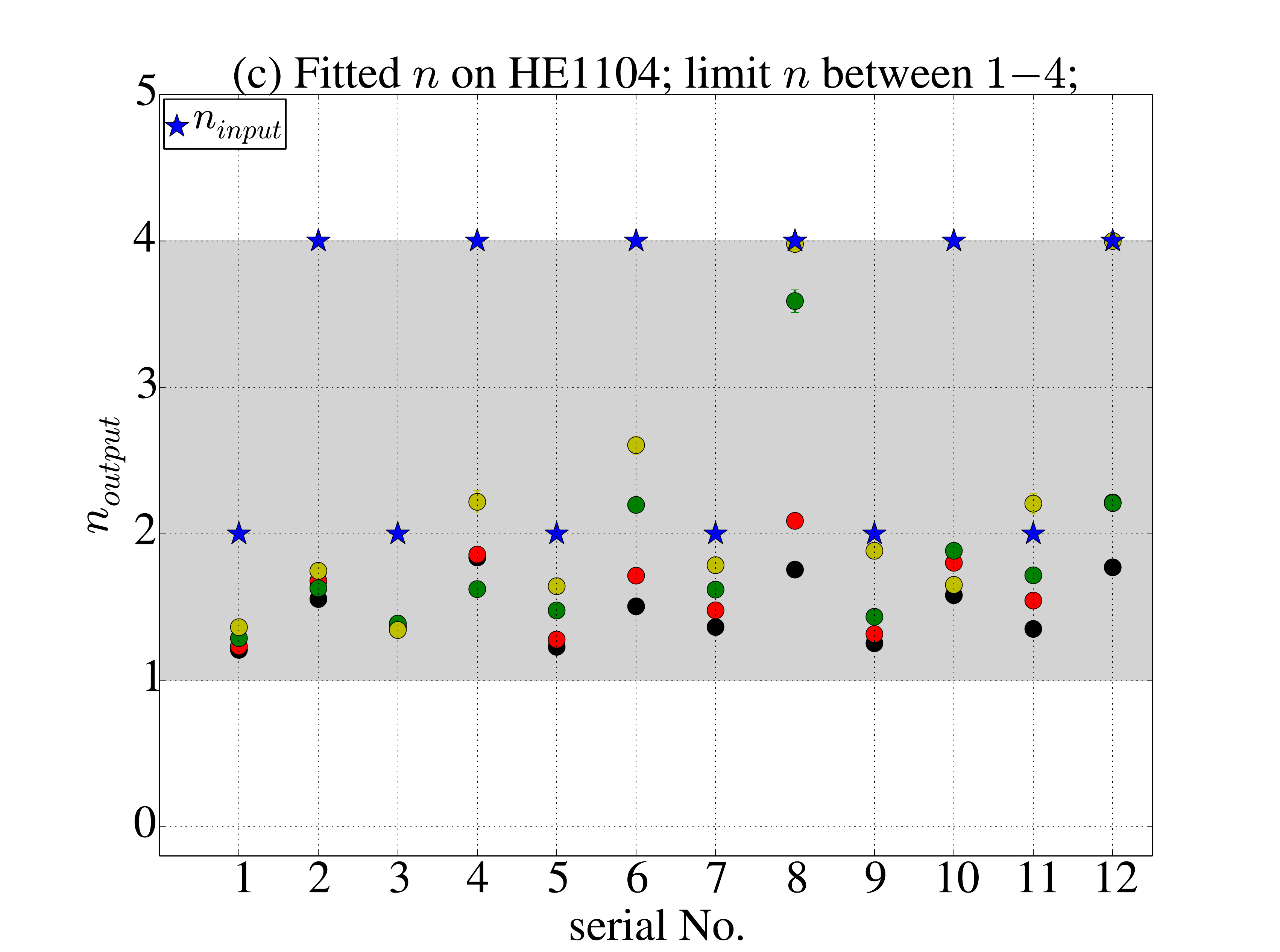}}&
{\includegraphics[width=9cm]{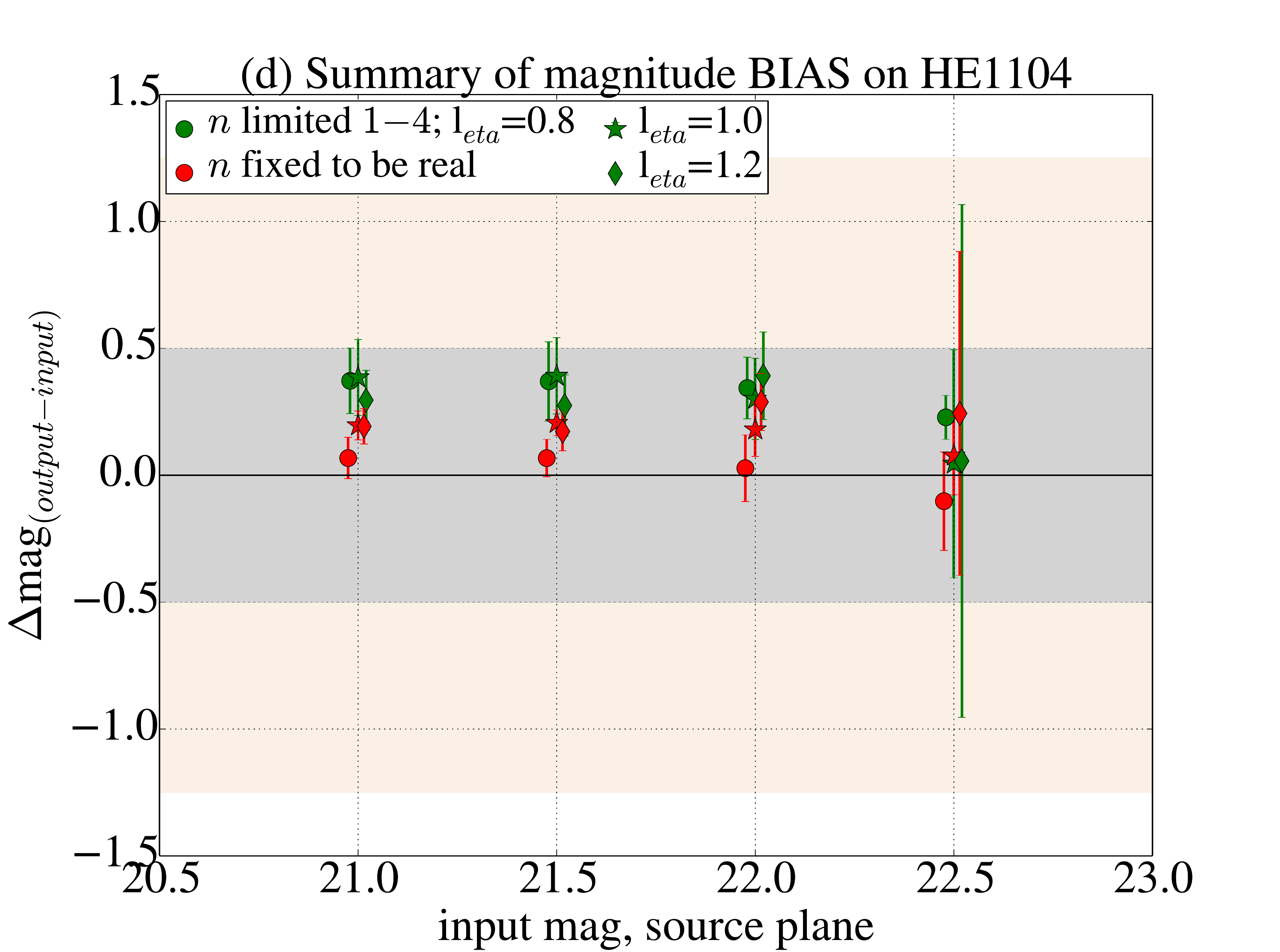}}\\
\end{tabular}
\caption{\label{fig:bias_HE1104}
Summary of the difference between input (true) and inferred model
parameters describing the lensed host galaxy light profile for HE1104.
For panels (a), (b) and (c), the serial numbers correspond to
different parameter configurations, as listed in Table \ref{serial}.
The $y$ axis shows the modelling results obtained with {\sc
Galfit}. Different colors represent different input magnitudes. The
filled circles are the results when allowing $n$ to float in the range
$1-4$; filled stars are obtained by fixing $n$ to the input value.
For panel (d), different colors represent different priors on $n$;
different shapes correspond to different $l_{eta}$ inputs.}
\end{figure*}

The results are summarized in Fig.~\ref{fig:bias_HE1104}. Panels (a),
(b), and (c) summarize the bias, i.e. the difference between inferred
and input value of each parameter, allowing $n$ to float in the range
$1-4$ (filled circles). The bias when $n$ is kept fixed to the input
value is represented as filled stars. The serial number representing
different parameter configurations in each panel is explained in
Table~\ref{serial}. The random uncertainties are estimated as the
scatter in the {\sc Galfit} inferred parameters for the 30 source
reconstructions taken from the \glee\ MCMC chain output. (This source
of random error dominates over the random uncertainties associated
with each pixel value, since it includes uncertainties in the lensing
parameters.) As expected, given the high signal to noise ratio of the
data, the random uncertainties are very small and the error bars are
barely visible in Panels (a), (b), and (c).

The main result is that the inferred magnitudes tend to be somewhat
larger (i.e. fainter) than the input value. The bias is found to be
larger than the estimated noise (illustrated by the error bars).
Crucially however, the bias is lower than $0.5$ magnitude, which is
within our target accuracy (i.e. $1.25$ magnitude) to ensure a useful
measurement of the \mbh $-L$ correlation. ($0.5$ and $1.25$ magnitude
area are filled with two colors in the figures.)

It is interesting to note that the bias does not depend strongly on
the effective radius of the simulated host galaxy, though lower
surface brightness galaxies are more difficult to detect, and a
smaller fraction of the entire galaxy is actually reconstructed. This
means the mask region we selected is large enough to enclose a
sufficient portion of the light. Moreover, the inferred S\'ersic
index $n$ tends to be smaller
than the input value. This bias is likely
due to regularization that tends to suppress sharp features and
therefore \glee\ prefers a flatter profile, whenever possible.

As we will discuss later, HE1104 is the hardest system to model among
the systems with WFC3-IR imaging, owing to the fact that the lens
configuration is a double - and thus we have considerably less
information than in quads - and that the AGN images outshine the host
galaxy by $\sim2-4$ magnitudes. In spite of all these challenges, the
simulations here show us the residual uncertainty is within 0.5
magnitudes ($<$0.2 dex), whether fixing $n$ or not, which is much
smaller than the uncertainty on \mbh\ and thus acceptable for our
purposes.

\begin{table}
\begin{center} \scriptsize
  \begin{threeparttable}
    \caption{Combination of parameters for each configuration}\label{serial}
     \begin{tabular}{cccc|cccc}
     \hline\hline
$Serial$&$n$&$S_{re}$&$l_{eta}$&$Serial$&$n$&$S_{re}$&$l_{eta}$\\
\hline
$No.1:$&$ 2$&$0.39$&$0.8$&$No.2:$&$4$&$0.39$&$0.8$\\
$No.3:$&$ 2$&$0.78$&$0.8$&$No.4:$&$4$&$0.78$&$0.8$\\
$No.5:$&$ 2$&$0.39$&$1.0$&$No.6:$&$4$&$0.39$&$1.0$\\
$No.7:$&$ 2$&$0.78$&$1.0$&$No.8:$&$4$&$0.78$&$1.0$\\
$No.9:$&$ 2$&$0.39$&$1.2$&$No.10:$&$4$&$0.39$&$1.2$\\
$No.11:$&$ 2$&$0.78$&$1.2$&$No.12:$&$4$&$0.78$&$1.2$\\
     \hline
     \end{tabular}
     \begin{tablenotes}
      \small 
       \item {Note:$-$ List of parameter sets for each configuration. We use WFC3-IR images as an example. For WFC3-UVIS and ACS images the corresponding values for $S_{re}$ are slightly different and given in Table~\ref{para_config}.}
     \end{tablenotes}
  \end{threeparttable}
\end{center}
\end{table}

\subsubsection{HE0435}
\label{sssec:he0435}
\begin{figure*}
\centering
\begin{tabular}{c c}
{\includegraphics[width=9cm]{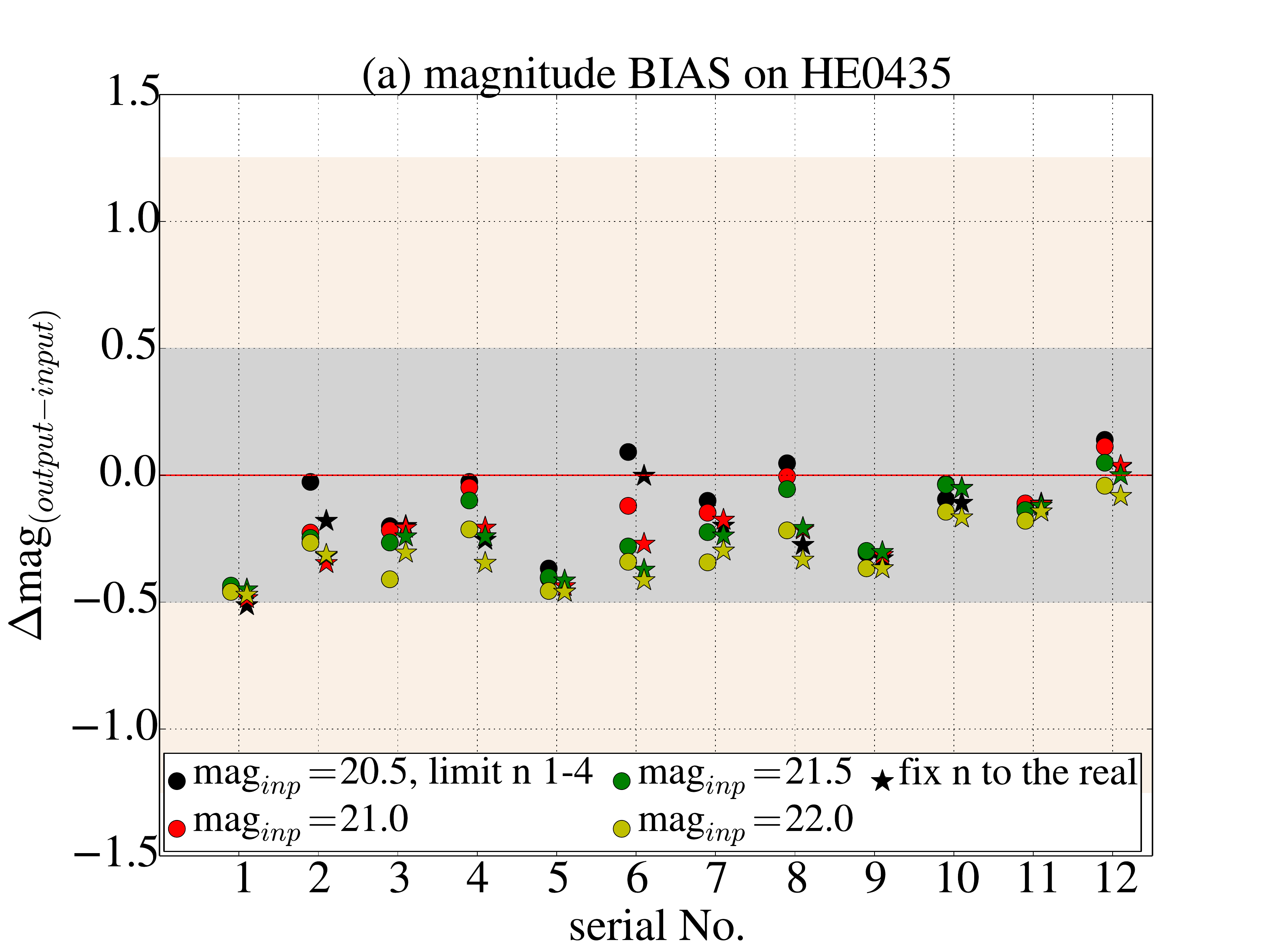}}&
{\includegraphics[width=9cm]{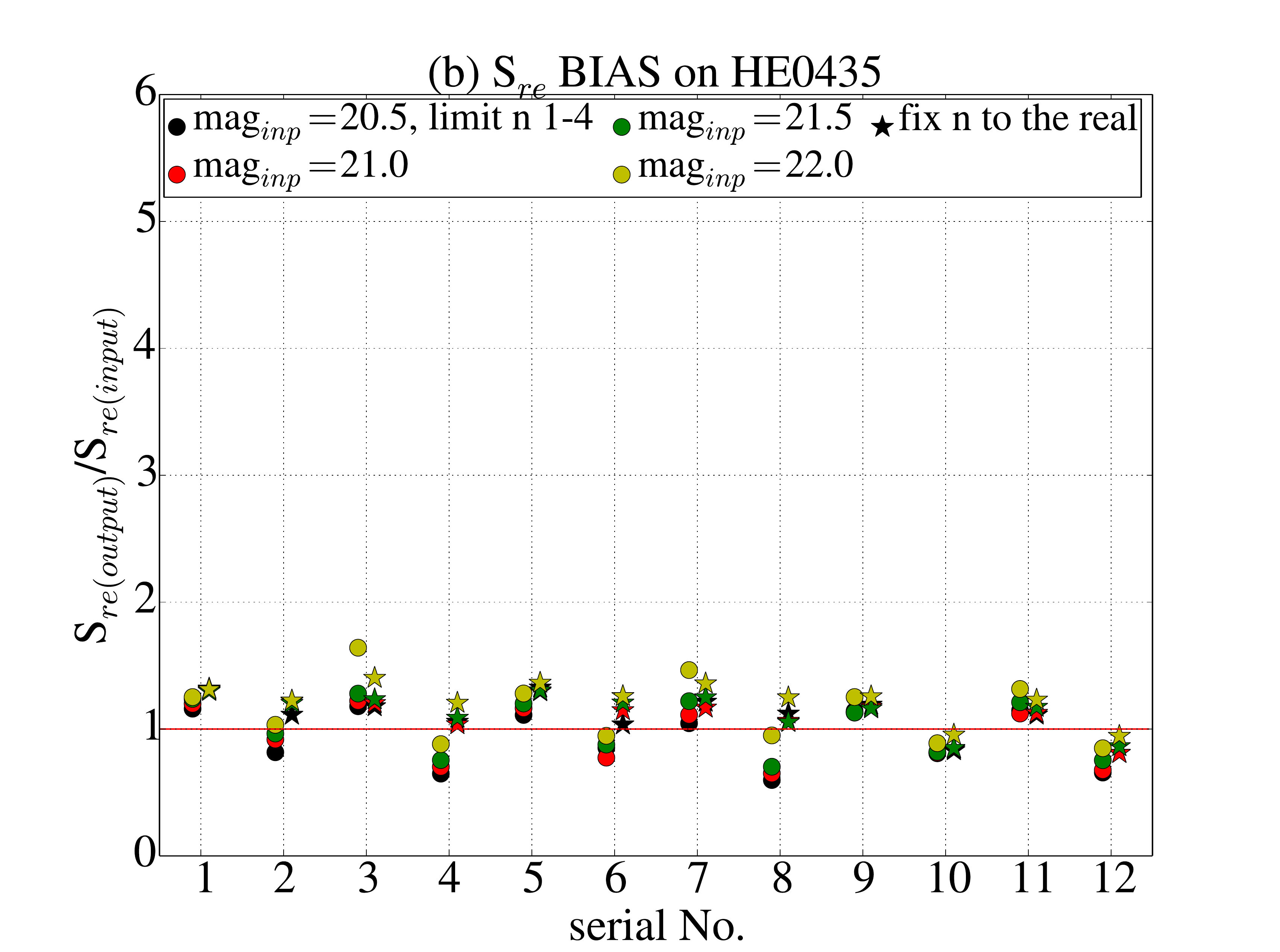}}\\
{\includegraphics[width=9cm]{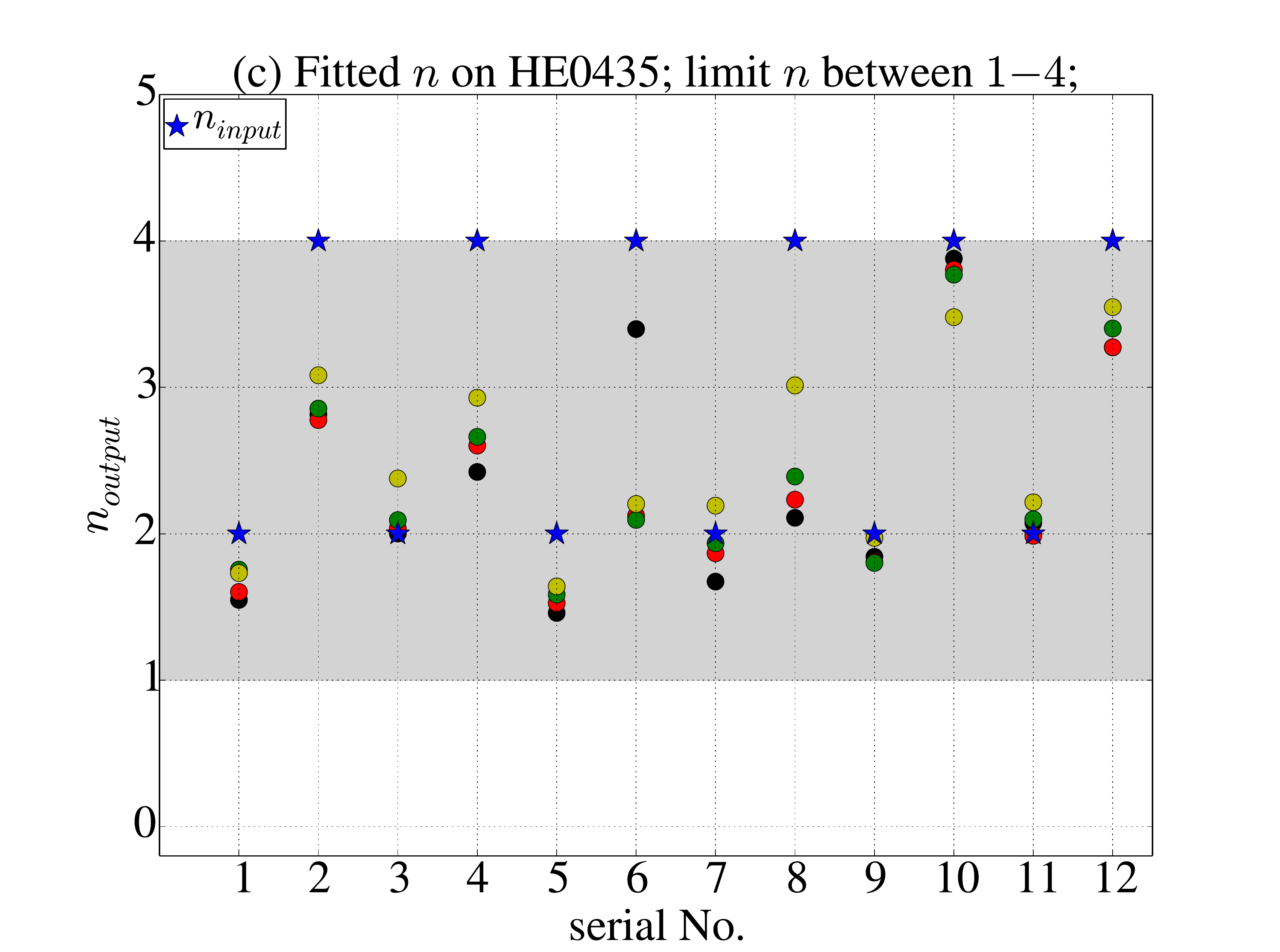}}&
{\includegraphics[width=9cm]{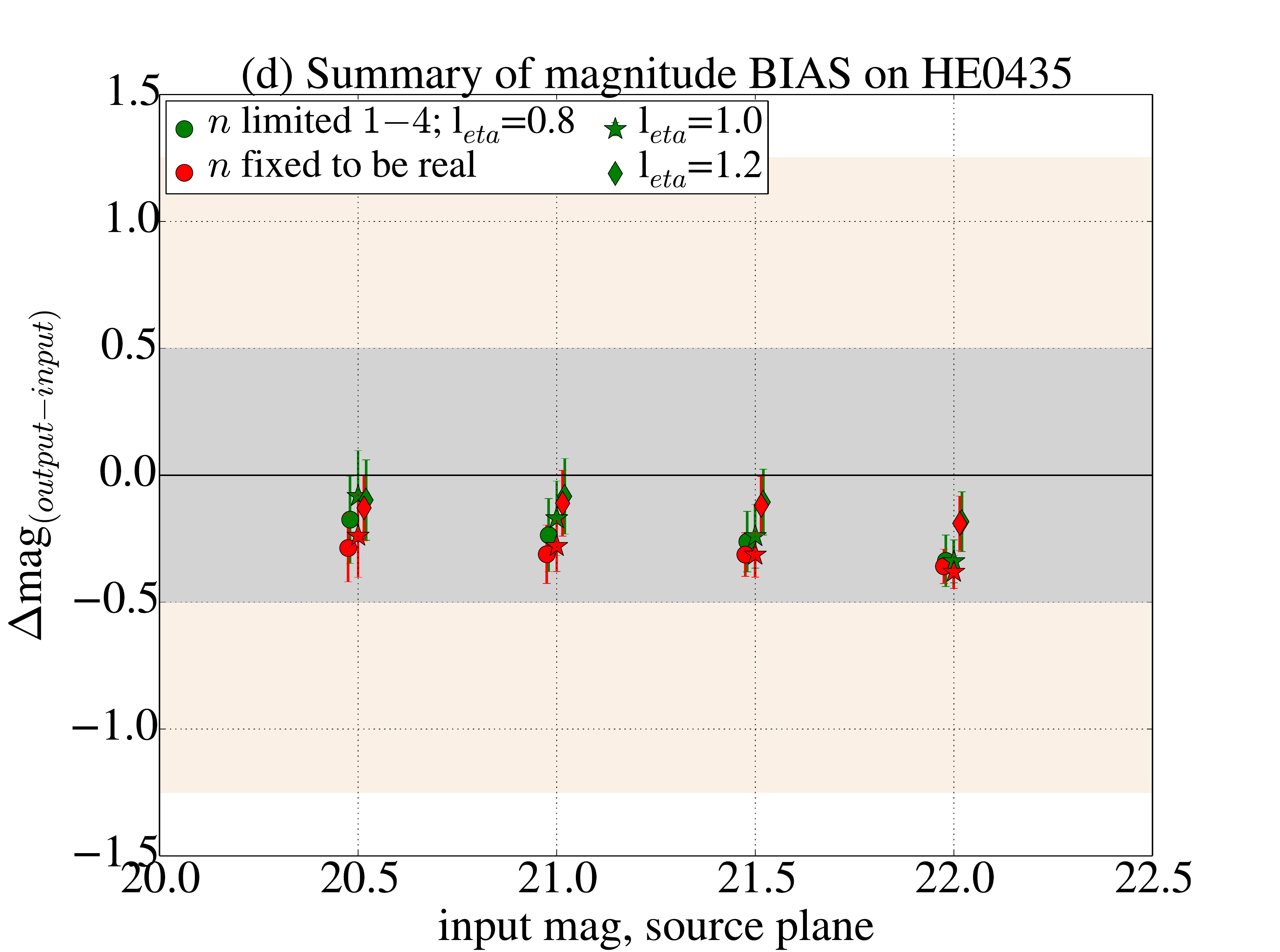}}\\
\end{tabular}
\caption{\label{fig:bias_HE0435} Results for HE0435. See caption of Fig.~\ref{fig:bias_HE1104} for details.}
\end{figure*}

The results for HE0435 are shown in Fig.~\ref{fig:bias_HE0435}, using
the same format as in the previous section. We note that the bias is
reduced 
relative to HE1104, for two reasons: i) the system has
four lensed images and there is thus more information available to
reconstruct the host galaxy surface brightness; ii) the contrast
between host galaxy magnitude and AGN magnitude is reduced to $\sim
-1-2$ magnitudes.

\subsubsection{WFI2033}
\label{sssec:wfi2033}
\begin{figure*}
\centering
\begin{tabular}{c c}
{\includegraphics[width=9cm]{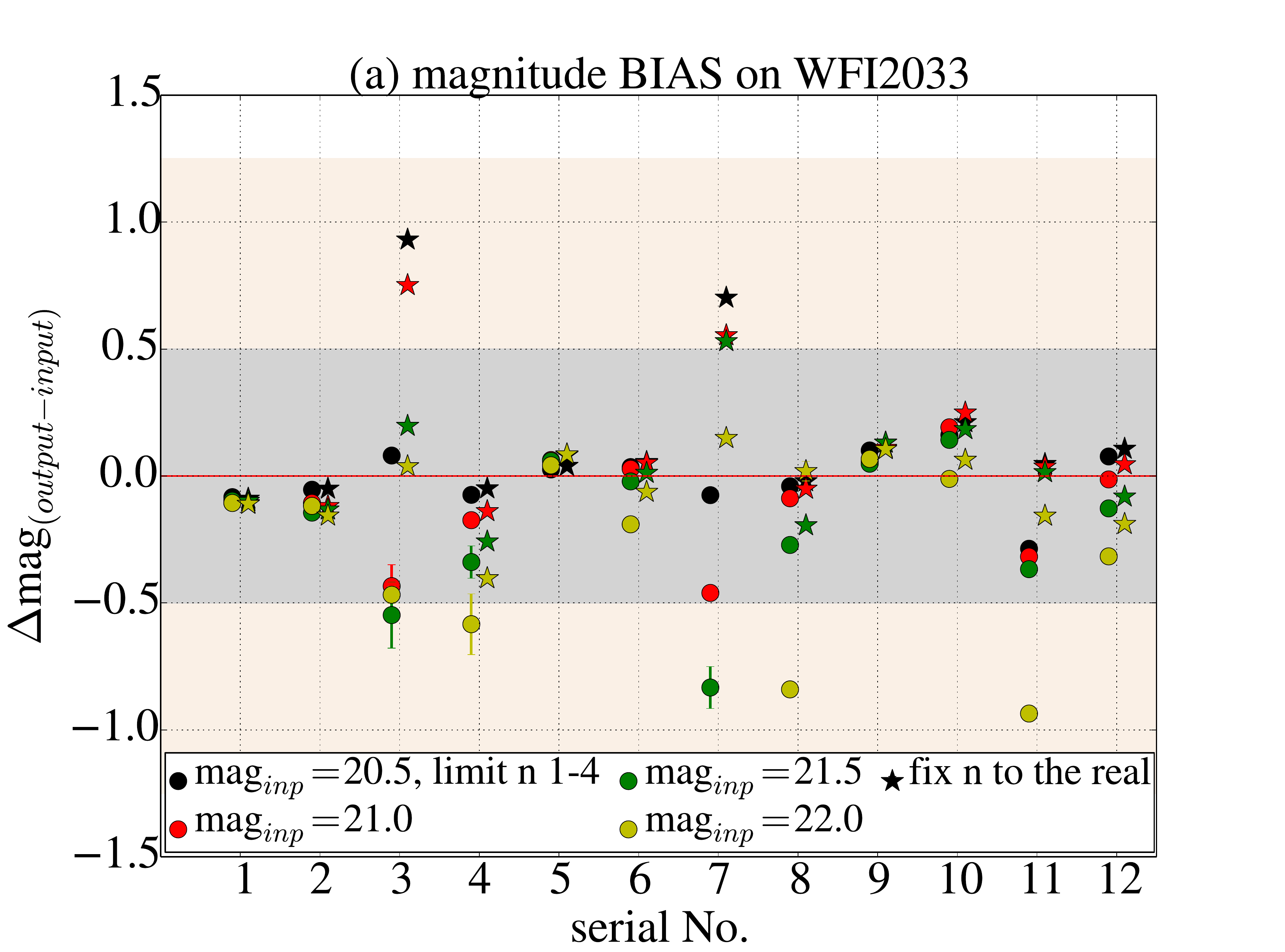}}&
{\includegraphics[width=9cm]{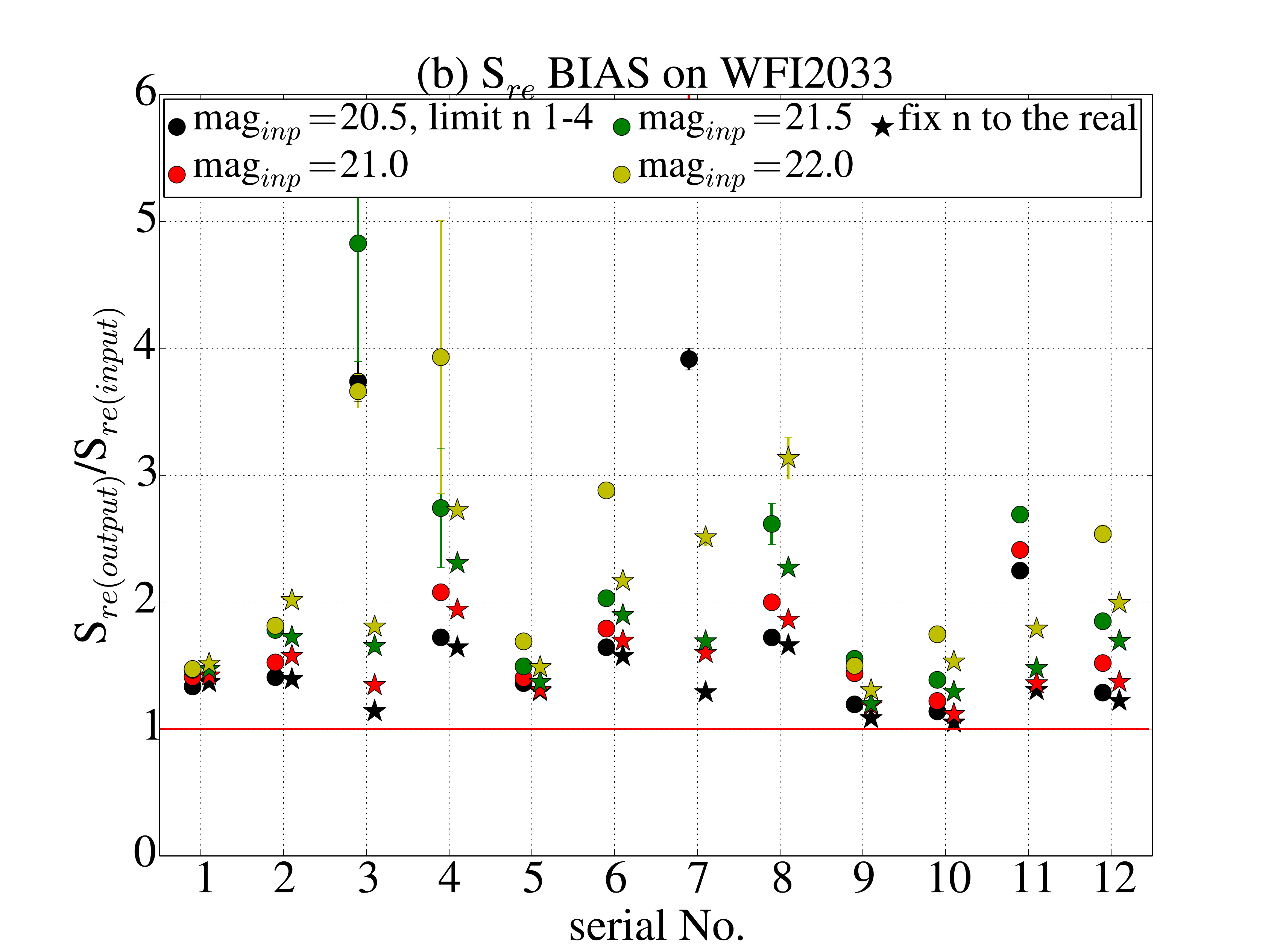}}\\
{\includegraphics[width=9cm]{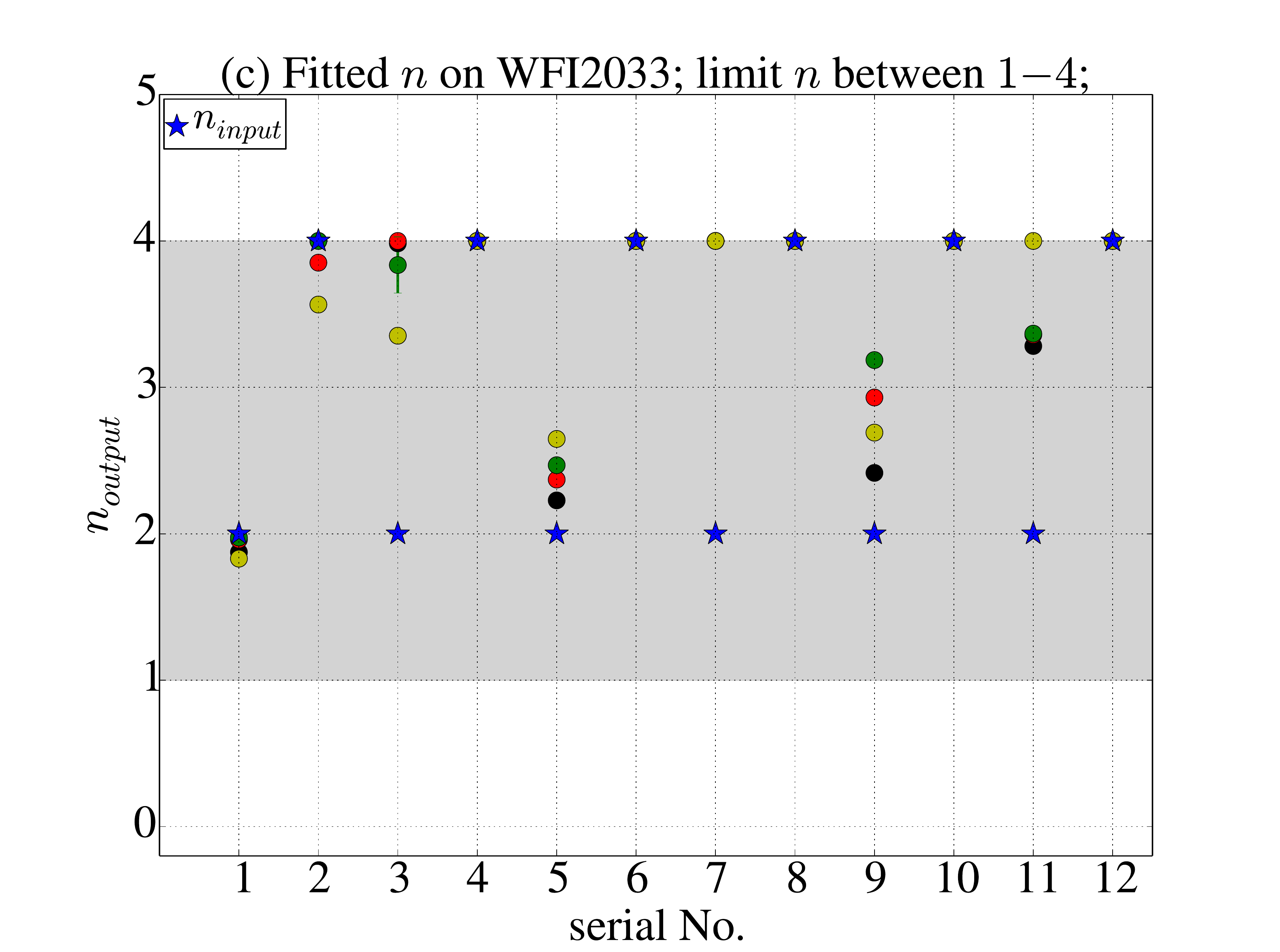}}&
{\includegraphics[width=9cm]{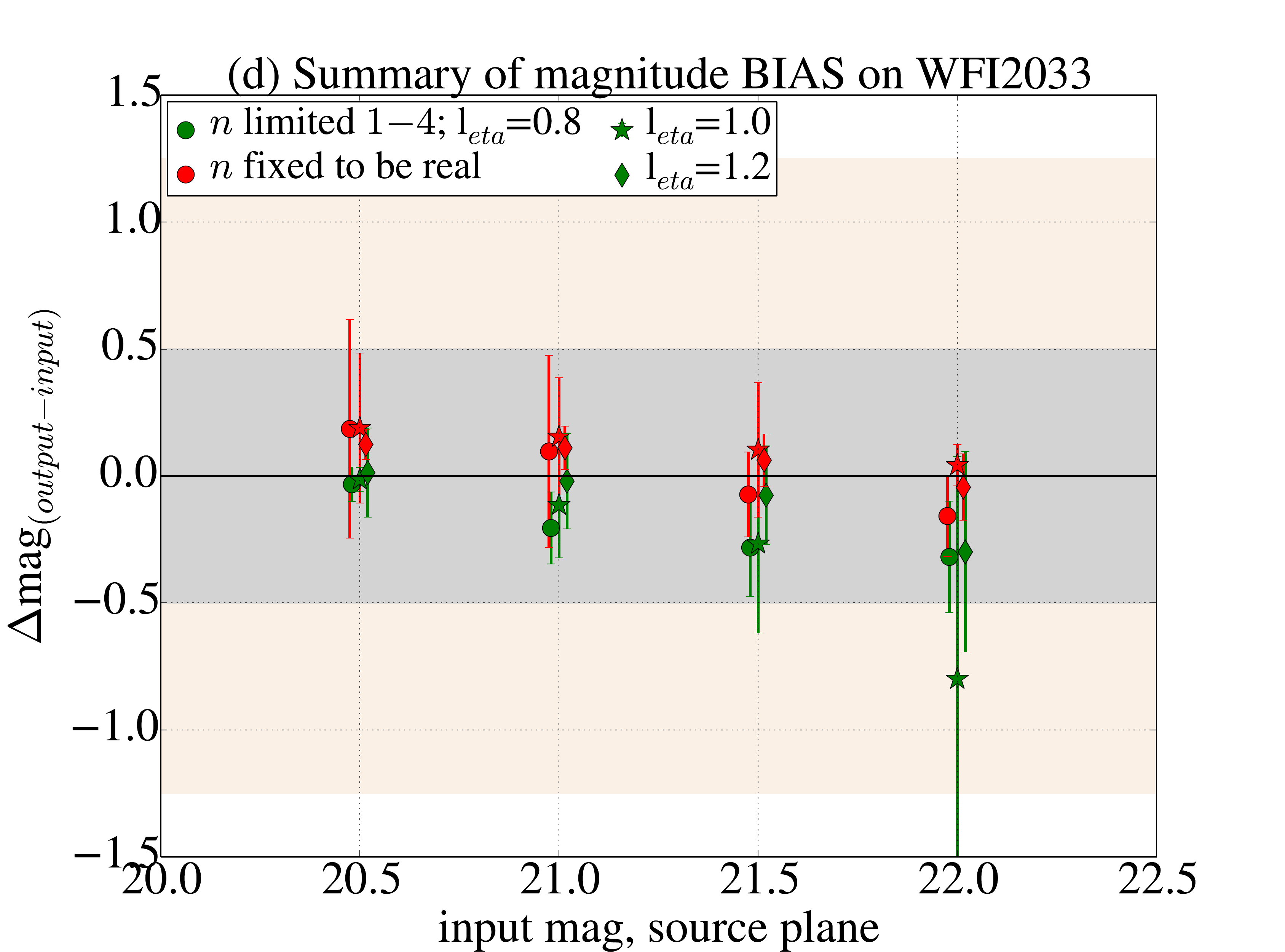}}\\
\end{tabular}
\caption{\label{fig:bias_WFI2033} Results for WFI2033. See caption of Fig.~\ref{fig:bias_HE1104} for details.}
\end{figure*}

Fig.~\ref{fig:bias_WFI2033} summarizes the results for WFI2033. The
host galaxy magnitude can be reliably inferred with a much smaller bias
than our target accuracy.

\subsubsection{SDSS1206}
\label{sssec:sdss1206}
\begin{figure*}
\centering
\begin{tabular}{c c}
{\includegraphics[width=9cm]{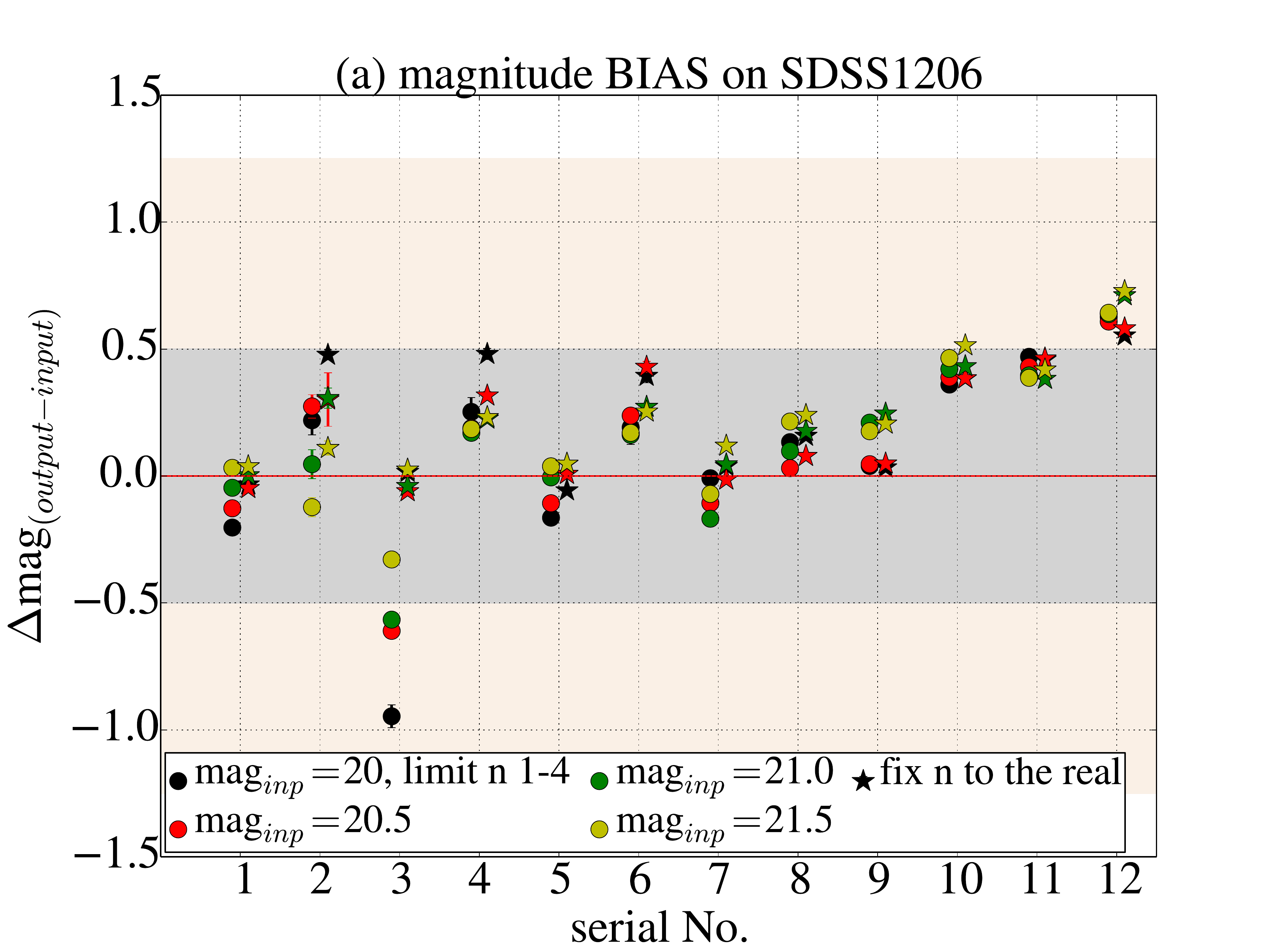}}&
{\includegraphics[width=9cm]{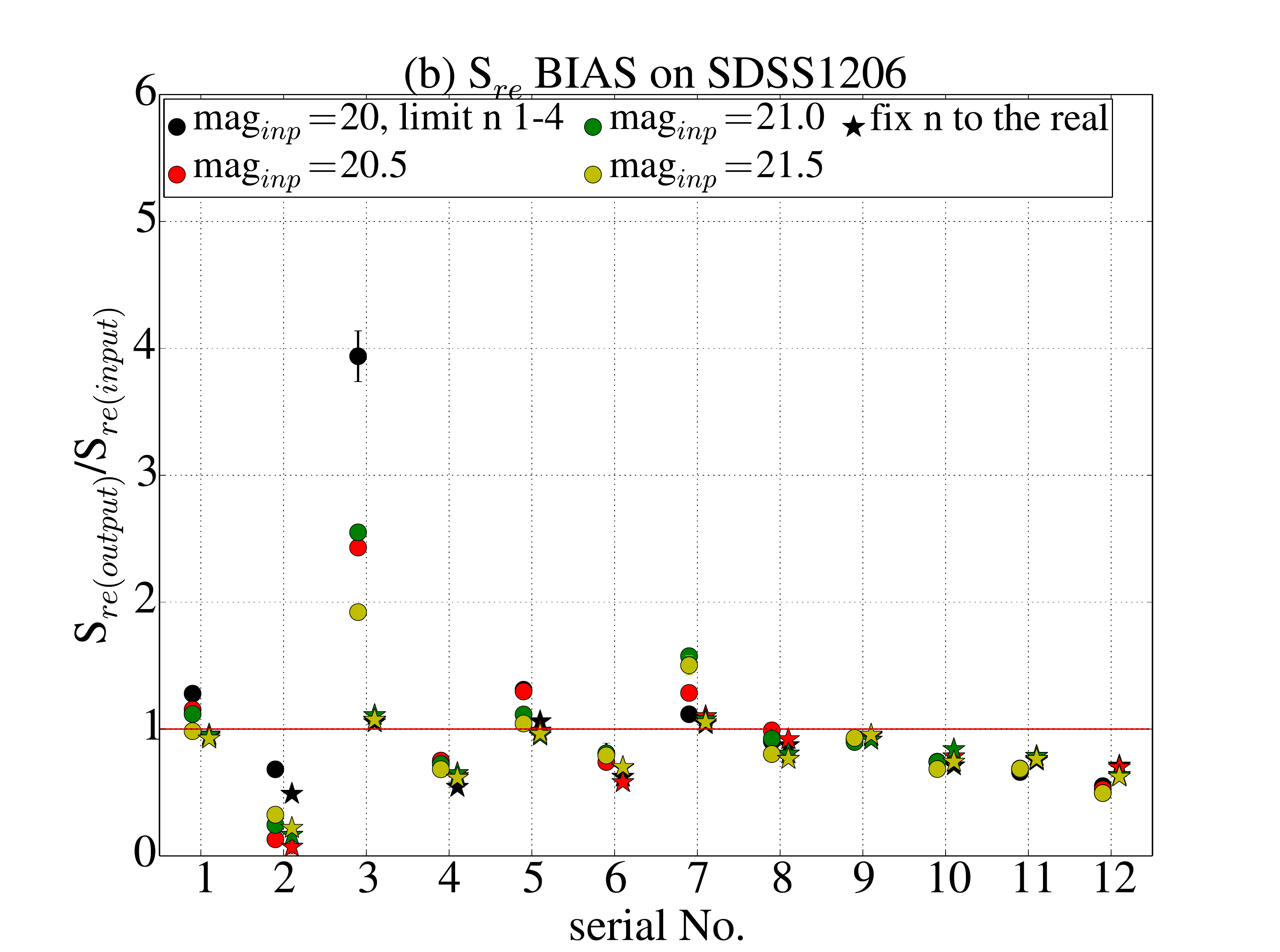}}\\
{\includegraphics[width=9cm]{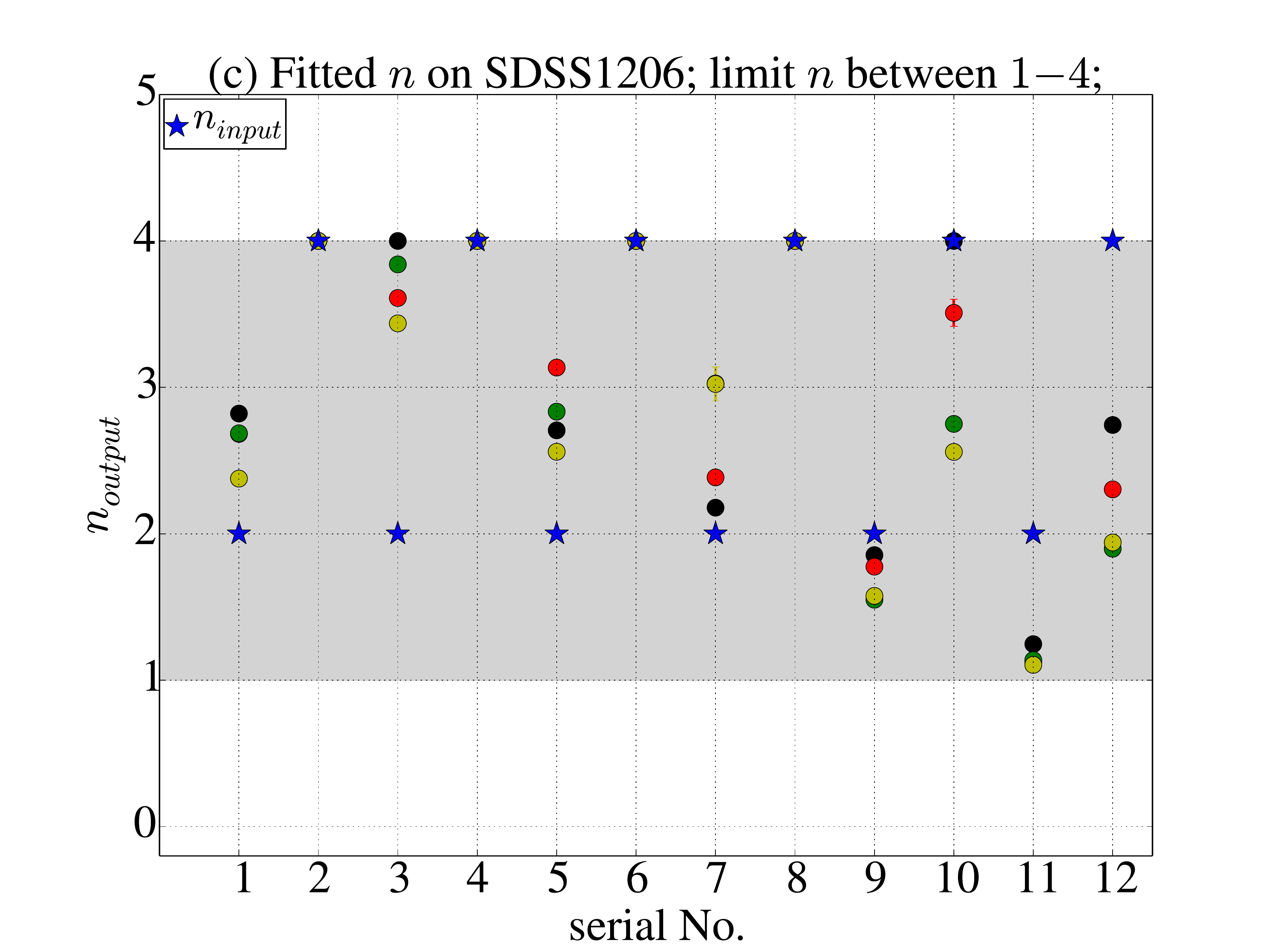}}&
{\includegraphics[width=9cm]{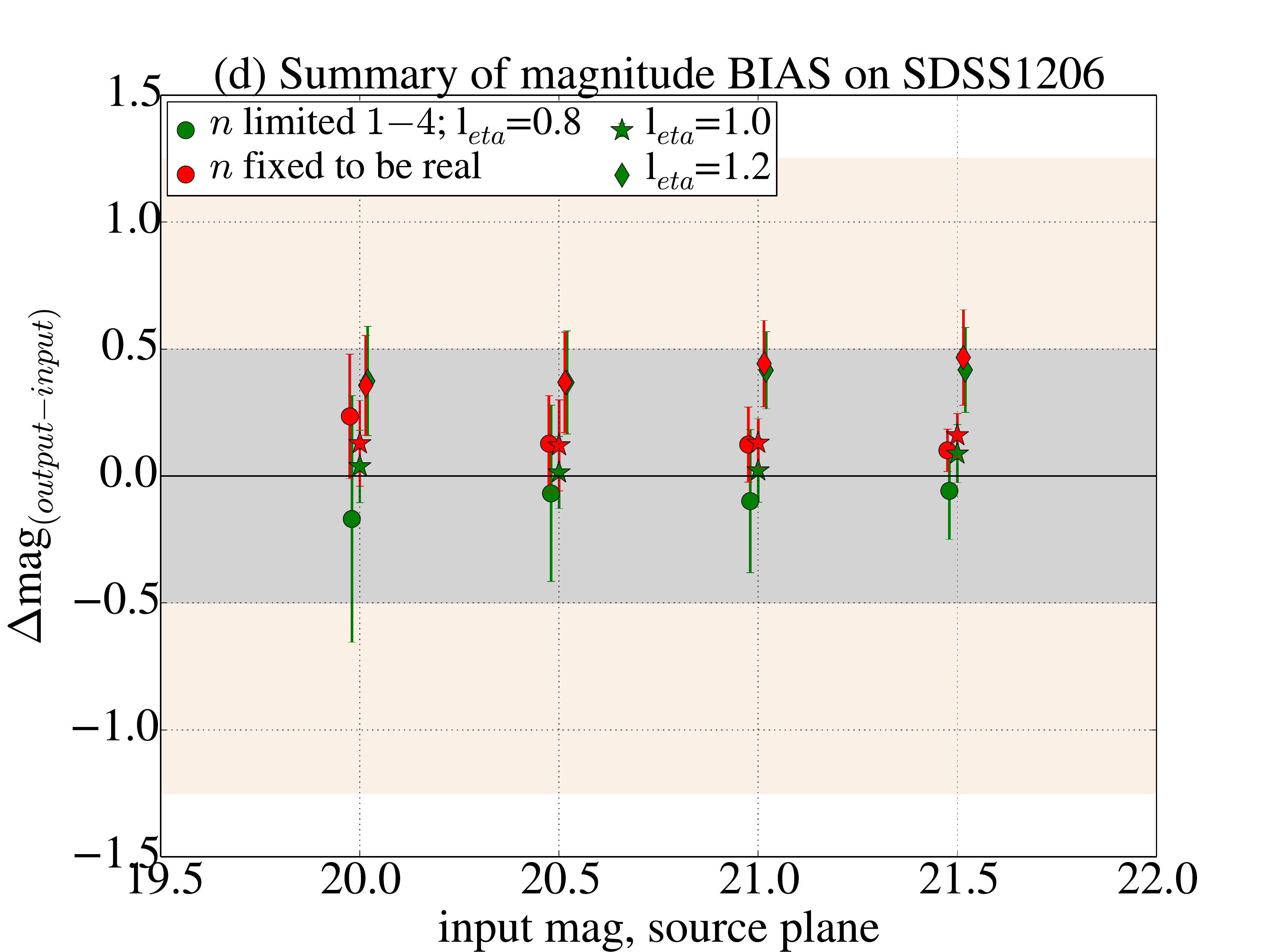}}\\
\end{tabular}
\caption{\label{fig:bias_SDSS1206} Results for SDSSJ1206. See caption of Fig.~\ref{fig:bias_HE1104} for details.}
\end{figure*}

SDSS1206 is a particularly interesting system for the study of the
lensed host. While most of the host galaxy is quadruply imaged, the
AGN falls just outside the inner caustic and is thus only doubly
imaged \citep{Agn++16}. As expected, the accuracy of the inferred
magnitudes and effective radius are similar to that obtained for
quadruply imaged AGNs and the bias is smaller than our target
accuracy. The results are summarized in Fig.~\ref{fig:bias_SDSS1206}.

\subsubsection{SDSS0246}
\label{sssec:sdss0246}
\begin{figure*}
\centering
\begin{tabular}{c c}
{\includegraphics[width=9cm]{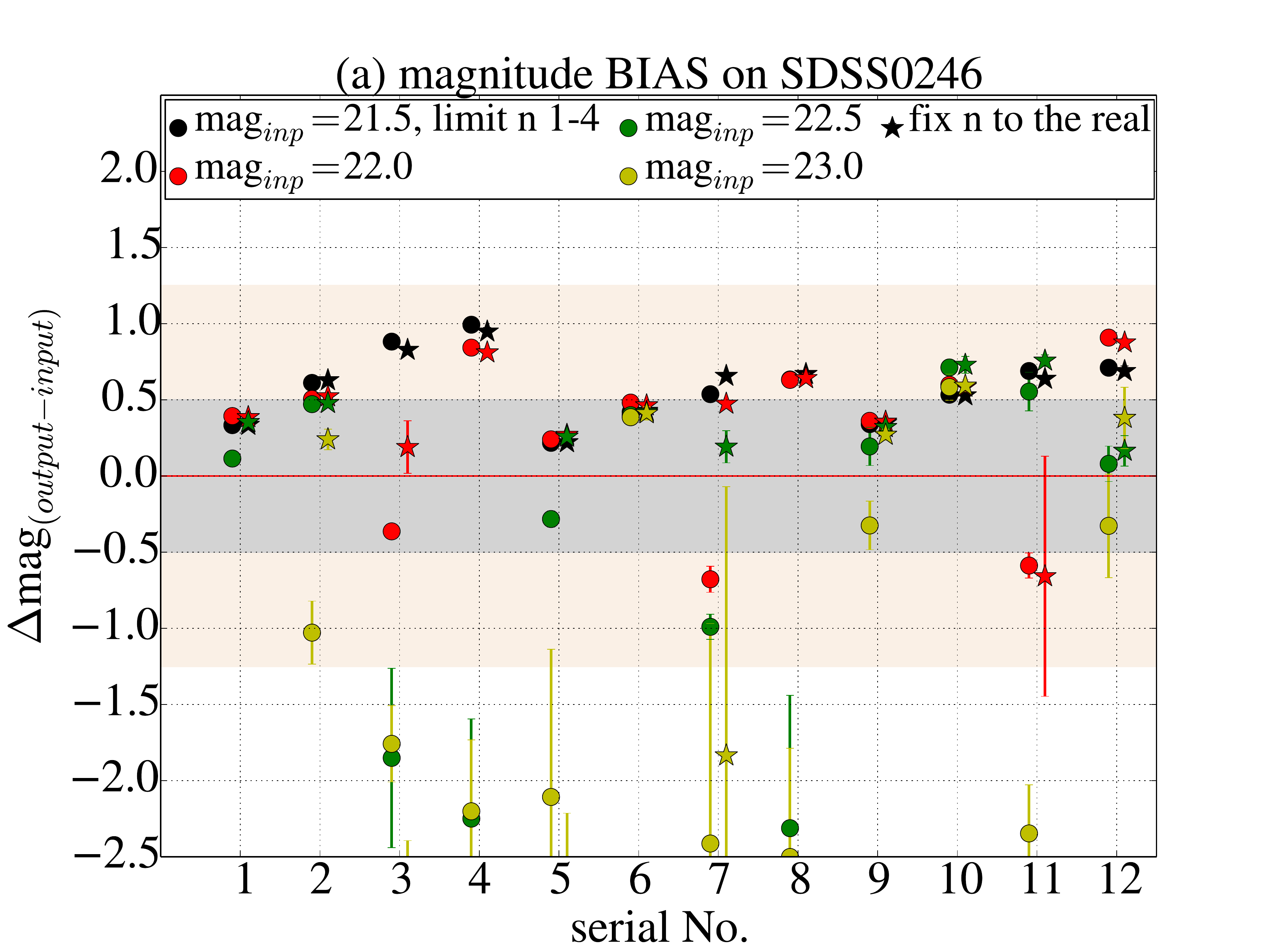}}&
{\includegraphics[width=9cm]{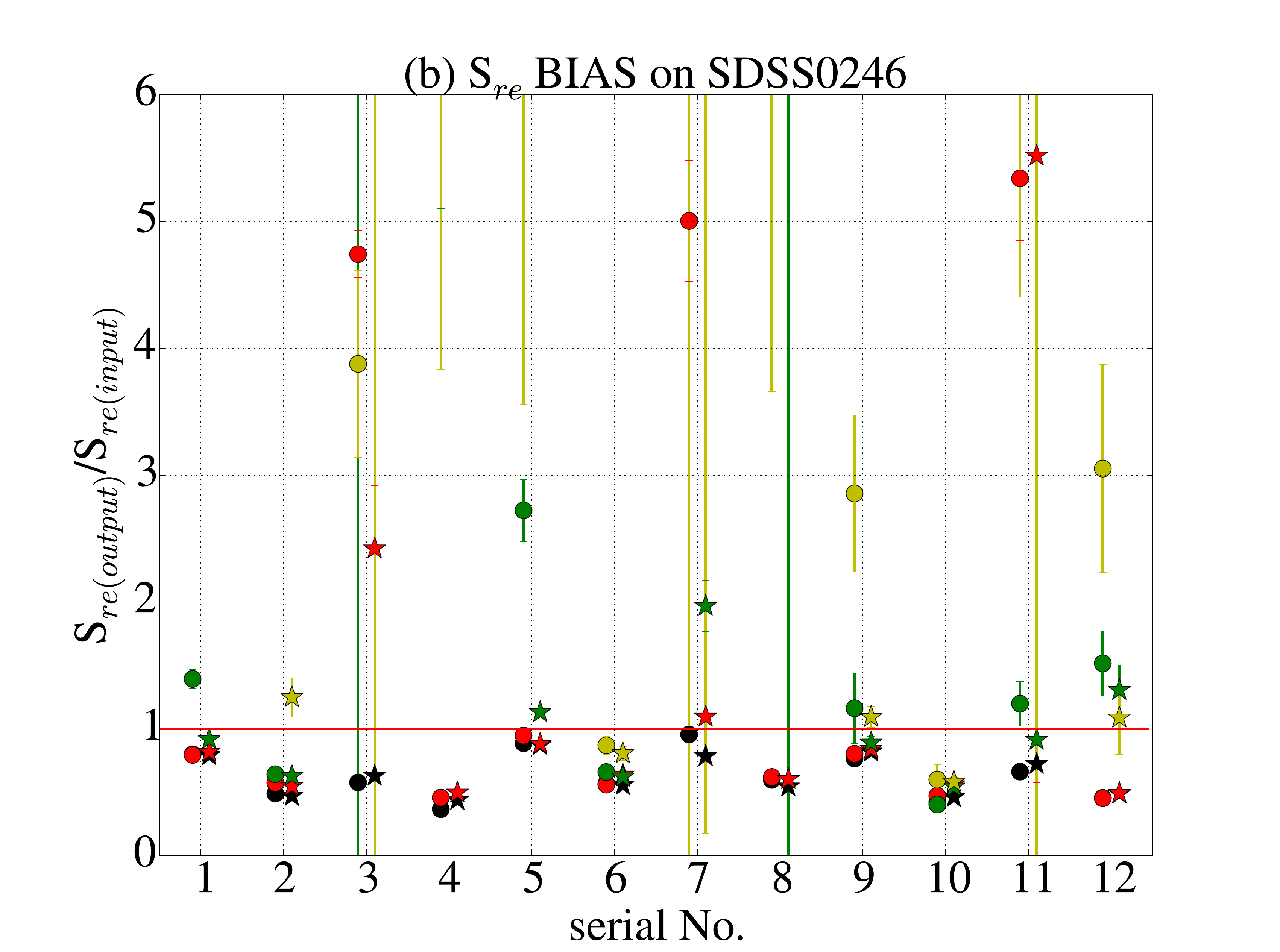}}\\
{\includegraphics[width=9cm]{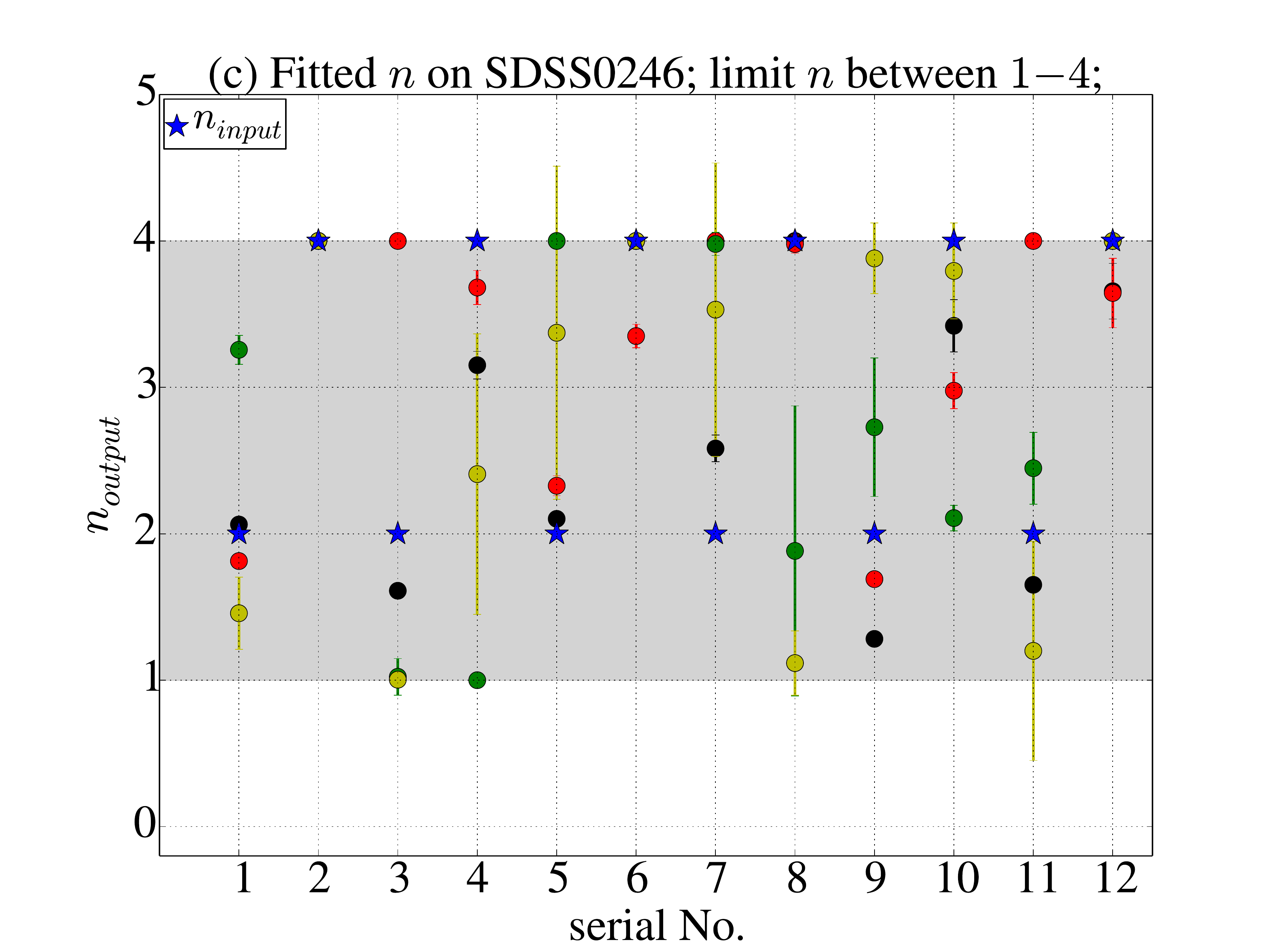}}&
{\includegraphics[width=9cm]{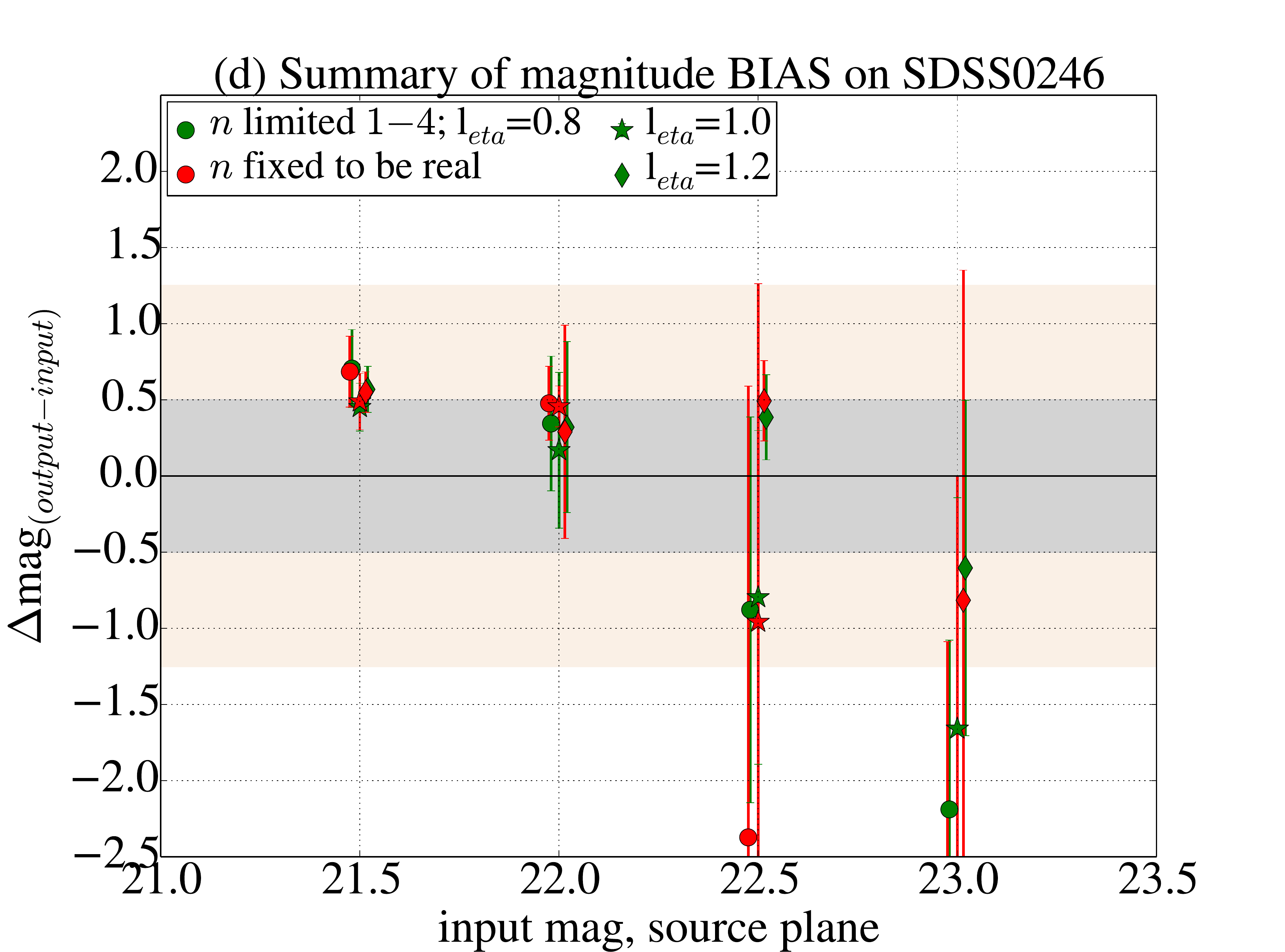}}\\
\end{tabular}
\caption{\label{fig:bias_SDSS0246} Results for SDSSJ0246. See caption of Fig.~\ref{fig:bias_HE1104} for details.}
\end{figure*}

The system SDSS0246 was imaged with WFC3-UVIS, and it has a much
fainter arc compared with other systems. Even though the pixels are
smaller than with WFC3-IR and the resolution is higher, the available
region for modelling the arc in the mask is quite limited in angular
size. In spite of this serious challenge, \glee\ successfully
reconstructed the host image within our target accuracy, at least for
the brighter range of source magnitudes considered here. The final
results are summarized in Fig.~\ref{fig:bias_SDSS0246}.


Not surprisingly, the bias is larger than for other systems. As we can
see, for brighter host set (21.5, 22.0 and part of 22.5 mag), the
inferred magnitudes are accurate within $1$ mag, i.e. within our
target of 1.25 mags.

For the fainter host configuration (e.g. 22.5 and 23 Mag with large
$S_{re}$), the inferred magnitudes are much smaller than the
input. The reasons can be easily understood in the following way. The
reconstructions for these images are too faint with an irregular light
distribution. In this case, {\sc Galfit} tries to fit an
unrealistically large $S_{re}$ (see Fig.~\ref{fig:bias_SDSS0246}-(b))
which corresponds to overestimating the brightness of the
source. Indeed, this kind of modelling bias can be offset when the
$S_{re}$ are fixed to the truth or limited within $1''$. However, the
final accuracy of our magnitude inference will be driven by our
knowledge of the prior on $S_{re}$. Our conclusion is that if the host
galaxy is inferred to be as faint or fainter than 22.5 in the final
analysis, the inference cannot be considered reliable given the
current data quality.

\subsubsection{HS2209}
\label{sssec:sdss2209}

\begin{figure*}
\centering
\begin{tabular}{c c}
{\includegraphics[width=9cm]{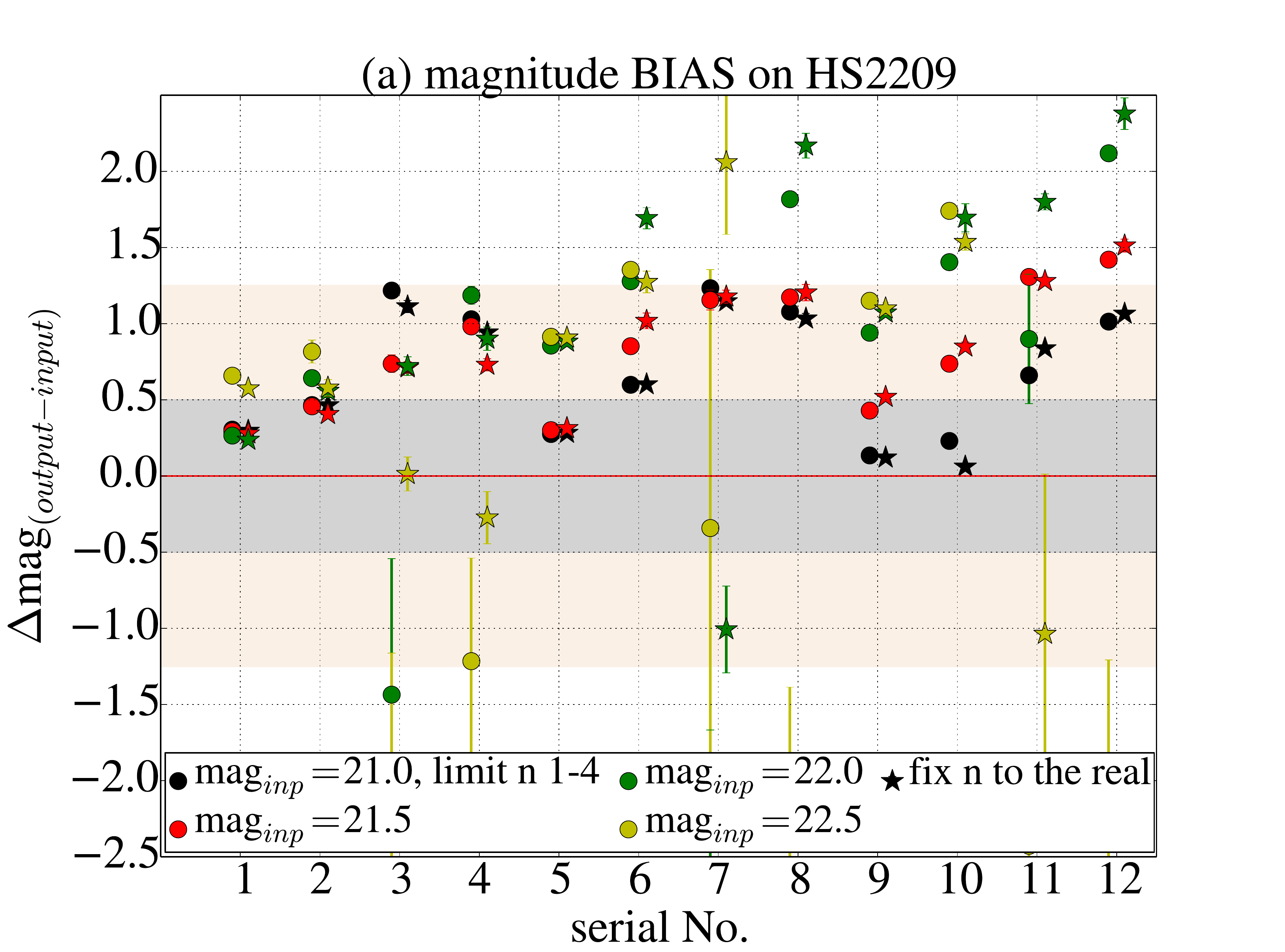}}&
{\includegraphics[width=9cm]{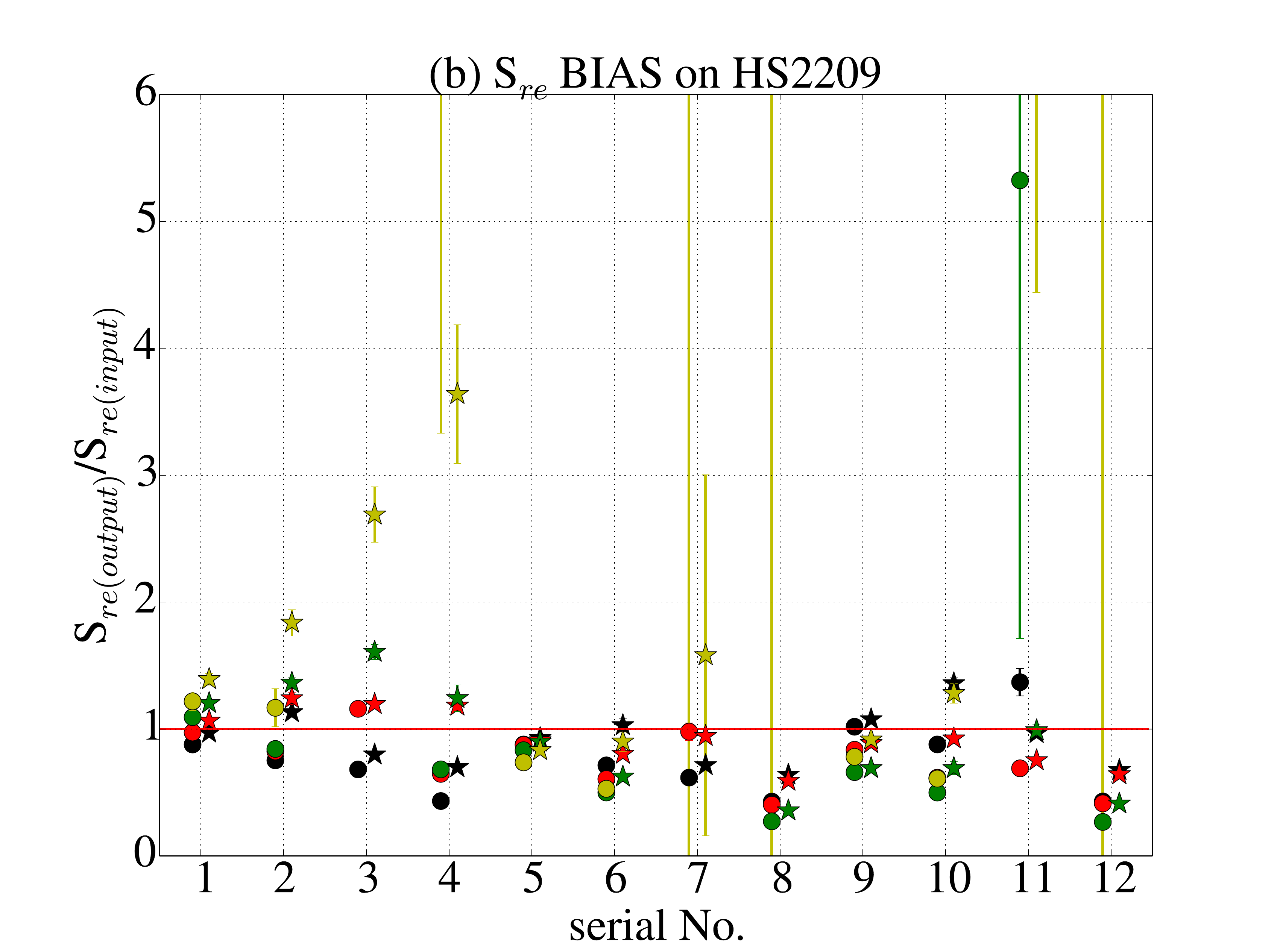}}\\
{\includegraphics[width=9cm]{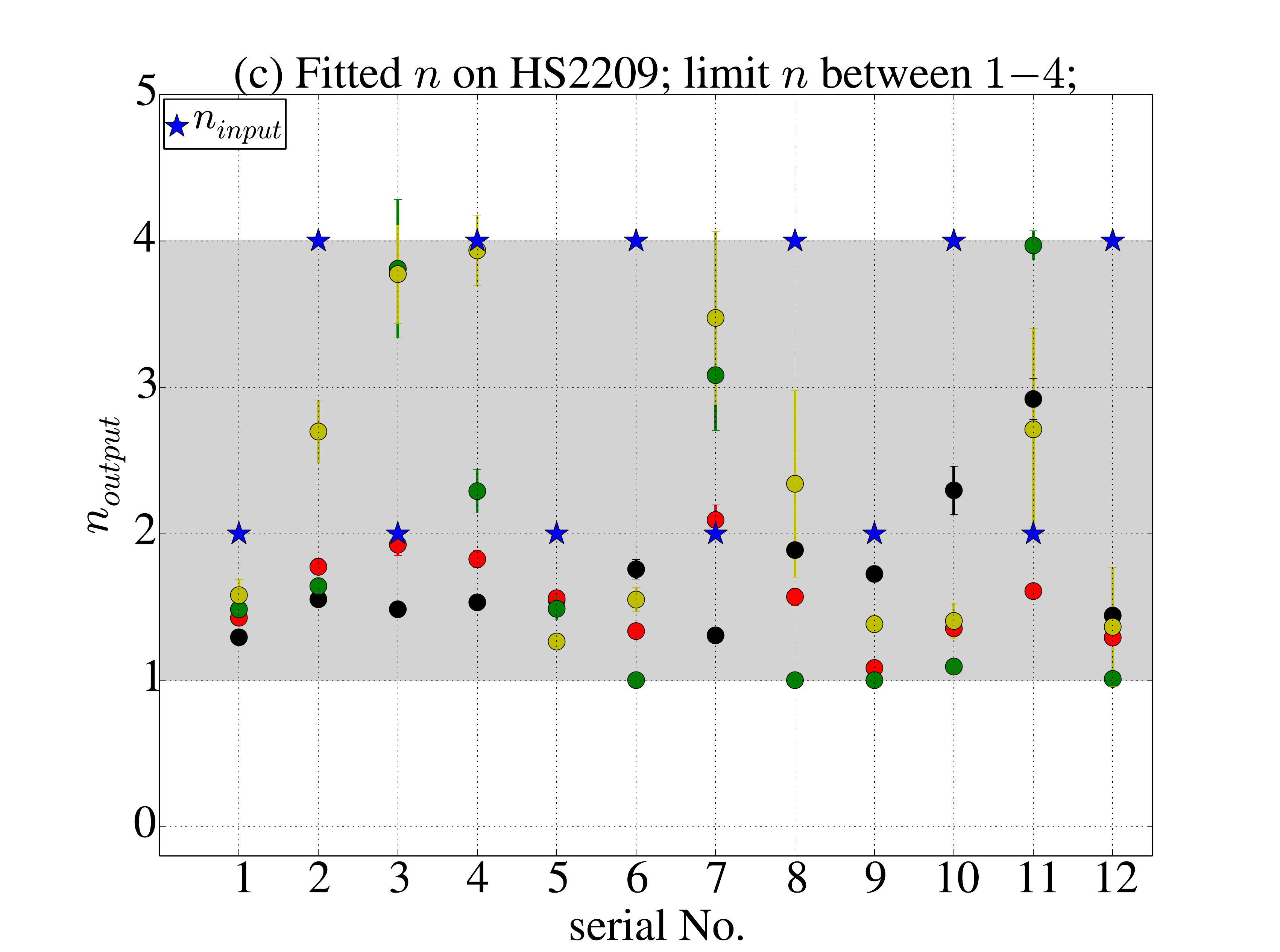}}&
{\includegraphics[width=9cm]{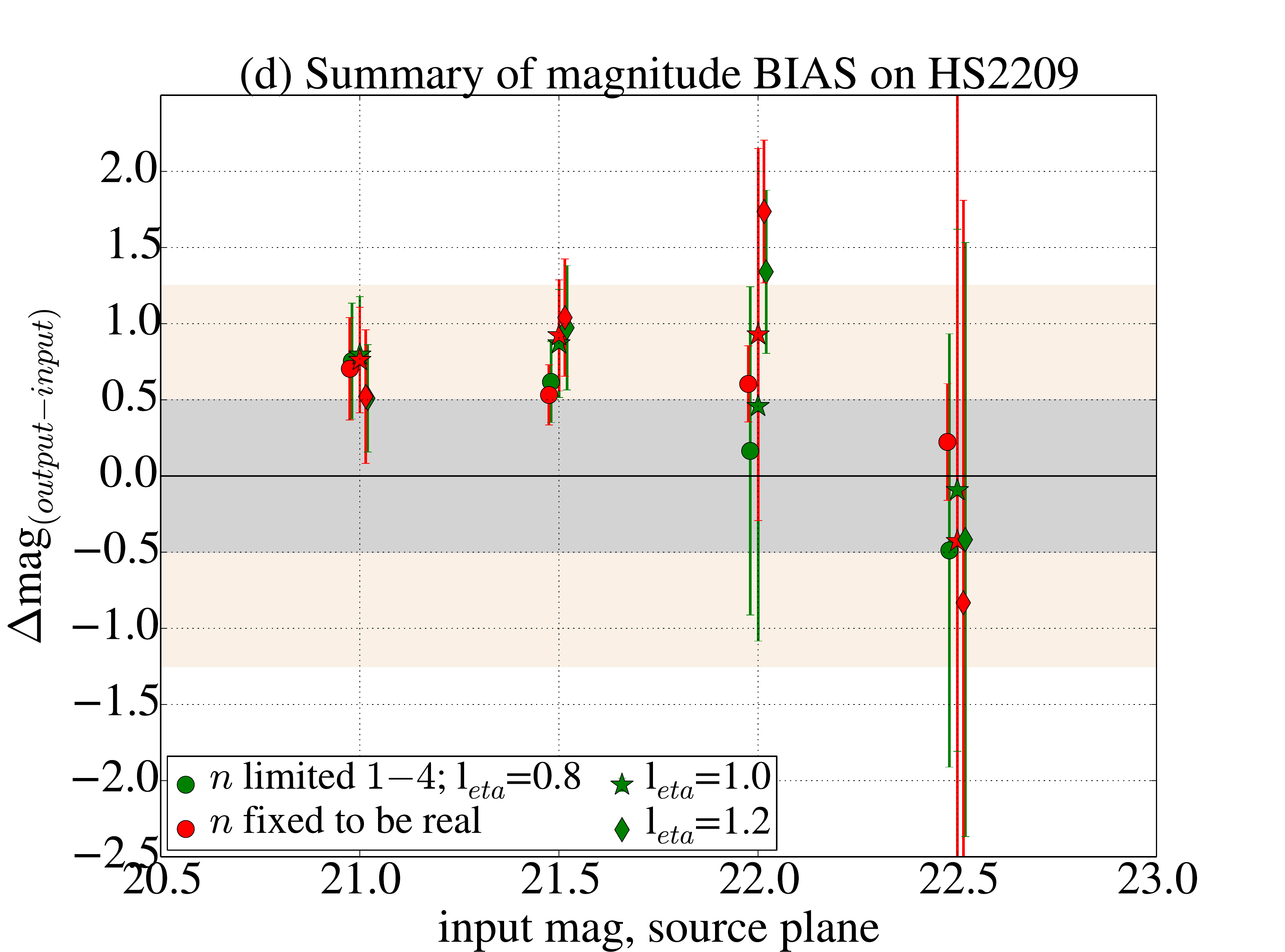}}\\
\end{tabular}
\caption{\label{fig:bias_HS2209} Results for HS2209. See caption of Fig.~\ref{fig:bias_HE1104} for details.}
\end{figure*}

It is worth noting that HS2209 is the most challenging system in our
sample with the AGN brightness approaching $15$ mag, in the image
plane. The results of HS2209 is summarized in
Fig.~\ref{fig:bias_HS2209} with generally larger bias than
SDSS0246. Nevertheless, in most cases, the bias levels are within our
target.

\subsubsection{HE0047}
\label{sssec:he0047}
\begin{figure*}
\centering
\begin{tabular}{c c}
{\includegraphics[width=9cm]{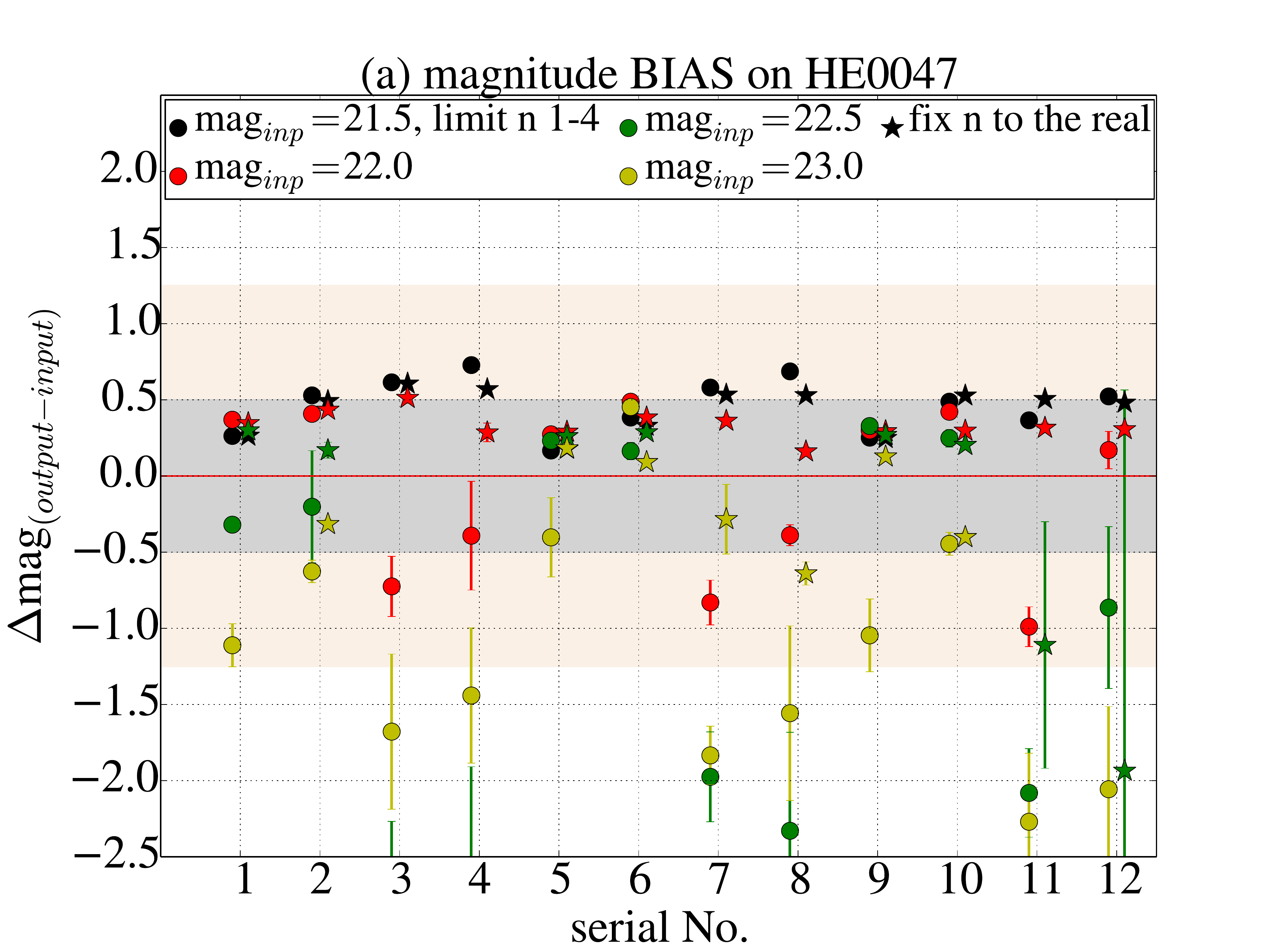}}&
{\includegraphics[width=9cm]{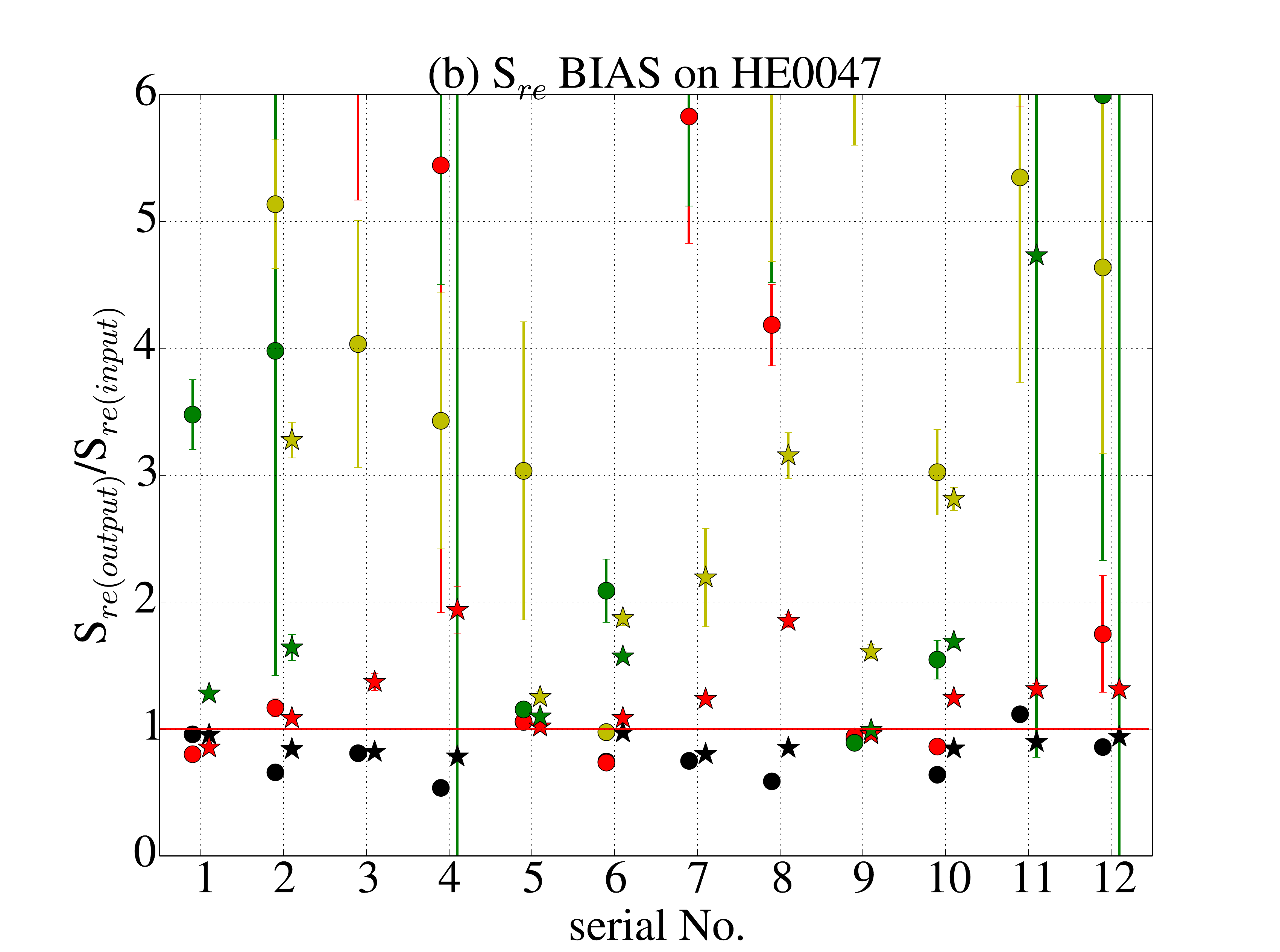}}\\
{\includegraphics[width=9cm]{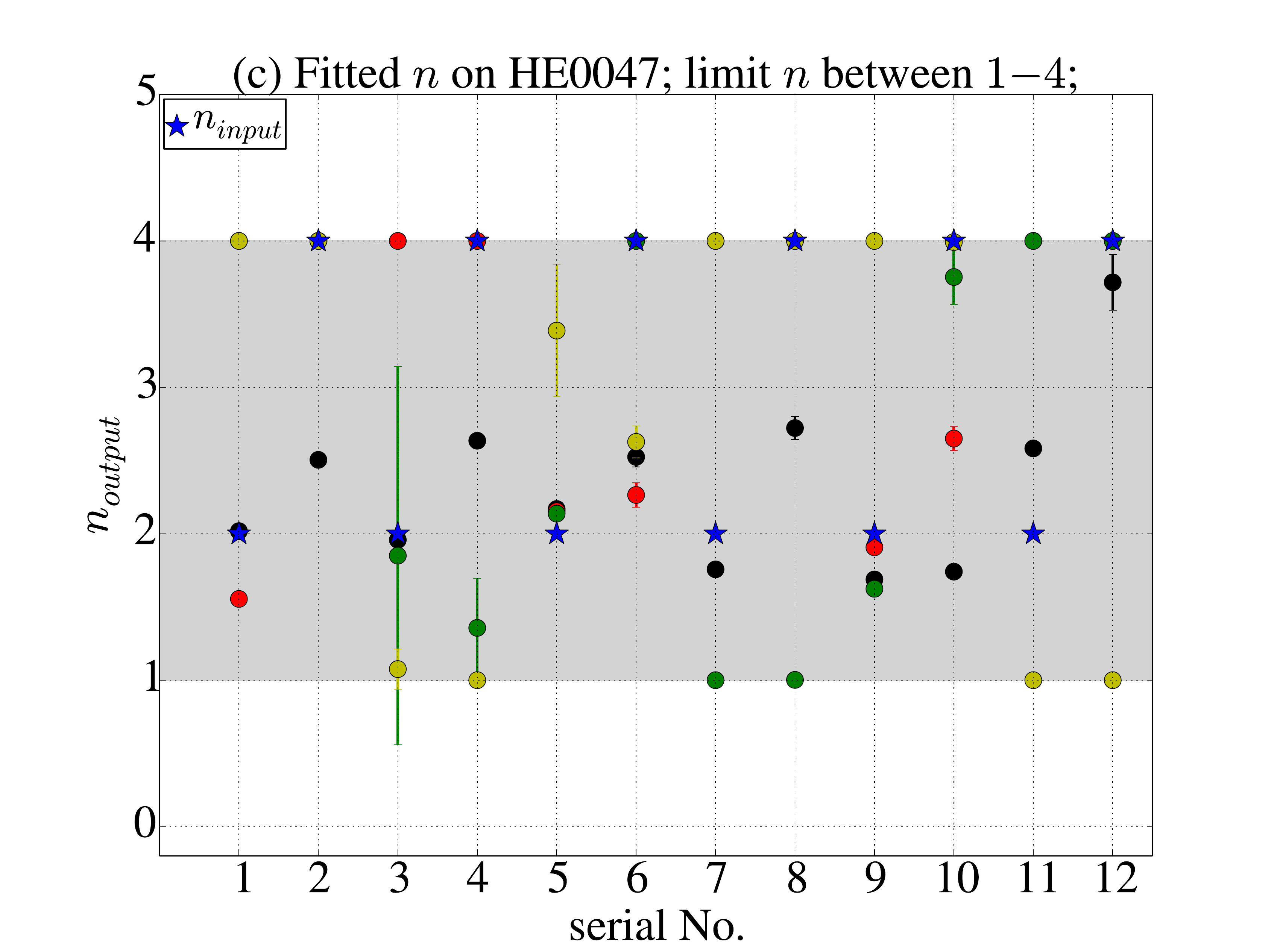}}&
{\includegraphics[width=9cm]{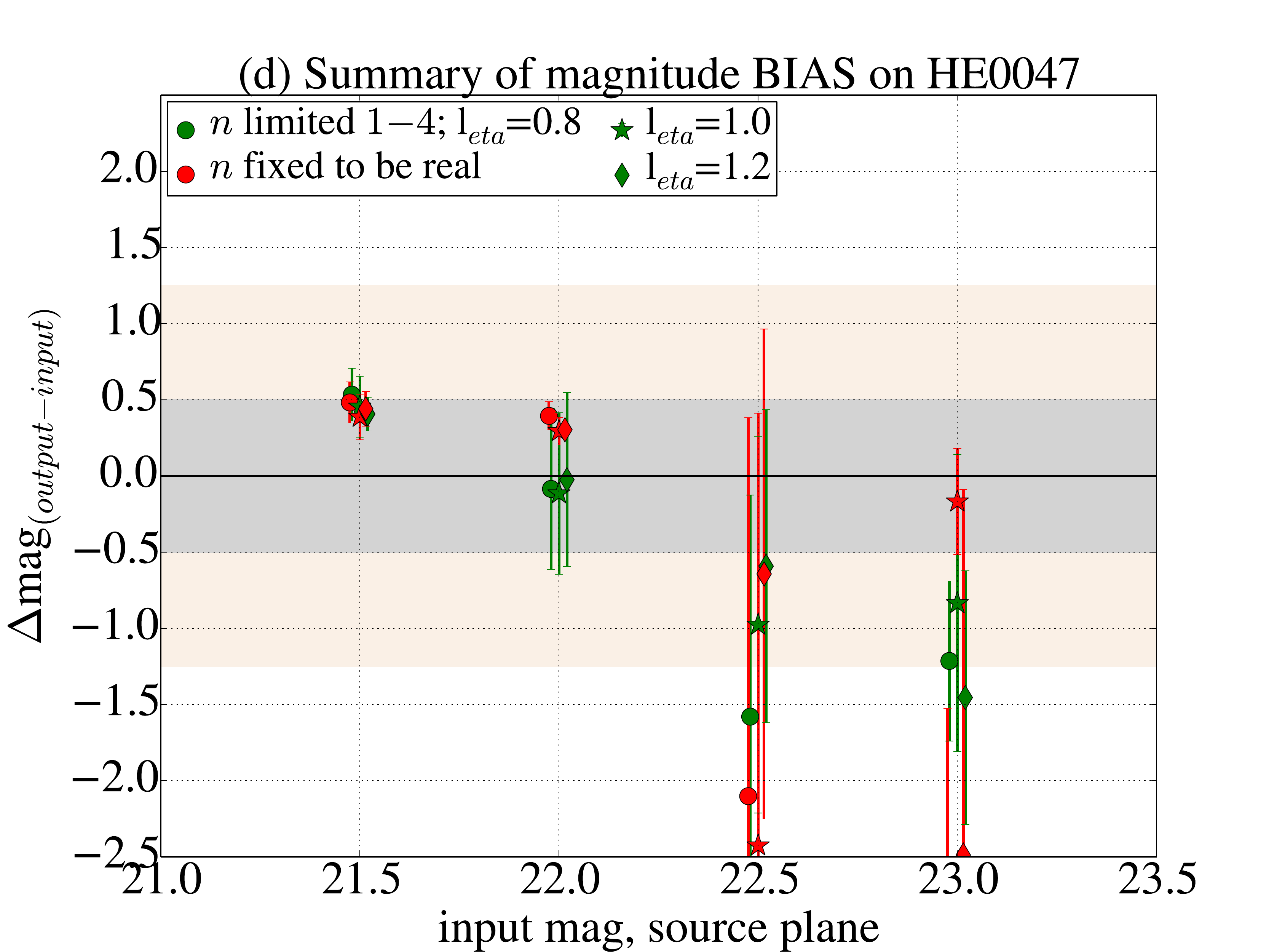}}\\
\end{tabular}
\caption{\label{fig:bias_HE0047} Results for HE0047. See caption of Fig.~\ref{fig:bias_HE1104} for details.}
\end{figure*}

The results for HE0047 are summarized in
Fig.~\ref{fig:bias_HE0047}. They are similar to those obtained for the
other doubles SDSS0246 and HS2209, and show a larger offset between
input and output model parameters than is seen in the quads. This is
easily understood because the quads have more information and in
general larger magnifications.

\subsubsection{RXJ1131}
\label{sssec:rxj1131}
\begin{figure*}
\centering
\begin{tabular}{c c}
{\includegraphics[width=9cm]{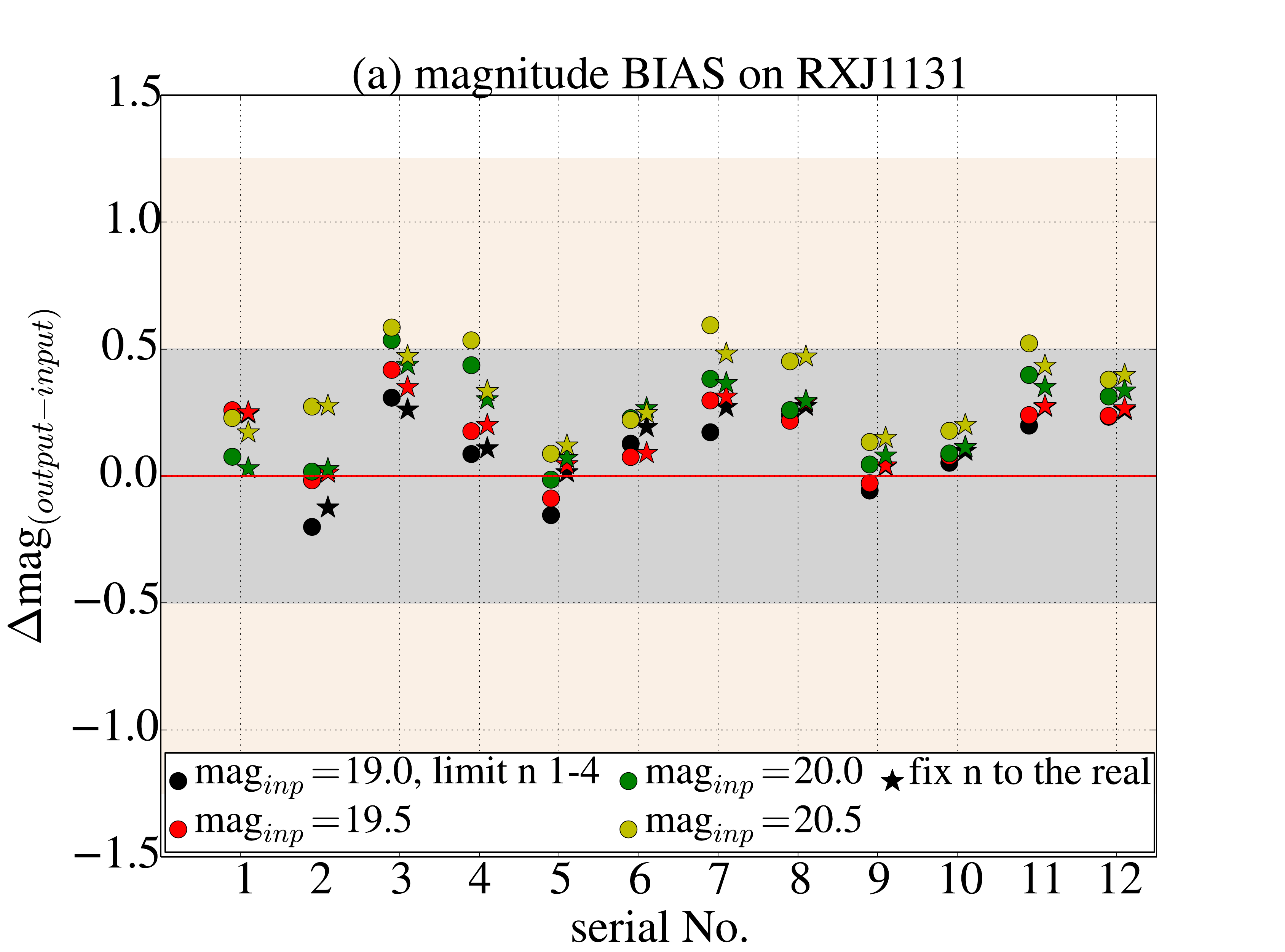}}&
{\includegraphics[width=9cm]{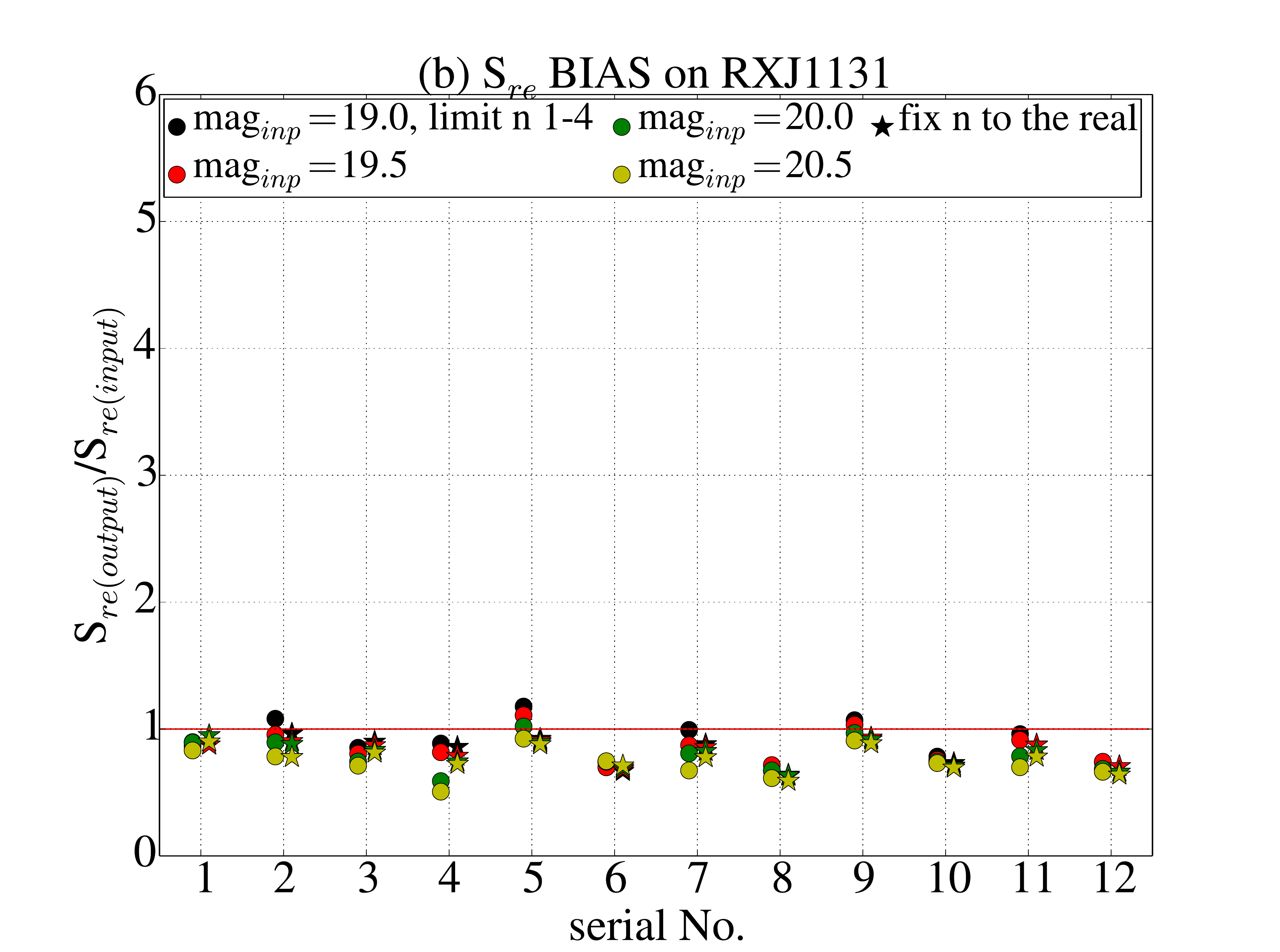}}\\
{\includegraphics[width=9cm]{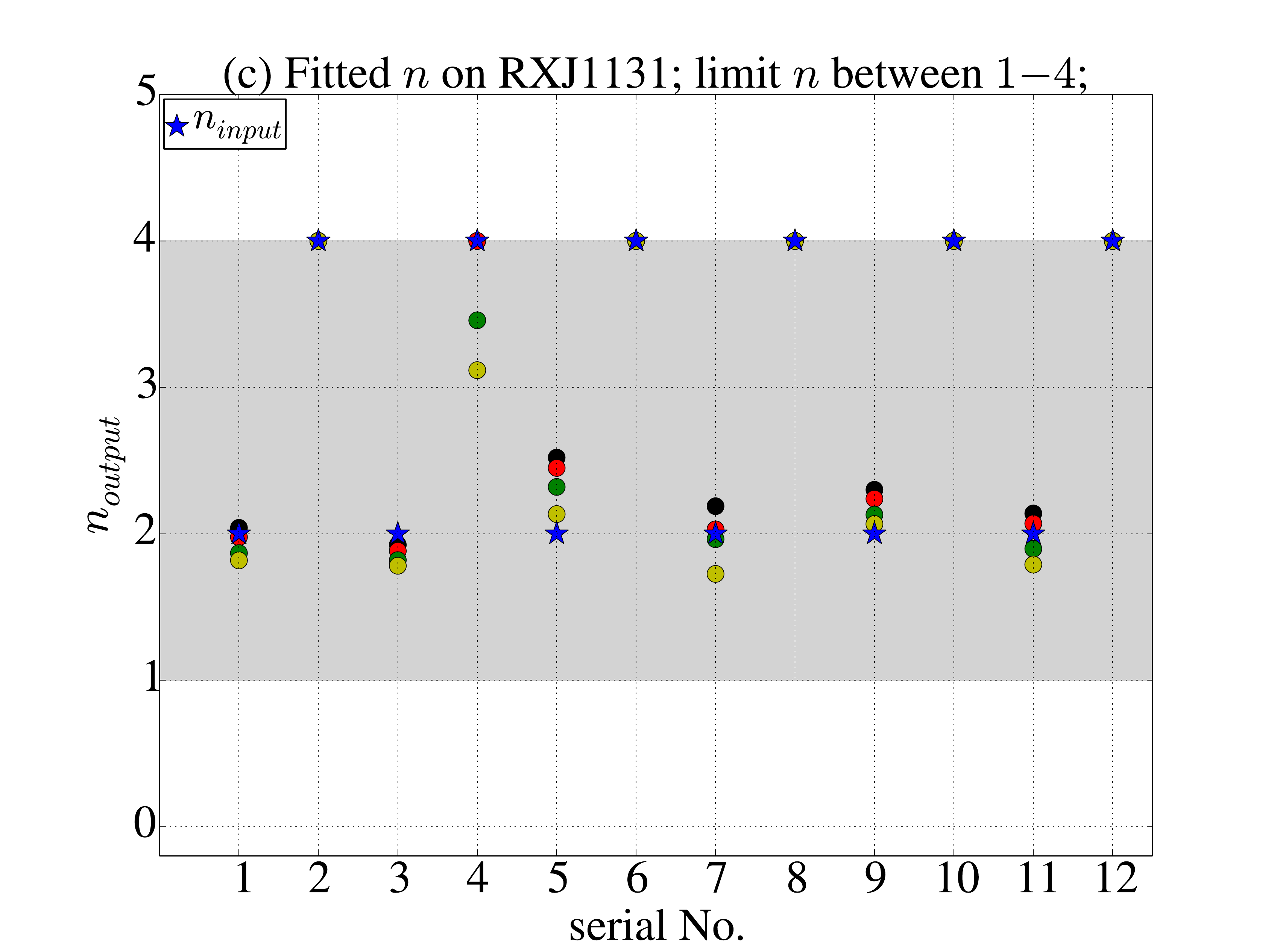}}&
{\includegraphics[width=9cm]{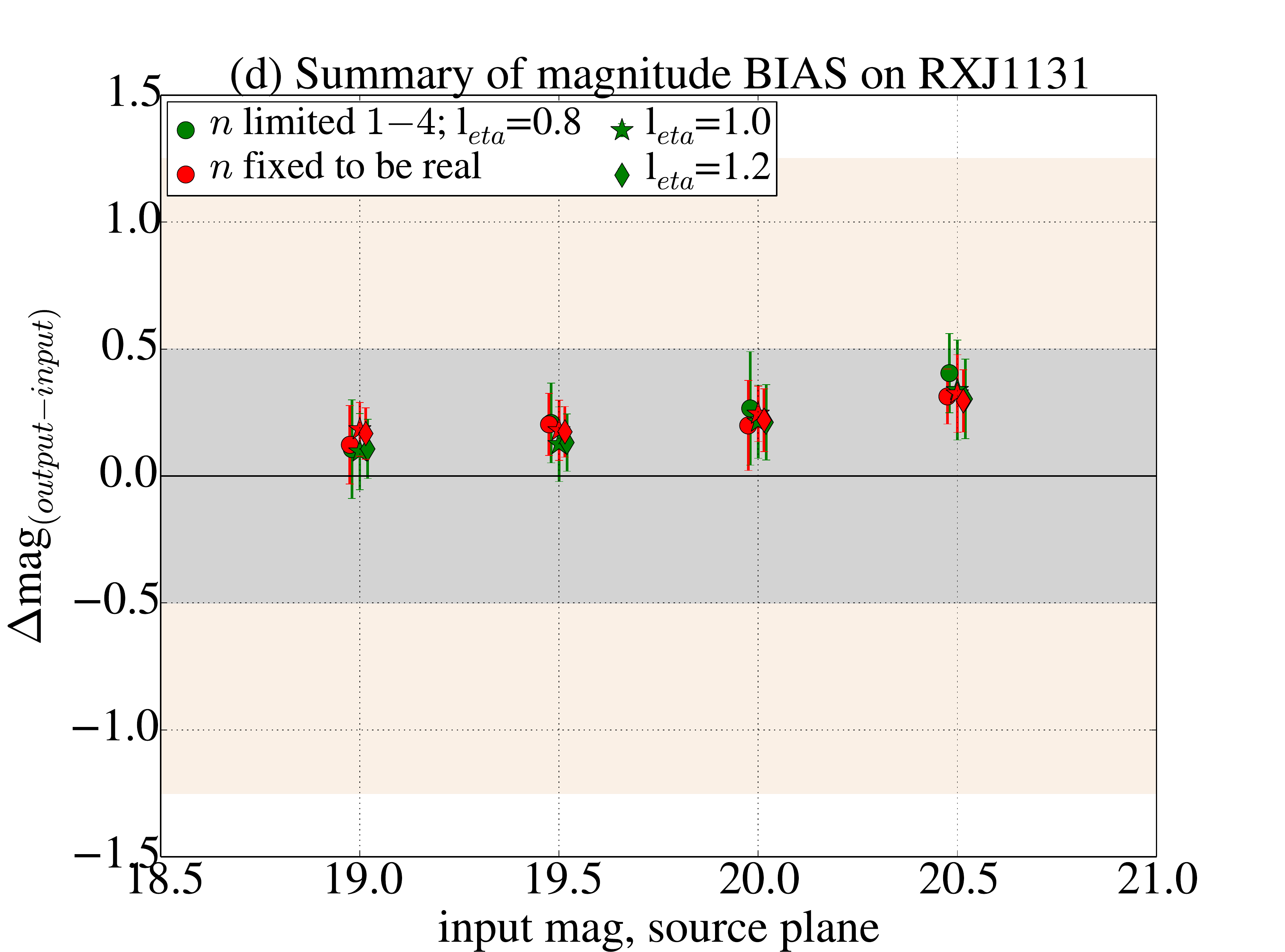}}\\
\end{tabular}
\caption{\label{fig:bias_RXJ1131} Results for RXJ1131. See caption of Fig.~\ref{fig:bias_HE1104} for details.}
\end{figure*}

RXJ1131 was observed with ACS through the F814W filter. A detailed
lens model with \glee\ has already been published by
\citet{Suy++13,Suy++14}. The host galaxy is much brighter than in the
other cases (see Table~\ref{para_config}). Thus, the results of
RXJ1131 have the smallest bias in the entire sample, as summarized in
Fig.~\ref{fig:bias_RXJ1131}.

\section{discussion}
\label{sec:disc}

Having analyzed each individual system we can now study the trends
across the sample to identify general lessons. We expect the magnitude
bias to depend on the total brightness of the lensed arc in image
plane, the contrast between the lensed arc and the AGN, and possibly
the effective surface brightness of the host galaxy. The latter
quantity is defined as the average surface brightness inside the
effective radius in the source plane, which should be approximately
independent of lensing magnification.

\begin{figure*}
\centering
\begin{tabular}{c c}
{\includegraphics[width=0.48\textwidth]{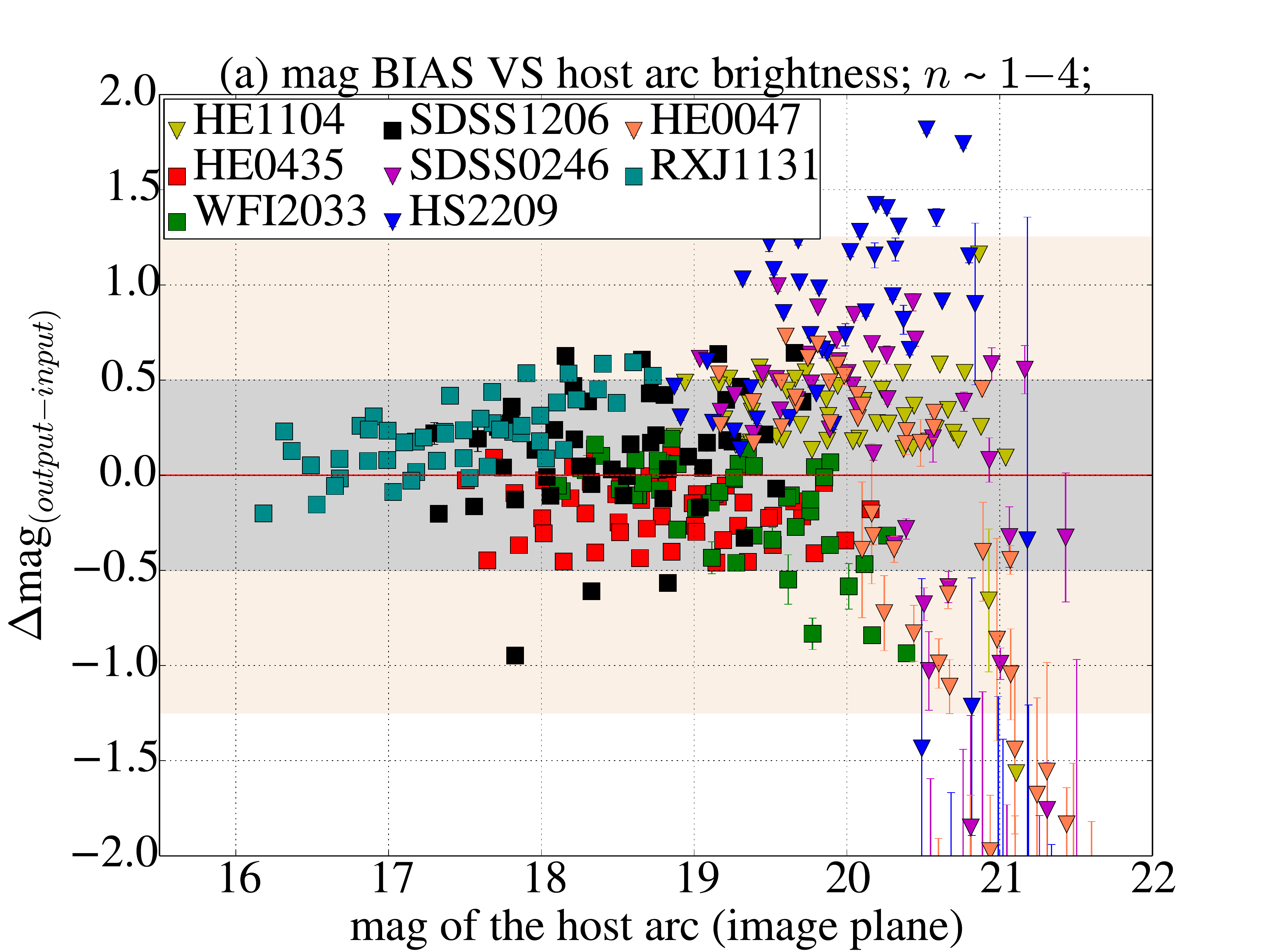}}&
{\includegraphics[width=0.48\textwidth]{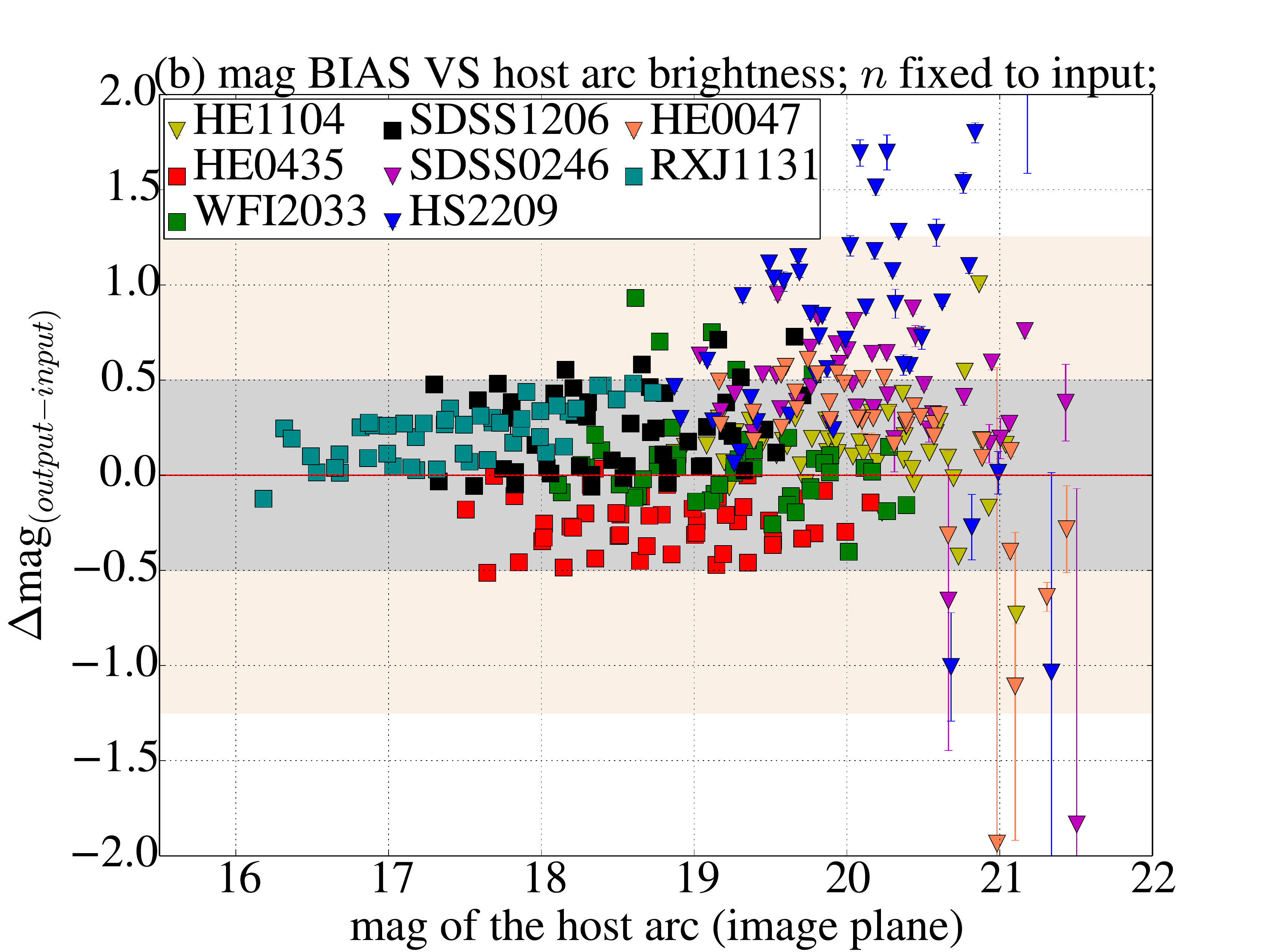}}\\
{\includegraphics[width=0.48\textwidth]{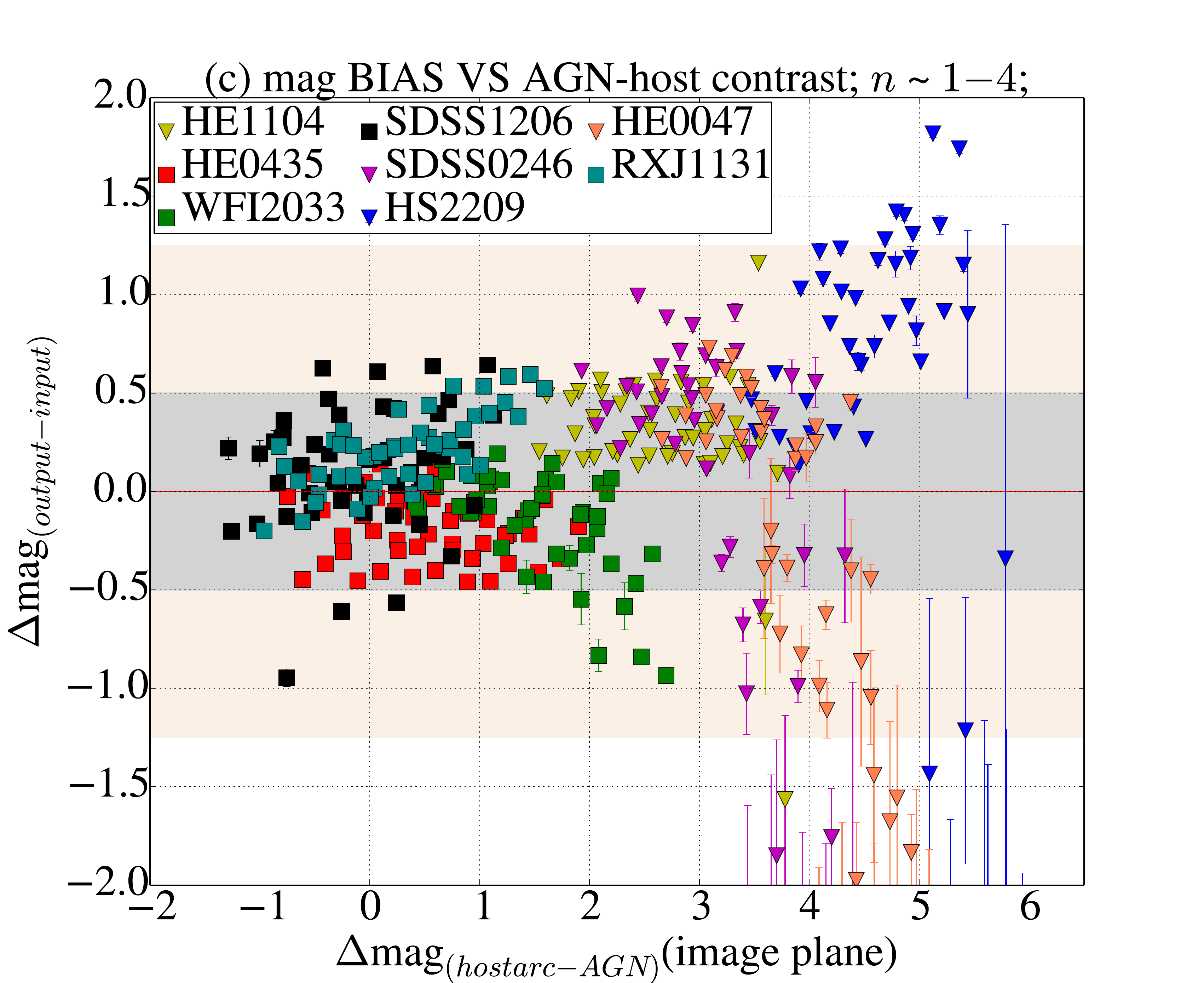}}&
{\includegraphics[width=0.48\textwidth]{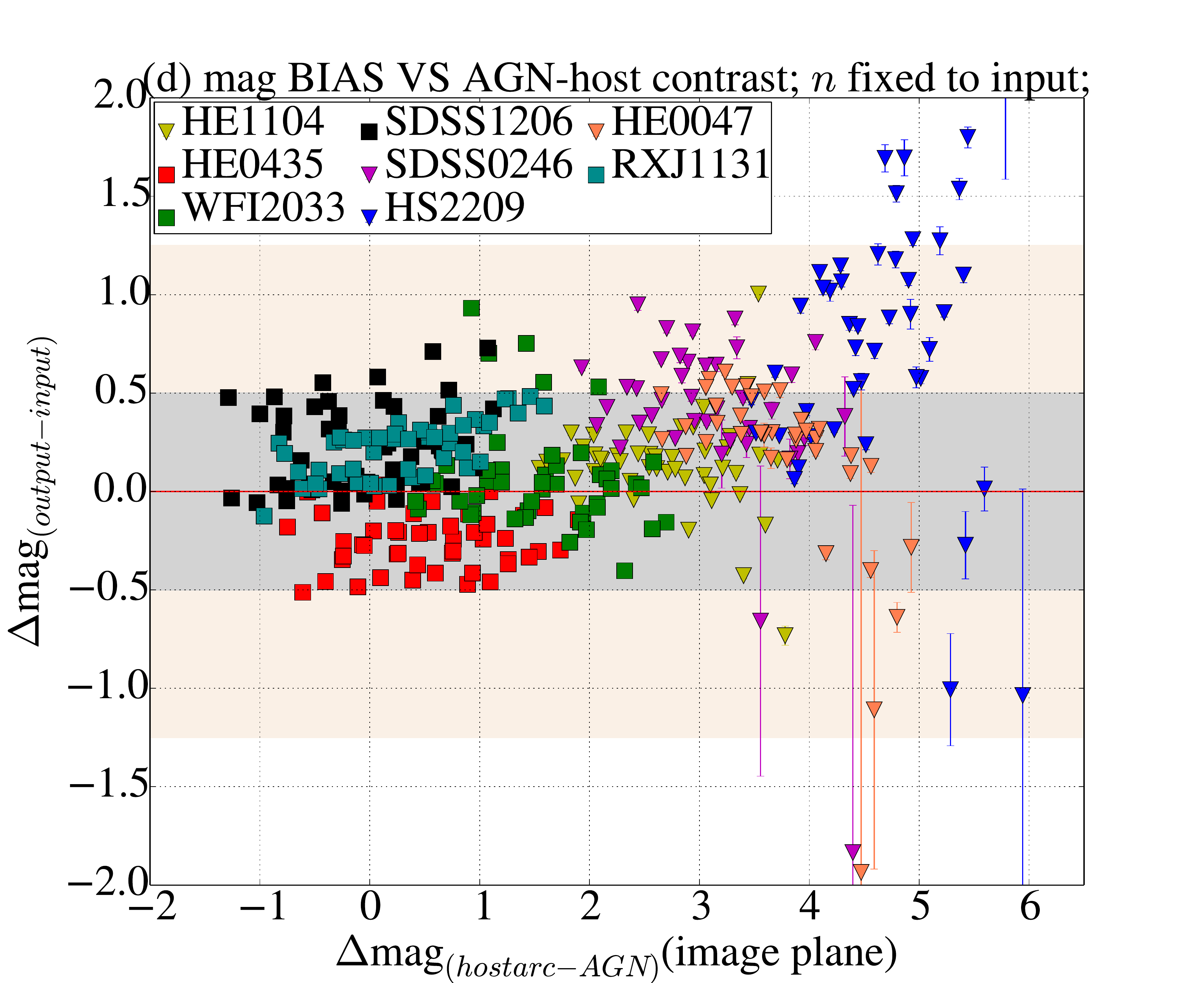}}\\
{\includegraphics[width=0.48\textwidth]{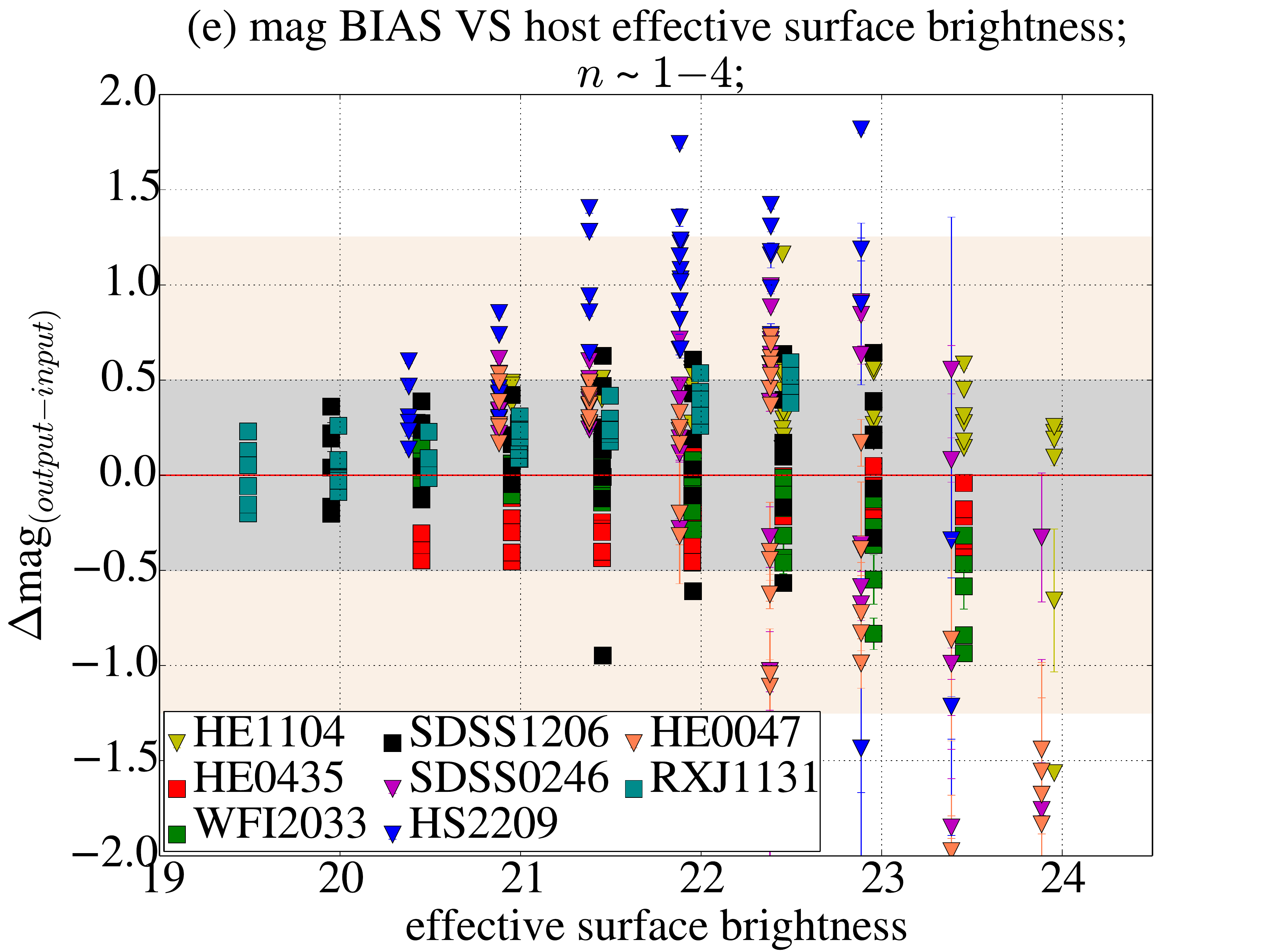}}&
{\includegraphics[width=0.48\textwidth]{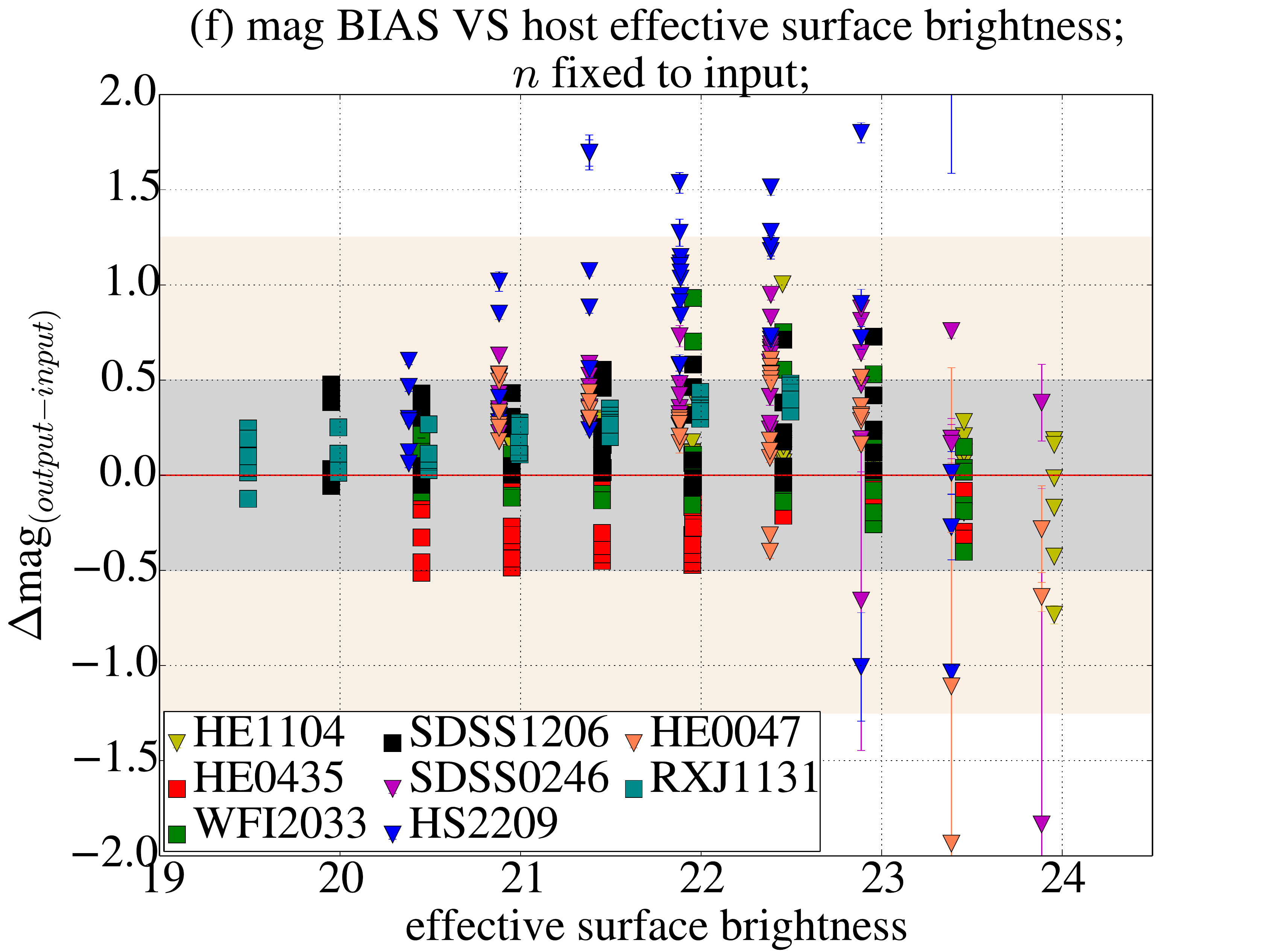}}\\
\end{tabular}
\caption{\label{fig:bias_hostarc} Magnitude bias across the sample as a function
of host arc magnitude in image plane (top row) and as a function of contrast
between the AGN and the host arc brightness (middle row), and effective surface brigthenss (bottom row). 
Different colors represent different systems in the sample. The square symbols
represent quadruply imaged sources while the triangles represent doubly-imaged system. In the left panels $n$ is left free within the range 1$-$4 during the fit; in the right panels it is kept fixed to the input value. 
} 
\end{figure*}

This intuition is supported by our results across the sample, as
illustrated in Fig.~\ref{fig:bias_hostarc} in which we relate the
magnitude bias level to the brightness of the host arc. (The total
magnitude of the host galaxy in the image plane is calculated by
re-simulating the images without the AGN and deflector light
(i.e. right panels in Fig.~\ref{fig:simufig}).) Indeed, the bias gets
larger than our target as magnitudes get fainter than $\sim20$, and it
is well within 0.5 mags for arc magnitudes brighter than $\sim$19. It
is also clear that the quads yield better results than the doubles, as
expected.  By comparing the six panels, it appears that magnitude of
the host galaxy and contrast between host and AGN are more important than
effective surface brightness of the host in driving the precision of
the recovery of the total magnitude. We note that except for RXJ1131
the exposure times are comparable for each system so that differences
across the sample are not driven by varying depth of the data, but
rather by the different properties of each system.

Besides the intrinsic bias from the source reconstruction and fitting,
additional bias can also be potentially introduced by the following
factors. First, the mask regions for modelling the arc are selected
manually. Of course, a different choice of mask region will produce a
different reconstructed image. However, we find that this difference
is negligible once the mask region is sufficiently large. Second, the
noise level in central AGN area was boosted to account for PSF
mismatch. This in general leads to underestimating the S\'ersic index
$n$. In practice, we find that this bias is not sufficient to affect
the inferred magnitudes beyond our target precisions, provided that
$n$ is allowed to be within a physically plausible range.

We note that in this work we have assumed that the PSF is known. The
analysis of the real systems to be presented in future work will have
to take into account additional uncertainties related on the host
galaxy magnitude arising from residual uncertainties in the PSF
reconstruction. This can be achieved by marginalizing over the
distribution of acceptable PSF reconstructions.

It is difficult to compare with \citet{Pen++06qsob} given the very
different data quality and inference methods. Three systems (HE0435,
HE1104, RXJ1131) are in common with their study and our simulations
suggest that indeed systematic errors should be smaller than their
reported 0.3 magnitudes uncertainty given our data quality.
Moreover, we find that their measurements agree with our estimated
range of magnitudes. The host magnitudes in the source plane for
HE1104 and HE0435 in \citet{Pen++06qsob} are measured to be 21.4 and
21.3 mag respectively (converting from Vega to AB), which are both within
our estimated range $21\sim22$ mag, taking into account our
approximate magnification\footnote{The approximate total magnification
in magnitudes (see Table~\ref{data_set}) is
estimated by calculating the magnification of the point source for an
isothermal mass model. For SDSS1206 the magnification of the host
galaxy is likely to be significantly larger, since it is quadruply
imaged.} (see Table~\ref{para_config}). The measured brightness of the AGN
in the source plane in their paper are 19.5 and 20.9 mag respectively
while our simulations range between $\sim$19.5 and 22.25. The
comparison for RXJ1131 is not straightforward because the analysis is
carried out in two different filters.

In this paper we have focused on the accuracy of the reconstruction of
the lensed host galaxy properties. A topic of great interest, but
beyond the scope of this paper, is how well the parameters of the
deflector are recovered. As shown by our analysis of HE0435
\citep{H0licow4} and RXJ1131 \citep{Suy++13}, the mass model is highly 
constrained by the high signal to noise ratio data available, especially when stellar velocity
dispersion, time delays, and other ancillary data are included \citep{D+K06a}.
Thus residual uncertainties in the lens model are unlikely to induce major
uncertainties in the lensed galaxy brightness. However, a full
assessment of this effect requires more extensive tests and
simulations and this is left for future work, when a blind lens
modelling challenge will be presented.

Finally, we conclude by noting that the selection function of our
sample of lensed quasars is non-trivial in terms of black hole mass
and host galaxy properties. Following the practice established for
non-lensed samples \citep[e.g.,][]{Tre++07,Lau++07,Park15}, it will have
to be modelled in order to reach general conclusions about the overall
population of AGN hosts from this sample.

\section{summary}
\label{sec:sum}
We carried out extensive and realistic simulations of lensed AGN
images taken with the \textit{Hubble Space Telescope}, in order to study the
fidelity of AGN host galaxy reconstruction using state of the art data and
lens reconstruction software. We simulated eight lensed AGN systems
(see Table \ref{data_set} for detailed information), with 48 sets of
parameters each covering the plausible physical range for the host
galaxy and lens galaxy properties. Reproducing as closely as possible
the analysis procedure of real data, we inferred the magnitude of the
AGN host galaxy and compared it to the input value. Our main
conclusions are:
\begin{enumerate}
\item Systematic errors in the reconstructed 
host galaxy luminosity are typically larger than random
errors. However, the level of systematic errors and biases are smaller
than the uncertainty of single epoch \mbh\ estimates ($\sim0.5$ dex,
i.e. 1.25 mag). In fact, for brighter host arc images
($F160W\lesssim19-20$ in the image plane) the level of bias and
systematic error is significantly smaller (within $0.5$ mag).
\item The magnitude bias depends on both the magnitude of the
host arc, and the contrast between the host arc and the AGN, as
expected.
\item The magnitude bias does not depend on the S\'ersic index of the host;
there is no significant difference between different S\'ersic index
(i.e. 2 and 4) inputs, in the range that may be expected for bright
AGN hosts.

\item The quadruply imaged systems tend to have smaller bias 
than the doubles, consistent with their larger magnification and
larger amount of available information.

\item For the fainter host galaxies (F160W$\gtrsim20$ mag in the 
image plane) systematic errors and bias can be larger than 1.25
magnitudes. Such cases may be difficult to analyze without introducing
informative priors to constrain the shape and size of the surface
brightness profile.

\end{enumerate}

We conclude that by using state of the art data and analysis tools the
magnitudes of lensed AGN host galaxies can be studied with
uncertainties that are smaller than those stemming from single epoch
black hole mass estimates. We will present the actual measurements in
forthcoming papers.

\section*{Acknowledgements}

We thank Thomas E. Collett for useful discussions and feedback on an
early draft of this manuscript.  We thank Roger Blandford, Vivien
Bonvin, Christopher D. Fassnacht, Yashar Hezaveh, Stefan Hilbert, Leon
Koopmans, John McKean, Georges Meylan, Danka Paraficz, Nicholas
Rumbaugh, Chiara Spiniello, Malte Tewes, Olga Tihhonova, and Simona
Vegetti for their contributions to the H0LiCOW project.
X.D. is supported by the China Scholarship
Council. T.T. acknowledges support by the Packard Foundations through
a Packard Research Fellowship and by the NSF through grants
AST-1450141 and AST-1412315.  S.H.S. gratefully acknowledges support
from the Max Planck Society through the Max Planck Research Group.
C.E.R. acknowledges support from the NSF grant AST-1312329.
D.S. acknowledges funding support from a {\it {Back to Belgium}} grant
from the Belgian Federal Science Policy (BELSPO).  A.S. acknowledges
support by World Premier International Research Center Initiative
(WPI), MEXT, Japan. K.C.W. is supported by an EACOA Fellowship awarded
by the East Asia Core Observatories Association, which consists of the
Academia Sinica Institute of Astronomy and Astrophysics, the National
Astronomical Observatory of Japan, the National Astronomical Observatories
of the Chinese Academy of Sciences, and the Korea Astronomy and Space Science Institute.

Based on observations made with the NASA/ESA Hubble Space Telescope,
obtained at the Space Telescope Science Institute, which is operated
by the Association of Universities for Research in Astronomy, Inc.,
under NASA contract NAS 5-26555. These observations are associated
with programs \# 9744, 12889, 14254. Financial support was provided by
NASA through grants from the Space Telescope Science
Institute.




\bibliographystyle{mnras}
\input{paper_sim_I_submission2.bbl}








\bsp	
\label{lastpage}
\end{document}